\documentclass[apjl,iop]{emulateapj}
\usepackage{comment}
\usepackage{ifthen}
\usepackage[switch]{lineno}

\newcommand{\forloop}[5][1]%
{%
\setcounter{#2}{#3}%
\ifthenelse{#4}%
	{%
	#5%
	\addtocounter{#2}{#1}%
	\forloop[#1]{#2}{\value{#2}}{#4}{#5}%
	}%
	{%
	}%
}%

\newcommand{\ctbd}[1]{}

\newcommand{\lc}{light curve}
\newcommand{\lcs}{light curves}
\newcommand{\Lc}{Light curve}

\newcommand{\masy}{\ensuremath{\rm mas\,yr^{-1}}}
\newcommand{\kms}{\ensuremath{\rm km\,s^{-1}}}
\newcommand{\ms}{\ensuremath{\rm m\,s^{-1}}}

\newcommand{\gcmc}{\ensuremath{\rm g\,cm^{-3}}}

\newcommand{\teff}{\ensuremath{T_{\rm eff}}}
\newcommand{\logg}{\ensuremath{\log{g}}}
\newcommand{\vsini}{\ensuremath{v \sin{i}}}
\newcommand{\feh}{\ensuremath{\rm [Fe/H]}}

\newcommand{\vmac}{\ensuremath{v_{\rm mac}}}
\newcommand{\vmic}{\ensuremath{v_{\rm mic}}}

\newcommand{\rsun}{\ensuremath{R_\sun}}
\newcommand{\msun}{\ensuremath{M_\sun}}
\newcommand{\lsun}{\ensuremath{L_\sun}}

\newcommand{\rstar}{\ensuremath{R_\star}}
\newcommand{\mstar}{\ensuremath{M_\star}}
\newcommand{\lstar}{\ensuremath{L_\star}}

\newcommand{\teffstar}{\ensuremath{T_{\rm eff\star}}}
\newcommand{\rhostar}{\ensuremath{\rho_\star}}
\newcommand{\loggstar}{\ensuremath{\log{g_{\star}}}}

\newcommand{\mearth}{\ensuremath{M_\earth}}

\newcommand{\rpl}{\ensuremath{R_{p}}}
\newcommand{\mpl}{\ensuremath{M_{p}}}

\newcommand{\rhopl}{\ensuremath{\rho_{p}}}

\newcommand{\arstar}{\ensuremath{a/\rstar}}
\newcommand{\zrstar}{\ensuremath{\zeta/\rstar}}

\newcommand{\rjup}{\ensuremath{R_{\rm J}}}
\newcommand{\mjup}{\ensuremath{M_{\rm J}}}

\newcommand{\reffigl}[1]{Figure~\ref{fig:#1}}
\newcommand{\refsecl}[1]{\mbox{Section \ref{sec:#1}}}

\newcommand{\reftabl}[1]{Table~\ref{tab:#1}}

\newcommand{\hj}{hot Jupiter }
\newcommand{\hjs}{hot Jupiters }

\newcommand{\loopand}{\ifnum\value{planetcounter}=2 and \else\fi}
\newcommand{\loopcomma}{\ifnum\value{planetcounter}<2 ,\else. \fi}
\newcommand{\loopcommanoperiod}{\ifnum\value{planetcounter}<2 ,\else \space\fi}
\newcommand{\loopcommanospace}{\ifnum\value{planetcounter}<2 ,\else \fi}

\newcommand{\hatcurhtrxxxxxA}{HATS610-015}                             
\newcommand{\hatcurfieldxxxxxA}{\ensuremath{string}}                   
\newcommand{\hatcurCCraxxxxxA}{\ensuremath{11^{\mathrm h}36^{\mathrm m}02.16{\mathrm s}}}                            
\newcommand{\hatcurCCdecxxxxxA}{\ensuremath{-29{\arcdeg}32{\arcmin}35.9{\arcsec}}}                           
\newcommand{\hatcurCCmagxxxxxA}{13.455}                                
\newcommand{\hatcurCCtwomassxxxxxA}{2MASS~11360233-2932359}            
\newcommand{\hatcurCCgscxxxxxA}{GSC~6664-00373}                        
\newcommand{\hatcurCCtassmvxxxxxA}{\ensuremath{13.455\pm0.040}}        
\newcommand{\hatcurCCtassmvshortxxxxxA}{\ensuremath{13.5}}             
\newcommand{\hatcurCCtassmBxxxxxA}{\ensuremath{14.496\pm0.040}}        
\newcommand{\hatcurCCtassmBshortxxxxxA}{\ensuremath{14.5}}             
\newcommand{\hatcurCCtassmIxxxxxA}{\ensuremath{100\pm1000}}            
\newcommand{\hatcurCCtassmIshortxxxxxA}{\ensuremath{100.0}}            
\newcommand{\hatcurCCtassmgxxxxxA}{\ensuremath{13.943\pm0.030}}        
\newcommand{\hatcurCCtassmgshortxxxxxA}{\ensuremath{13.9}}             
\newcommand{\hatcurCCtassmrxxxxxA}{\ensuremath{13.056\pm0.030}}        
\newcommand{\hatcurCCtassmrshortxxxxxA}{\ensuremath{13.1}}             
\newcommand{\hatcurCCtassmixxxxxA}{\ensuremath{12.82\pm0.14}}          
\newcommand{\hatcurCCtassmishortxxxxxA}{\ensuremath{12.8}}             
\newcommand{\hatcurCCtwomassJmagxxxxxA}{\ensuremath{11.556\pm0.023}}   
\newcommand{\hatcurCCtwomassHmagxxxxxA}{\ensuremath{11.006\pm0.022}}   
\newcommand{\hatcurCCtwomassKmagxxxxxA}{\ensuremath{10.942\pm0.019}}   
\newcommand{\hatcurCCcitJmagxxxxxA}{\ensuremath{11.560\pm0.024}}       
\newcommand{\hatcurCCcitHmagxxxxxA}{\ensuremath{11.001\pm0.023}}       
\newcommand{\hatcurCCcitKmagxxxxxA}{\ensuremath{10.966\pm0.020}}       
\newcommand{\hatcurCCbbJmagxxxxxA}{\ensuremath{11.629\pm0.026}}        
\newcommand{\hatcurCCbbHmagxxxxxA}{\ensuremath{11.023\pm0.024}}        
\newcommand{\hatcurCCbbKmagxxxxxA}{\ensuremath{10.986\pm0.020}}        
\newcommand{\hatcurCCesoJmagxxxxxA}{\ensuremath{11.634\pm0.028}}       
\newcommand{\hatcurCCesoHmagxxxxxA}{\ensuremath{11.016\pm0.026}}       
\newcommand{\hatcurCCesoKmagxxxxxA}{\ensuremath{10.984\pm0.021}}       
\newcommand{\hatcurCCesoJHmagxxxxxA}{\ensuremath{0.618\pm0.036}}       
\newcommand{\hatcurCCesoJKmagxxxxxA}{\ensuremath{0.651\pm0.034}}       
\newcommand{\hatcurCCesoHKmagxxxxxA}{\ensuremath{0.032\pm0.033}}       
\newcommand{\hatcurLCdipxxxxxA}{\ensuremath{23.8}}                     
\newcommand{\hatcurLCrprstarxxxxxA}{\ensuremath{0.1445\pm0.0019}}      
\newcommand{\hatcurLCbsqxxxxxA}{\ensuremath{0.332_{-0.029}^{+0.029}}}  
\newcommand{\hatcurLCimpxxxxxA}{\ensuremath{0.576_{-0.026}^{+0.025}}}  
\newcommand{\hatcurLCzetaxxxxxA}{\ensuremath{26.06\pm0.19}}            
\newcommand{\hatcurLCdurxxxxxA}{\ensuremath{0.09287\pm0.00092}}        
\newcommand{\hatcurLCdurshortxxxxxA}{\ensuremath{0.0929}}              
\newcommand{\hatcurLCdurhrxxxxxA}{\ensuremath{2.229\pm0.022}}          
\newcommand{\hatcurLCdurhrshortxxxxxA}{\ensuremath{2.229}}             
\newcommand{\hatcurLCqxxxxxA}{\ensuremath{0.01970\pm0.00020}}          
\newcommand{\hatcurLCqshortxxxxxA}{\ensuremath{0.020}}                 
\newcommand{\hatcurLCingdurxxxxxA}{\ensuremath{0.01673\pm0.00088}}     
\newcommand{\hatcurLCPxxxxxA}{\ensuremath{4.7228112\pm0.0000051}}      
\newcommand{\hatcurLCPprecxxxxxA}{\ensuremath{4.7228112}}              
\newcommand{\hatcurLCPshortxxxxxA}{\ensuremath{4.7228}}                
\newcommand{\hatcurLCTxxxxxA}{\ensuremath{2457064.41181\pm0.00024}}    
\newcommand{\hatcurLCTAxxxxxA}{\ensuremath{2455671.1825\pm0.0015}}     
\newcommand{\hatcurLCTBxxxxxA}{\ensuremath{2457111.63992\pm0.00026}}   
\newcommand{\hatcurLChatnetmxxxxxA}{\ensuremath{13.130850\pm0.000069}} 
\newcommand{\hatcurLCiblendxxxxxA}{\ensuremath{0.861\pm0.030}}         
\newcommand{\hatcurLCrhoxxxxxA}{\ensuremath{3.89\pm0.47}}              
\newcommand{\hatcurSMEiteffxxxxxA}{\ensuremath{4771\pm80}}             
\newcommand{\hatcurSMEizfehxxxxxA}{\ensuremath{0.000\pm0.060}}         
\newcommand{\hatcurSMEizfehshortxxxxxA}{\ensuremath{0.00}}             
\newcommand{\hatcurSMEiloggxxxxxA}{\ensuremath{4.58\pm0.16}}           
\newcommand{\hatcurSMEivsinxxxxxA}{\ensuremath{0.5\pm1.7}}             
\newcommand{\hatcurSMEivmacxxxxxA}{\ensuremath{0.0}}                   
\newcommand{\hatcurSMEivmicxxxxxA}{\ensuremath{0.0}}                   
\newcommand{\hatcurSMEiiteffxxxxxA}{\ensuremath{4803\pm55}}            
\newcommand{\hatcurSMEiizfehxxxxxA}{\ensuremath{0.000\pm0.040}}        
\newcommand{\hatcurSMEiizfehshortxxxxxA}{\ensuremath{0.0}}             
\newcommand{\hatcurSMEiiloggxxxxxA}{\ensuremath{4.664\pm0.021}}        
\newcommand{\hatcurSMEiivsinxxxxxA}{\ensuremath{0.50\pm0.50}}          
\newcommand{\hatcurLBizxxxxxA}{\ensuremath{0.3472}}                    
\newcommand{\hatcurLBiizxxxxxA}{\ensuremath{0.2502}}                   
\newcommand{\hatcurLBiixxxxxA}{\ensuremath{0.4355}}                    
\newcommand{\hatcurLBiiixxxxxA}{\ensuremath{0.2255}}                   
\newcommand{\hatcurLBiIxxxxxA}{\ensuremath{0.4055}}                    
\newcommand{\hatcurLBiiIxxxxxA}{\ensuremath{0.2342}}                   
\newcommand{\hatcurLBigxxxxxA}{\ensuremath{0.8412}}                    
\newcommand{\hatcurLBiigxxxxxA}{\ensuremath{-0.0025}}                  
\newcommand{\hatcurLBirxxxxxA}{\ensuremath{0.5808}}                    
\newcommand{\hatcurLBiirxxxxxA}{\ensuremath{0.1721}}                   
\newcommand{\hatcurLBiRxxxxxA}{\ensuremath{0.5404}}                    
\newcommand{\hatcurLBiiRxxxxxA}{\ensuremath{0.1879}}                   
\newcommand{\hatcurLBikepxxxxxA}{\ensuremath{0.1000}}                  
\newcommand{\hatcurLBiikepxxxxxA}{\ensuremath{0.1000}}                 
\newcommand{\hatcurISOmxxxxxA}{\ensuremath{0.758\pm0.018}}             
\newcommand{\hatcurISOmshortxxxxxA}{\ensuremath{0.76}}                 
\newcommand{\hatcurISOmlongxxxxxA}{\ensuremath{0.758\pm0.018}}         
\newcommand{\hatcurISOrxxxxxA}{\ensuremath{0.676\pm0.017}}             
\newcommand{\hatcurISOrshortxxxxxA}{\ensuremath{0.68}}                 
\newcommand{\hatcurISOrlongxxxxxA}{\ensuremath{0.676\pm0.017}}         
\newcommand{\hatcurISOrhoxxxxxA}{\ensuremath{3.47\pm0.22}}             
\newcommand{\hatcurISOrholongxxxxxA}{\ensuremath{3.47\pm0.22}}         
\newcommand{\hatcurISOloggxxxxxA}{\ensuremath{4.660\pm0.018}}          
\newcommand{\hatcurISOlumxxxxxA}{\ensuremath{0.214\pm0.018}}           
\newcommand{\hatcurISOlumshortxxxxxA}{\ensuremath{0.21}}               
\newcommand{\hatcurISOmvxxxxxA}{\ensuremath{6.79\pm0.12}}              
\newcommand{\hatcurISOvixxxxxA}{\ensuremath{0.999\pm0.028}}            
\newcommand{\hatcurISOagexxxxxA}{\ensuremath{1.7_{-1.4}^{+4.4}}}       
\newcommand{\hatcurISOsigmaxxxxxA}{\ensuremath{0.00900\pm0.00081}}     
\newcommand{\hatcurISOMJxxxxxA}{\ensuremath{5.033\pm0.077}}            
\newcommand{\hatcurISOMHxxxxxA}{\ensuremath{4.497\pm0.068}}            
\newcommand{\hatcurISOMKxxxxxA}{\ensuremath{4.406\pm0.066}}            
\newcommand{\hatcurISOJKxxxxxA}{\ensuremath{0.630\pm0.010}}            
\newcommand{\hatcurISOspecxxxxxA}{K}                                   
\newcommand{\hatcurRVKxxxxxA}{\ensuremath{411\pm17}}                   
\newcommand{\hatcurRVrkxxxxxA}{\ensuremath{0\pm0}}                     
\newcommand{\hatcurRVrhxxxxxA}{\ensuremath{0\pm0}}                     
\newcommand{\hatcurRVkxxxxxA}{\ensuremath{0\pm0}}                      
\newcommand{\hatcurRVhxxxxxA}{\ensuremath{0\pm0}}                      
\newcommand{\hatcurRVtronexxxxxA}{\ensuremath{0\pm0}}                  
\newcommand{\hatcurRVtrtwoxxxxxA}{\ensuremath{0\pm0}}                  
\newcommand{\hatcurRVgammaAxxxxxA}{\ensuremath{-7442\pm28}}            
\newcommand{\hatcurRVjitterAxxxxxA}{\ensuremath{27\pm40}}              
\newcommand{\hatcurRVjittertwosiglimAxxxxxA}{\ensuremath{<126.6}}      
\newcommand{\hatcurRVfitrmsAxxxxxA}{\ensuremath{0.0}}                  
\newcommand{\hatcurRVgammaBxxxxxA}{\ensuremath{-7422\pm22}}            
\newcommand{\hatcurRVjitterBxxxxxA}{\ensuremath{41\pm24}}              
\newcommand{\hatcurRVjittertwosiglimBxxxxxA}{\ensuremath{<88.9}}       
\newcommand{\hatcurRVfitrmsBxxxxxA}{\ensuremath{0.0}}                  
\newcommand{\hatcurRVgammaCxxxxxA}{\ensuremath{-7374\pm45}}            
\newcommand{\hatcurRVjitterCxxxxxA}{\ensuremath{60\pm59}}              
\newcommand{\hatcurRVjittertwosiglimCxxxxxA}{\ensuremath{<192.3}}      
\newcommand{\hatcurRVfitrmsCxxxxxA}{\ensuremath{0.0}}                  
\newcommand{\hatcurRVeccenxxxxxA}{\ensuremath{0\pm0}}                  
\newcommand{\hatcurRVeccentwosiglimxxxxxA}{\ensuremath{<0.000}}        
\newcommand{\hatcurRVomegaxxxxxA}{\ensuremath{0\pm0}}                  
\newcommand{\hatcurPPixxxxxA}{\ensuremath{87.93\pm0.13}}               
\newcommand{\hatcurPPgxxxxxA}{\ensuremath{77.5\pm5.6}}                 
\newcommand{\hatcurPPloggxxxxxA}{\ensuremath{3.889\pm0.031}}           
\newcommand{\hatcurPParxxxxxA}{\ensuremath{16.00\pm0.34}}              
\newcommand{\hatcurPParelxxxxxA}{\ensuremath{0.05023\pm0.00041}}       
\newcommand{\hatcurPPrhoxxxxxA}{\ensuremath{4.08\pm0.41}}              
\newcommand{\hatcurPPmxxxxxA}{\ensuremath{2.83\pm0.12}}                
\newcommand{\hatcurPPmshortxxxxxA}{\ensuremath{2.83}}                  
\newcommand{\hatcurPPmlongxxxxxA}{\ensuremath{2.83\pm0.12}}            
\newcommand{\hatcurPPmexxxxxA}{\ensuremath{899\pm40}}                  
\newcommand{\hatcurPPmeshortxxxxxA}{\ensuremath{898.5}}                
\newcommand{\hatcurPPmelongxxxxxA}{\ensuremath{899\pm40}}              
\newcommand{\hatcurPPrxxxxxA}{\ensuremath{0.950\pm0.031}}              
\newcommand{\hatcurPPrshortxxxxxA}{\ensuremath{0.95}}                  
\newcommand{\hatcurPPrlongxxxxxA}{\ensuremath{0.950\pm0.031}}          
\newcommand{\hatcurPPrexxxxxA}{\ensuremath{10.65\pm0.35}}              
\newcommand{\hatcurPPreshortxxxxxA}{\ensuremath{10.7}}                 
\newcommand{\hatcurPPrelongxxxxxA}{\ensuremath{10.65\pm0.35}}          
\newcommand{\hatcurPPmrcorrxxxxxA}{\ensuremath{0.21}}                  
\newcommand{\hatcurPPteffxxxxxA}{\ensuremath{846\pm15}}                
\newcommand{\hatcurPPthetaxxxxxA}{\ensuremath{0.393\pm0.020}}          
\newcommand{\hatcurPPfluxperixxxxxA}{\ensuremath{1.158\pm0.085}}       
\newcommand{\hatcurPPfluxperidimxxxxxA}{\ensuremath{8}}                
\newcommand{\hatcurPPfluxapxxxxxA}{\ensuremath{1.158\pm0.085}}         
\newcommand{\hatcurPPfluxapdimxxxxxA}{\ensuremath{8}}                  
\newcommand{\hatcurPPfluxavgxxxxxA}{\ensuremath{1.158\pm0.085}}        
\newcommand{\hatcurPPfluxavgdimxxxxxA}{\ensuremath{8}}                 
\newcommand{\hatcurPPfluxavglogxxxxxA}{\ensuremath{8.064\pm0.032}}     
\newcommand{\hatcurXsecphasexxxxxA}{\ensuremath{0\pm0}}                
\newcommand{\hatcurXsecondaryxxxxxA}{\ensuremath{2457066.77321\pm0.00024}} 
\newcommand{\hatcurXsecdurxxxxxA}{\ensuremath{0.09287\pm0.00092}}      
\newcommand{\hatcurXsecingdurxxxxxA}{\ensuremath{0.01673\pm0.00088}}   
\newcommand{\hatcurPPphiconjxxxxxA}{\ensuremath{0\pm0}}                
\newcommand{\hatcurPPperixxxxxA}{\ensuremath{2457063.23110\pm0.00024}} 
\newcommand{\hatcurPPaequivxxxxxA}{\ensuremath{0.1086\pm0.0040}}       
\newcommand{\hatcurPPtcircxxxxxA}{\ensuremath{7700\pm1200}}            
\newcommand{\hatcurPPtinfallxxxxxA}{\ensuremath{12700\pm1500}}         
\newcommand{\hatcurXdistxxxxxA}{\ensuremath{206.6\pm6.6}}              
\newcommand{\hatcurXAvxxxxxA}{\ensuremath{0.121\pm0.080}}              
\newcommand{\hatcurXdistredxxxxxA}{\ensuremath{203.6\pm6.0}}           
\newcommand{\hatcurXEBVxxxxxA}{\ensuremath{0.039\pm0.026}}             
\newcommand{\hatcurXmvisoredxxxxxA}{\ensuremath{13.455\pm0.038}}       
\newcommand{\hatcurXmiisoredxxxxxA}{\ensuremath{12.390\pm0.020}}       
\newcommand{\hatcurXmjisoredxxxxxA}{\ensuremath{11.612\pm0.014}}       
\newcommand{\hatcurXmhisoredxxxxxA}{\ensuremath{11.065\pm0.014}}       
\newcommand{\hatcurXmkisoredxxxxxA}{\ensuremath{10.964\pm0.015}}       
\newcommand{\hatcurXviisoredxxxxxA}{\ensuremath{1.064\pm0.027}}        
\newcommand{\hatcurXvkisoredxxxxxA}{\ensuremath{2.490\pm0.043}}        
\newcommand{\hatcurXjhisoredxxxxxA}{\ensuremath{0.5470\pm0.0069}}      
\newcommand{\hatcurXjkisoredxxxxxA}{\ensuremath{0.6470\pm0.0086}}      
\newcommand{\hatcurCCpmraxxxxxA}{\ensuremath{27.1\pm1.1}}              
\newcommand{\hatcurCCpmdecxxxxxA}{\ensuremath{-8.7\pm1.4}}             
\newcommand{\hatcurCCpmxxxxxA}{\ensuremath{28.5\pm1.8}}                

\newcommand{\hatcurhtrxxxxxB}{HATS747-017}                             
\newcommand{\hatcurfieldxxxxxB}{\ensuremath{string}}                   
\newcommand{\hatcurCCraxxxxxB}{\ensuremath{19^{\mathrm h}05^{\mathrm m}27.96{\mathrm s}}}                            
\newcommand{\hatcurCCdecxxxxxB}{\ensuremath{-50{\arcdeg}04{\arcmin}02.5{\arcsec}}}                           
\newcommand{\hatcurCCmagxxxxxB}{13.901}                                
\newcommand{\hatcurCCtwomassxxxxxB}{2MASS~19052800-5004024}            
\newcommand{\hatcurCCgscxxxxxB}{GSC~8382-01464}                        
\newcommand{\hatcurCCtassmvxxxxxB}{\ensuremath{13.901\pm0.010}}        
\newcommand{\hatcurCCtassmvshortxxxxxB}{\ensuremath{13.9}}             
\newcommand{\hatcurCCtassmBxxxxxB}{\ensuremath{14.625\pm0.010}}        
\newcommand{\hatcurCCtassmBshortxxxxxB}{\ensuremath{14.6}}             
\newcommand{\hatcurCCtassmIxxxxxB}{\ensuremath{nff\pmnff}}             
\newcommand{\hatcurCCtassmIshortxxxxxB}{\ensuremath{0.0}}              
\newcommand{\hatcurCCtassmgxxxxxB}{\ensuremath{14.246\pm0.010}}        
\newcommand{\hatcurCCtassmgshortxxxxxB}{\ensuremath{14.2}}             
\newcommand{\hatcurCCtassmrxxxxxB}{\ensuremath{13.735\pm0.010}}        
\newcommand{\hatcurCCtassmrshortxxxxxB}{\ensuremath{13.7}}             
\newcommand{\hatcurCCtassmixxxxxB}{\ensuremath{13.427\pm0.010}}        
\newcommand{\hatcurCCtassmishortxxxxxB}{\ensuremath{13.4}}             
\newcommand{\hatcurCCtwomassJmagxxxxxB}{\ensuremath{12.636\pm0.025}}   
\newcommand{\hatcurCCtwomassHmagxxxxxB}{\ensuremath{12.293\pm0.025}}   
\newcommand{\hatcurCCtwomassKmagxxxxxB}{\ensuremath{12.262\pm0.030}}   
\newcommand{\hatcurCCcitJmagxxxxxB}{\ensuremath{12.652\pm0.025}}       
\newcommand{\hatcurCCcitHmagxxxxxB}{\ensuremath{12.289\pm0.025}}       
\newcommand{\hatcurCCcitKmagxxxxxB}{\ensuremath{12.286\pm0.030}}       
\newcommand{\hatcurCCbbJmagxxxxxB}{\ensuremath{12.702\pm0.027}}        
\newcommand{\hatcurCCbbHmagxxxxxB}{\ensuremath{12.309\pm0.026}}        
\newcommand{\hatcurCCbbKmagxxxxxB}{\ensuremath{12.306\pm0.030}}        
\newcommand{\hatcurCCesoJmagxxxxxB}{\ensuremath{12.704\pm0.028}}       
\newcommand{\hatcurCCesoHmagxxxxxB}{\ensuremath{12.302\pm0.029}}       
\newcommand{\hatcurCCesoKmagxxxxxB}{\ensuremath{12.305\pm0.030}}       
\newcommand{\hatcurCCesoJHmagxxxxxB}{\ensuremath{0.4030\pm0.0090}}     
\newcommand{\hatcurCCesoJKmagxxxxxB}{\ensuremath{0.399\pm0.042}}       
\newcommand{\hatcurCCesoHKmagxxxxxB}{\ensuremath{-0.003\pm0.042}}      
\newcommand{\hatcurLCdipxxxxxB}{\ensuremath{12.8}}                     
\newcommand{\hatcurLCrprstarxxxxxB}{\ensuremath{0.159\pm0.020}}        
\newcommand{\hatcurLCbsqxxxxxB}{\ensuremath{0.901_{-0.090}^{+0.057}}}  
\newcommand{\hatcurLCimpxxxxxB}{\ensuremath{0.949_{-0.049}^{+0.029}}}  
\newcommand{\hatcurLCbsqsiglowerlimxxxxxB}{\ensuremath{>0.771}}        
\newcommand{\hatcurLCimpsiglowerlimxxxxxB}{\ensuremath{>0.878}}        
\newcommand{\hatcurLCzetaxxxxxB}{\ensuremath{56_{-13}^{+28}}}          
\newcommand{\hatcurLCdurxxxxxB}{\ensuremath{0.0752\pm0.0031}}          
\newcommand{\hatcurLCdurshortxxxxxB}{\ensuremath{0.0752}}              
\newcommand{\hatcurLCdurhrxxxxxB}{\ensuremath{1.804\pm0.073}}          
\newcommand{\hatcurLCdurhrshortxxxxxB}{\ensuremath{1.804}}             
\newcommand{\hatcurLCqxxxxxB}{\ensuremath{0.0348\pm0.0014}}            
\newcommand{\hatcurLCqshortxxxxxB}{\ensuremath{0.035}}                 
\newcommand{\hatcurLCingdurxxxxxB}{\ensuremath{0.094\pm0.013}}         
\newcommand{\hatcurLCPxxxxxB}{\ensuremath{2.1605156\pm0.0000045}}      
\newcommand{\hatcurLCPprecxxxxxB}{\ensuremath{2.1605156}}              
\newcommand{\hatcurLCPshortxxxxxB}{\ensuremath{2.1605}}                
\newcommand{\hatcurLCTxxxxxB}{\ensuremath{2457072.85266\pm0.00070}}    
\newcommand{\hatcurLCTAxxxxxB}{\ensuremath{2456366.3642\pm0.0016}}     
\newcommand{\hatcurLCTBxxxxxB}{\ensuremath{2457282.42267\pm0.00087}}   
\newcommand{\hatcurLChatnetmxxxxxB}{\ensuremath{13.730540\pm0.000078}} 
\newcommand{\hatcurLCiblendxxxxxB}{\ensuremath{0.750\pm0.062}}         
\newcommand{\hatcurLCrhoxxxxxB}{\ensuremath{0.92_{-0.11}^{+0.20}}}     
\newcommand{\hatcurSMEiteffxxxxxB}{\ensuremath{5950\pm130}}            
\newcommand{\hatcurSMEizfehxxxxxB}{\ensuremath{0.360\pm0.060}}         
\newcommand{\hatcurSMEizfehshortxxxxxB}{\ensuremath{0.36}}             
\newcommand{\hatcurSMEiloggxxxxxB}{\ensuremath{4.61\pm0.19}}           
\newcommand{\hatcurSMEivsinxxxxxB}{\ensuremath{4.22\pm0.84}}           
\newcommand{\hatcurSMEivmacxxxxxB}{\ensuremath{0.0}}                   
\newcommand{\hatcurSMEivmicxxxxxB}{\ensuremath{0.0}}                   
\newcommand{\hatcurSMEiiteffxxxxxB}{\ensuremath{5780\pm120}}           
\newcommand{\hatcurSMEiizfehxxxxxB}{\ensuremath{0.280\pm0.070}}        
\newcommand{\hatcurSMEiizfehshortxxxxxB}{\ensuremath{0.28}}            
\newcommand{\hatcurSMEiiloggxxxxxB}{\ensuremath{4.291\pm0.037}}        
\newcommand{\hatcurSMEiivsinxxxxxB}{\ensuremath{4.62\pm0.49}}          
\newcommand{\hatcurLBizxxxxxB}{\ensuremath{0.2116}}                    
\newcommand{\hatcurLBiizxxxxxB}{\ensuremath{0.3342}}                   
\newcommand{\hatcurLBiixxxxxB}{\ensuremath{0.2774}}                    
\newcommand{\hatcurLBiiixxxxxB}{\ensuremath{0.3334}}                   
\newcommand{\hatcurLBiIxxxxxB}{\ensuremath{0.2550}}                    
\newcommand{\hatcurLBiiIxxxxxB}{\ensuremath{0.3347}}                   
\newcommand{\hatcurLBigxxxxxB}{\ensuremath{0.5783}}                    
\newcommand{\hatcurLBiigxxxxxB}{\ensuremath{0.2199}}                   
\newcommand{\hatcurLBirxxxxxB}{\ensuremath{0.3728}}                    
\newcommand{\hatcurLBiirxxxxxB}{\ensuremath{0.3215}}                   
\newcommand{\hatcurLBiRxxxxxB}{\ensuremath{0.3462}}                    
\newcommand{\hatcurLBiiRxxxxxB}{\ensuremath{0.3259}}                   
\newcommand{\hatcurLBikepxxxxxB}{\ensuremath{0.1000}}                  
\newcommand{\hatcurLBiikepxxxxxB}{\ensuremath{0.1000}}                 
\newcommand{\hatcurISOmxxxxxB}{\ensuremath{1.121\pm0.046}}             
\newcommand{\hatcurISOmshortxxxxxB}{\ensuremath{1.12}}                 
\newcommand{\hatcurISOmlongxxxxxB}{\ensuremath{1.121\pm0.046}}         
\newcommand{\hatcurISOrxxxxxB}{\ensuremath{1.199_{-0.081}^{+0.061}}}   
\newcommand{\hatcurISOrshortxxxxxB}{\ensuremath{1.20}}                 
\newcommand{\hatcurISOrlongxxxxxB}{\ensuremath{1.199_{-0.081}^{+0.061}}} 
\newcommand{\hatcurISOrhoxxxxxB}{\ensuremath{0.91_{-0.11}^{+0.20}}}    
\newcommand{\hatcurISOrholongxxxxxB}{\ensuremath{0.91_{-0.11}^{+0.20}}} 
\newcommand{\hatcurISOloggxxxxxB}{\ensuremath{4.328\pm0.044}}          
\newcommand{\hatcurISOlumxxxxxB}{\ensuremath{1.43\pm0.22}}             
\newcommand{\hatcurISOlumshortxxxxxB}{\ensuremath{1.43}}               
\newcommand{\hatcurISOmvxxxxxB}{\ensuremath{4.43\pm0.18}}              
\newcommand{\hatcurISOvixxxxxB}{\ensuremath{0.701\pm0.036}}            
\newcommand{\hatcurISOagexxxxxB}{\ensuremath{4.2\pm1.5}}               
\newcommand{\hatcurISOsigmaxxxxxB}{\ensuremath{0.00140\pm0.00017}}     
\newcommand{\hatcurISOMJxxxxxB}{\ensuremath{3.29\pm0.15}}              
\newcommand{\hatcurISOMHxxxxxB}{\ensuremath{2.96\pm0.14}}              
\newcommand{\hatcurISOMKxxxxxB}{\ensuremath{2.90\pm0.13}}              
\newcommand{\hatcurISOJKxxxxxB}{\ensuremath{0.20\pm0.19}}              
\newcommand{\hatcurISOspecxxxxxB}{G}                                   
\newcommand{\hatcurRVKxxxxxB}{\ensuremath{212.3\pm8.6}}                
\newcommand{\hatcurRVrkxxxxxB}{\ensuremath{0\pm0}}                     
\newcommand{\hatcurRVrhxxxxxB}{\ensuremath{0\pm0}}                     
\newcommand{\hatcurRVkxxxxxB}{\ensuremath{0\pm0}}                      
\newcommand{\hatcurRVhxxxxxB}{\ensuremath{0\pm0}}                      
\newcommand{\hatcurRVtronexxxxxB}{\ensuremath{0\pm0}}                  
\newcommand{\hatcurRVtrtwoxxxxxB}{\ensuremath{0\pm0}}                  
\newcommand{\hatcurRVgammaxxxxxB}{\ensuremath{-13372.0\pm6.1}}         
\newcommand{\hatcurRVjitterxxxxxB}{\ensuremath{0.0\pm6.6}}             
\newcommand{\hatcurRVjittertwosiglimxxxxxB}{\ensuremath{<14.4}}        
\newcommand{\hatcurRVfitrmsxxxxxB}{\ensuremath{.1fym}}                 %
\newcommand{\hatcurRVeccenxxxxxB}{\ensuremath{0\pm0}}                  
\newcommand{\hatcurRVeccentwosiglimxxxxxB}{\ensuremath{<0.000}}        
\newcommand{\hatcurRVomegaxxxxxB}{\ensuremath{0\pm0}}                  
\newcommand{\hatcurPPixxxxxB}{\ensuremath{81.02_{-0.62}^{+0.93}}}      
\newcommand{\hatcurPPitwosigupperlimxxxxxB}{\ensuremath{<83.5}}        
\newcommand{\hatcurPPgxxxxxB}{\ensuremath{10.6_{-2.6}^{+6.5}}}         
\newcommand{\hatcurPPloggxxxxxB}{\ensuremath{3.02_{-0.12}^{+0.21}}}    
\newcommand{\hatcurPParxxxxxB}{\ensuremath{6.08_{-0.26}^{+0.41}}}      
\newcommand{\hatcurPParelxxxxxB}{\ensuremath{0.03397\pm0.00047}}       
\newcommand{\hatcurPPrhoxxxxxB}{\ensuremath{0.29_{-0.10}^{+0.30}}}     
\newcommand{\hatcurPPmxxxxxB}{\ensuremath{1.470\pm0.072}}              
\newcommand{\hatcurPPmshortxxxxxB}{\ensuremath{1.47}}                  
\newcommand{\hatcurPPmlongxxxxxB}{\ensuremath{1.470\pm0.072}}          
\newcommand{\hatcurPPmexxxxxB}{\ensuremath{467\pm23}}                  
\newcommand{\hatcurPPmeshortxxxxxB}{\ensuremath{467.1}}                
\newcommand{\hatcurPPmelongxxxxxB}{\ensuremath{467\pm23}}              
\newcommand{\hatcurPPrxxxxxB}{\ensuremath{1.86_{-0.40}^{+0.30}}}       
\newcommand{\hatcurPPrshortxxxxxB}{\ensuremath{1.86}}                  
\newcommand{\hatcurPPrlongxxxxxB}{\ensuremath{1.86_{-0.40}^{+0.30}}}   
\newcommand{\hatcurPPrtwosiglowerlimxxxxxB}{\ensuremath{>1.31}}        
\newcommand{\hatcurPPrexxxxxB}{\ensuremath{20.8_{-4.5}^{+3.4}}}        
\newcommand{\hatcurPPreshortxxxxxB}{\ensuremath{20.8}}                 
\newcommand{\hatcurPPrelongxxxxxB}{\ensuremath{20.8_{-4.5}^{+3.4}}}    
\newcommand{\hatcurPPmrcorrxxxxxB}{\ensuremath{0.23}}                  
\newcommand{\hatcurPPteffxxxxxB}{\ensuremath{1654\pm54}}               
\newcommand{\hatcurPPthetaxxxxxB}{\ensuremath{0.0475_{-0.0067}^{+0.0131}}} 
\newcommand{\hatcurPPfluxperixxxxxB}{\ensuremath{1.69\pm0.22}}         
\newcommand{\hatcurPPfluxperidimxxxxxB}{\ensuremath{9}}                
\newcommand{\hatcurPPfluxapxxxxxB}{\ensuremath{1.69\pm0.22}}           
\newcommand{\hatcurPPfluxapdimxxxxxB}{\ensuremath{9}}                  
\newcommand{\hatcurPPfluxavgxxxxxB}{\ensuremath{1.69\pm0.22}}          
\newcommand{\hatcurPPfluxavgdimxxxxxB}{\ensuremath{9}}                 
\newcommand{\hatcurPPfluxavglogxxxxxB}{\ensuremath{9.228\pm0.058}}     
\newcommand{\hatcurXsecphasexxxxxB}{\ensuremath{0\pm0}}                
\newcommand{\hatcurXsecondaryxxxxxB}{\ensuremath{2457073.93291\pm0.00071}} 
\newcommand{\hatcurXsecdurxxxxxB}{\ensuremath{0.0902\pm0.0095}}        
\newcommand{\hatcurXsecingdurxxxxxB}{\ensuremath{0.0451\pm0.0053}}     
\newcommand{\hatcurPPphiconjxxxxxB}{\ensuremath{0\pm0}}                
\newcommand{\hatcurPPperixxxxxB}{\ensuremath{2457072.31253\pm0.00070}} 
\newcommand{\hatcurPPaequivxxxxxB}{\ensuremath{0.0284_{-0.0016}^{+0.0021}}} 
\newcommand{\hatcurPPtcircxxxxxB}{\ensuremath{6.2_{-3.2}^{+14.3}}}     
\newcommand{\hatcurPPtinfallxxxxxB}{\ensuremath{131_{-26}^{+50}}}      
\newcommand{\hatcurXdistxxxxxB}{\ensuremath{760\pm47}}                 
\newcommand{\hatcurXAvxxxxxB}{\ensuremath{0.106\pm0.075}}              
\newcommand{\hatcurXdistredxxxxxB}{\ensuremath{747\pm46}}              
\newcommand{\hatcurXEBVxxxxxB}{\ensuremath{0.034\pm0.024}}             
\newcommand{\hatcurXmvisoredxxxxxB}{\ensuremath{13.902\pm0.011}}       
\newcommand{\hatcurXmiisoredxxxxxB}{\ensuremath{13.145\pm0.012}}       
\newcommand{\hatcurXmjisoredxxxxxB}{\ensuremath{12.682\pm0.017}}       
\newcommand{\hatcurXmhisoredxxxxxB}{\ensuremath{12.342\pm0.023}}       
\newcommand{\hatcurXmkisoredxxxxxB}{\ensuremath{12.280\pm0.024}}       
\newcommand{\hatcurXviisoredxxxxxB}{\ensuremath{0.757\pm0.014}}        
\newcommand{\hatcurXvkisoredxxxxxB}{\ensuremath{1.622\pm0.029}}        
\newcommand{\hatcurXjhisoredxxxxxB}{\ensuremath{0.341\pm0.013}}        
\newcommand{\hatcurXjkisoredxxxxxB}{\ensuremath{0.4020_{-0.0080}^{+0.0110}}} 
\newcommand{\hatcurCCpmraxxxxxB}{\ensuremath{3.3\pm1.4}}               
\newcommand{\hatcurCCpmdecxxxxxB}{\ensuremath{-1.1\pm1.5}}             
\newcommand{\hatcurCCpmxxxxxB}{\ensuremath{3.5\pm2.1}}                 

\newcommand{\hatcurhtrxxxxxC}{HATS776-001}                      
\newcommand{\hatcurfieldxxxxxC}{\ensuremath{string}}            
\newcommand{\hatcurCCraxxxxxC}{\ensuremath{17^{\mathrm h}55^{\mathrm m}33.60{\mathrm s}}}                     
\newcommand{\hatcurCCdecxxxxxC}{\ensuremath{-61{\arcdeg}44{\arcmin}50.3{\arcsec}}}                    
\newcommand{\hatcurCCmagxxxxxC}{12.830}                         
\newcommand{\hatcurCCtwomassxxxxxC}{2MASS~17553376-6144503}     
\newcommand{\hatcurCCgscxxxxxC}{GSC~9054-00129}                 
\newcommand{\hatcurCCtassmvxxxxxC}{\ensuremath{12.830\pm0.010}} 
\newcommand{\hatcurCCtassmvshortxxxxxC}{\ensuremath{12.8}}      
\newcommand{\hatcurCCtassmBxxxxxC}{\ensuremath{13.404\pm0.010}} 
\newcommand{\hatcurCCtassmBshortxxxxxC}{\ensuremath{13.4}}      
\newcommand{\hatcurCCtassmIxxxxxC}{\ensuremath{100\pm1000}}     
\newcommand{\hatcurCCtassmIshortxxxxxC}{\ensuremath{100.0}}     
\newcommand{\hatcurCCtassmgxxxxxC}{\ensuremath{13.071\pm0.010}} 
\newcommand{\hatcurCCtassmgshortxxxxxC}{\ensuremath{13.1}}      
\newcommand{\hatcurCCtassmrxxxxxC}{\ensuremath{12.643\pm0.010}} 
\newcommand{\hatcurCCtassmrshortxxxxxC}{\ensuremath{12.6}}      
\newcommand{\hatcurCCtassmixxxxxC}{\ensuremath{12.518\pm0.090}} 
\newcommand{\hatcurCCtassmishortxxxxxC}{\ensuremath{12.5}}      
\newcommand{\hatcurCCtwomassJmagxxxxxC}{\ensuremath{11.678\pm0.022}} 
\newcommand{\hatcurCCtwomassHmagxxxxxC}{\ensuremath{11.447\pm0.025}} 
\newcommand{\hatcurCCtwomassKmagxxxxxC}{\ensuremath{11.382\pm0.023}} 
\newcommand{\hatcurCCcitJmagxxxxxC}{\ensuremath{11.699\pm0.022}} 
\newcommand{\hatcurCCcitHmagxxxxxC}{\ensuremath{11.442\pm0.025}} 
\newcommand{\hatcurCCcitKmagxxxxxC}{\ensuremath{11.406\pm0.023}} 
\newcommand{\hatcurCCbbJmagxxxxxC}{\ensuremath{11.742\pm0.024}} 
\newcommand{\hatcurCCbbHmagxxxxxC}{\ensuremath{11.463\pm0.026}} 
\newcommand{\hatcurCCbbKmagxxxxxC}{\ensuremath{11.426\pm0.023}} 
\newcommand{\hatcurCCesoJmagxxxxxC}{\ensuremath{11.743\pm0.025}} 
\newcommand{\hatcurCCesoHmagxxxxxC}{\ensuremath{11.458\pm0.029}} 
\newcommand{\hatcurCCesoKmagxxxxxC}{\ensuremath{11.425\pm0.024}} 
\newcommand{\hatcurCCesoJHmagxxxxxC}{\ensuremath{0.285\pm0.036}} 
\newcommand{\hatcurCCesoJKmagxxxxxC}{\ensuremath{0.319\pm0.034}} 
\newcommand{\hatcurCCesoHKmagxxxxxC}{\ensuremath{0.034\pm0.038}} 
\newcommand{\hatcurLCdipxxxxxC}{\ensuremath{17.3}}              
\newcommand{\hatcurLCrprstarxxxxxC}{\ensuremath{0.1307\pm0.0030}} 
\newcommand{\hatcurLCbsqxxxxxC}{\ensuremath{0.076_{-0.045}^{+0.050}}} 
\newcommand{\hatcurLCimpxxxxxC}{\ensuremath{0.276_{-0.101}^{+0.079}}} 
\newcommand{\hatcurLCzetaxxxxxC}{\ensuremath{22.64\pm0.17}}     
\newcommand{\hatcurLCdurxxxxxC}{\ensuremath{0.1008\pm0.0010}}   
\newcommand{\hatcurLCdurshortxxxxxC}{\ensuremath{0.1008}}       
\newcommand{\hatcurLCdurhrxxxxxC}{\ensuremath{2.419\pm0.024}}   
\newcommand{\hatcurLCdurhrshortxxxxxC}{\ensuremath{2.419}}      
\newcommand{\hatcurLCqxxxxxC}{\ensuremath{0.07480\pm0.00074}}   
\newcommand{\hatcurLCqshortxxxxxC}{\ensuremath{0.075}}          
\newcommand{\hatcurLCingdurxxxxxC}{\ensuremath{0.01241\pm0.00081}} 
\newcommand{\hatcurLCPxxxxxC}{\ensuremath{1.3484954\pm0.0000013}} 
\newcommand{\hatcurLCPprecxxxxxC}{\ensuremath{1.3484954}}       
\newcommand{\hatcurLCPshortxxxxxC}{\ensuremath{1.3485}}         
\newcommand{\hatcurLCTxxxxxC}{\ensuremath{2457038.47327\pm0.00038}} 
\newcommand{\hatcurLCTAxxxxxC}{\ensuremath{2455696.7203\pm0.0013}} 
\newcommand{\hatcurLCTBxxxxxC}{\ensuremath{2457181.41379\pm0.00043}} 
\newcommand{\hatcurLChatnetmxxxxxC}{\ensuremath{12.60998\pm0.00013}} 
\newcommand{\hatcurLCiblendxxxxxC}{\ensuremath{0.787\pm0.042}}  
\newcommand{\hatcurLCrhoxxxxxC}{\ensuremath{1.096_{-0.085}^{+0.059}}} 
\newcommand{\hatcurSMEiteffxxxxxC}{\ensuremath{6300\pm100}}     
\newcommand{\hatcurSMEizfehxxxxxC}{\ensuremath{-0.040\pm0.060}} 
\newcommand{\hatcurSMEizfehshortxxxxxC}{\ensuremath{-0.04}}     
\newcommand{\hatcurSMEiloggxxxxxC}{\ensuremath{4.20\pm0.15}}    
\newcommand{\hatcurSMEivsinxxxxxC}{\ensuremath{9.52\pm0.28}}    
\newcommand{\hatcurSMEivmacxxxxxC}{\ensuremath{0.0}}            
\newcommand{\hatcurSMEivmicxxxxxC}{\ensuremath{0.0}}            
\newcommand{\hatcurSMEiiteffxxxxxC}{\ensuremath{6346\pm81}}     
\newcommand{\hatcurSMEiizfehxxxxxC}{\ensuremath{0.000\pm0.050}} 
\newcommand{\hatcurSMEiizfehshortxxxxxC}{\ensuremath{0.00}}     
\newcommand{\hatcurSMEiiloggxxxxxC}{\ensuremath{4\pm0}}         
\newcommand{\hatcurSMEiivsinxxxxxC}{\ensuremath{9.44\pm0.21}}   
\newcommand{\hatcurLBizxxxxxC}{\ensuremath{0.1405}}             
\newcommand{\hatcurLBiizxxxxxC}{\ensuremath{0.3577}}            
\newcommand{\hatcurLBiixxxxxC}{\ensuremath{0.1919}}             
\newcommand{\hatcurLBiiixxxxxC}{\ensuremath{0.3654}}            
\newcommand{\hatcurLBiIxxxxxC}{\ensuremath{0.1738}}             
\newcommand{\hatcurLBiiIxxxxxC}{\ensuremath{0.3639}}            
\newcommand{\hatcurLBigxxxxxC}{\ensuremath{0.4251}}             
\newcommand{\hatcurLBiigxxxxxC}{\ensuremath{0.3251}}            
\newcommand{\hatcurLBirxxxxxC}{\ensuremath{0.2638}}             
\newcommand{\hatcurLBiirxxxxxC}{\ensuremath{0.3753}}            
\newcommand{\hatcurLBiRxxxxxC}{\ensuremath{0.2434}}             
\newcommand{\hatcurLBiiRxxxxxC}{\ensuremath{0.3740}}            
\newcommand{\hatcurLBikepxxxxxC}{\ensuremath{0.1000}}           
\newcommand{\hatcurLBiikepxxxxxC}{\ensuremath{0.1000}}          
\newcommand{\hatcurISOmxxxxxC}{\ensuremath{1.212\pm0.033}}      
\newcommand{\hatcurISOmshortxxxxxC}{\ensuremath{1.21}}          
\newcommand{\hatcurISOmlongxxxxxC}{\ensuremath{1.212\pm0.033}}  
\newcommand{\hatcurISOrxxxxxC}{\ensuremath{1.172\pm0.033}}      
\newcommand{\hatcurISOrshortxxxxxC}{\ensuremath{1.17}}          
\newcommand{\hatcurISOrlongxxxxxC}{\ensuremath{1.172\pm0.033}}  
\newcommand{\hatcurISOrhoxxxxxC}{\ensuremath{1.059\pm0.075}}    
\newcommand{\hatcurISOrholongxxxxxC}{\ensuremath{1.059\pm0.075}} 
\newcommand{\hatcurISOloggxxxxxC}{\ensuremath{4.384\pm0.021}}   
\newcommand{\hatcurISOlumxxxxxC}{\ensuremath{1.96\pm0.18}}      
\newcommand{\hatcurISOlumshortxxxxxC}{\ensuremath{1.96}}        
\newcommand{\hatcurISOmvxxxxxC}{\ensuremath{4.03\pm0.10}}       
\newcommand{\hatcurISOvixxxxxC}{\ensuremath{0.547\pm0.018}}     
\newcommand{\hatcurISOagexxxxxC}{\ensuremath{0.88_{-0.45}^{+0.67}}} 
\newcommand{\hatcurISOsigmaxxxxxC}{\ensuremath{0.00240\pm0.00024}} 
\newcommand{\hatcurISOMJxxxxxC}{\ensuremath{3.142\pm0.076}}     
\newcommand{\hatcurISOMHxxxxxC}{\ensuremath{2.887\pm0.066}}     
\newcommand{\hatcurISOMKxxxxxC}{\ensuremath{2.844\pm0.065}}     
\newcommand{\hatcurISOJKxxxxxC}{\ensuremath{0.300\pm0.010}}     
\newcommand{\hatcurISOspecxxxxxC}{F}                            
\newcommand{\hatcurRVKxxxxxC}{\ensuremath{396\pm29}}            
\newcommand{\hatcurRVrkxxxxxC}{\ensuremath{0\pm0}}              
\newcommand{\hatcurRVrhxxxxxC}{\ensuremath{0\pm0}}              
\newcommand{\hatcurRVkxxxxxC}{\ensuremath{0\pm0}}               
\newcommand{\hatcurRVhxxxxxC}{\ensuremath{0\pm0}}               
\newcommand{\hatcurRVtronexxxxxC}{\ensuremath{0\pm0}}           
\newcommand{\hatcurRVtrtwoxxxxxC}{\ensuremath{0\pm0}}           
\newcommand{\hatcurRVgammaAxxxxxC}{\ensuremath{-3259\pm41}}     
\newcommand{\hatcurRVjitterAxxxxxC}{\ensuremath{8\pm52}}        
\newcommand{\hatcurRVjittertwosiglimAxxxxxC}{\ensuremath{<136.7}} 
\newcommand{\hatcurRVfitrmsAxxxxxC}{\ensuremath{0.0}}           
\newcommand{\hatcurRVgammaBxxxxxC}{\ensuremath{-3236\pm85}}     
\newcommand{\hatcurRVjitterBxxxxxC}{\ensuremath{70\pm140}}      
\newcommand{\hatcurRVjittertwosiglimBxxxxxC}{\ensuremath{<486.9}} 
\newcommand{\hatcurRVfitrmsBxxxxxC}{\ensuremath{0.0}}           
\newcommand{\hatcurRVgammaCxxxxxC}{\ensuremath{-3370\pm120}}    
\newcommand{\hatcurRVjitterCxxxxxC}{\ensuremath{60\pm140}}      
\newcommand{\hatcurRVjittertwosiglimCxxxxxC}{\ensuremath{<399.8}} 
\newcommand{\hatcurRVfitrmsCxxxxxC}{\ensuremath{0.0}}           
\newcommand{\hatcurRVgammaDxxxxxC}{\ensuremath{-3284\pm35}}     
\newcommand{\hatcurRVjitterDxxxxxC}{\ensuremath{72\pm55}}       
\newcommand{\hatcurRVjittertwosiglimDxxxxxC}{\ensuremath{<155.9}} 
\newcommand{\hatcurRVfitrmsDxxxxxC}{\ensuremath{.1fym}}         %
\newcommand{\hatcurRVeccenxxxxxC}{\ensuremath{0\pm0}}           
\newcommand{\hatcurRVeccentwosiglimxxxxxC}{\ensuremath{<0.000}} 
\newcommand{\hatcurRVomegaxxxxxC}{\ensuremath{0\pm0}}           
\newcommand{\hatcurPPixxxxxC}{\ensuremath{86.6\pm1.2}}          
\newcommand{\hatcurPPgxxxxxC}{\ensuremath{27.2\pm3.4}}          
\newcommand{\hatcurPPloggxxxxxC}{\ensuremath{3.435\pm0.054}}    
\newcommand{\hatcurPParxxxxxC}{\ensuremath{4.67_{-0.14}^{+0.10}}} 
\newcommand{\hatcurPParelxxxxxC}{\ensuremath{0.02547\pm0.00023}} 
\newcommand{\hatcurPPrhoxxxxxC}{\ensuremath{0.92\pm0.15}}       
\newcommand{\hatcurPPmxxxxxC}{\ensuremath{2.44\pm0.18}}         
\newcommand{\hatcurPPmshortxxxxxC}{\ensuremath{2.44}}           
\newcommand{\hatcurPPmlongxxxxxC}{\ensuremath{2.44\pm0.18}}     
\newcommand{\hatcurPPmexxxxxC}{\ensuremath{776\pm58}}           
\newcommand{\hatcurPPmeshortxxxxxC}{\ensuremath{776.0}}         
\newcommand{\hatcurPPmelongxxxxxC}{\ensuremath{776\pm58}}       
\newcommand{\hatcurPPrxxxxxC}{\ensuremath{1.487_{-0.054}^{+0.078}}} 
\newcommand{\hatcurPPrshortxxxxxC}{\ensuremath{1.49}}           
\newcommand{\hatcurPPrlongxxxxxC}{\ensuremath{1.487_{-0.054}^{+0.078}}} 
\newcommand{\hatcurPPrexxxxxC}{\ensuremath{16.67_{-0.61}^{+0.87}}} 
\newcommand{\hatcurPPreshortxxxxxC}{\ensuremath{16.7}}          
\newcommand{\hatcurPPrelongxxxxxC}{\ensuremath{16.67_{-0.61}^{+0.87}}} 
\newcommand{\hatcurPPmrcorrxxxxxC}{\ensuremath{-0.18}}          
\newcommand{\hatcurPPteffxxxxxC}{\ensuremath{2067\pm39}}        
\newcommand{\hatcurPPthetaxxxxxC}{\ensuremath{0.0684_{-0.0070}^{+0.0092}}} 
\newcommand{\hatcurPPfluxperixxxxxC}{\ensuremath{4.12\pm0.32}}  
\newcommand{\hatcurPPfluxperidimxxxxxC}{\ensuremath{9}}         
\newcommand{\hatcurPPfluxapxxxxxC}{\ensuremath{4.12\pm0.32}}    
\newcommand{\hatcurPPfluxapdimxxxxxC}{\ensuremath{9}}           
\newcommand{\hatcurPPfluxavgxxxxxC}{\ensuremath{4.12\pm0.32}}   
\newcommand{\hatcurPPfluxavgdimxxxxxC}{\ensuremath{9}}          
\newcommand{\hatcurPPfluxavglogxxxxxC}{\ensuremath{9.615\pm0.033}} 
\newcommand{\hatcurXsecphasexxxxxC}{\ensuremath{0\pm0}}         
\newcommand{\hatcurXsecondaryxxxxxC}{\ensuremath{2457039.14751\pm0.00038}} 
\newcommand{\hatcurXsecdurxxxxxC}{\ensuremath{0.1008\pm0.0010}} 
\newcommand{\hatcurXsecingdurxxxxxC}{\ensuremath{0.01241\pm0.00081}} 
\newcommand{\hatcurPPphiconjxxxxxC}{\ensuremath{0\pm0}}         
\newcommand{\hatcurPPperixxxxxC}{\ensuremath{2457038.13614\pm0.00038}} 
\newcommand{\hatcurPPaequivxxxxxC}{\ensuremath{0.01820\pm0.00069}} 
\newcommand{\hatcurPPtcircxxxxxC}{\ensuremath{4.20\pm0.96}}     
\newcommand{\hatcurPPtinfallxxxxxC}{\ensuremath{14.0\pm1.7}}    
\newcommand{\hatcurXdistxxxxxC}{\ensuremath{520\pm17}}          
\newcommand{\hatcurXAvxxxxxC}{\ensuremath{0.261\pm0.061}}       
\newcommand{\hatcurXdistredxxxxxC}{\ensuremath{510\pm15}}       
\newcommand{\hatcurXEBVxxxxxC}{\ensuremath{0.084\pm0.020}}      
\newcommand{\hatcurXmvisoredxxxxxC}{\ensuremath{12.8300\pm0.0100}} 
\newcommand{\hatcurXmiisoredxxxxxC}{\ensuremath{12.147\pm0.015}} 
\newcommand{\hatcurXmjisoredxxxxxC}{\ensuremath{11.752\pm0.014}} 
\newcommand{\hatcurXmhisoredxxxxxC}{\ensuremath{11.472\pm0.015}} 
\newcommand{\hatcurXmkisoredxxxxxC}{\ensuremath{11.410\pm0.016}} 
\newcommand{\hatcurXviisoredxxxxxC}{\ensuremath{0.683\pm0.016}} 
\newcommand{\hatcurXvkisoredxxxxxC}{\ensuremath{1.420\pm0.020}} 
\newcommand{\hatcurXjhisoredxxxxxC}{\ensuremath{0.2790\pm0.0081}} 
\newcommand{\hatcurXjkisoredxxxxxC}{\ensuremath{0.3420\pm0.0065}} 
\newcommand{\hatcurCCpmraxxxxxC}{\ensuremath{2.9\pm2.6}}        
\newcommand{\hatcurCCpmdecxxxxxC}{\ensuremath{-11.0\pm2.6}}     
\newcommand{\hatcurCCpmxxxxxC}{\ensuremath{11.4\pm3.7}}         

\newcommand{\hatcurCCbbHmag}[1]{\ifnum#1=22 %
\hatcurCCbbHmagxxxxxA
\else
\ifnum#1=23 %
\hatcurCCbbHmagxxxxxB
\else
\ifnum#1=24 %
\hatcurCCbbHmagxxxxxC
\else
??????\fi
\fi
\fi
}
\newcommand{\hatcurCCbbJmag}[1]{\ifnum#1=22 %
\hatcurCCbbJmagxxxxxA
\else
\ifnum#1=23 %
\hatcurCCbbJmagxxxxxB
\else
\ifnum#1=24 %
\hatcurCCbbJmagxxxxxC
\else
??????\fi
\fi
\fi
}
\newcommand{\hatcurCCbbKmag}[1]{\ifnum#1=22 %
\hatcurCCbbKmagxxxxxA
\else
\ifnum#1=23 %
\hatcurCCbbKmagxxxxxB
\else
\ifnum#1=24 %
\hatcurCCbbKmagxxxxxC
\else
??????\fi
\fi
\fi
}
\newcommand{\hatcurCCcitHmag}[1]{\ifnum#1=22 %
\hatcurCCcitHmagxxxxxA
\else
\ifnum#1=23 %
\hatcurCCcitHmagxxxxxB
\else
\ifnum#1=24 %
\hatcurCCcitHmagxxxxxC
\else
??????\fi
\fi
\fi
}
\newcommand{\hatcurCCcitJmag}[1]{\ifnum#1=22 %
\hatcurCCcitJmagxxxxxA
\else
\ifnum#1=23 %
\hatcurCCcitJmagxxxxxB
\else
\ifnum#1=24 %
\hatcurCCcitJmagxxxxxC
\else
??????\fi
\fi
\fi
}
\newcommand{\hatcurCCcitKmag}[1]{\ifnum#1=22 %
\hatcurCCcitKmagxxxxxA
\else
\ifnum#1=23 %
\hatcurCCcitKmagxxxxxB
\else
\ifnum#1=24 %
\hatcurCCcitKmagxxxxxC
\else
??????\fi
\fi
\fi
}
\newcommand{\hatcurCCdec}[1]{\ifnum#1=22 %
\hatcurCCdecxxxxxA
\else
\ifnum#1=23 %
\hatcurCCdecxxxxxB
\else
\ifnum#1=24 %
\hatcurCCdecxxxxxC
\else
??????\fi
\fi
\fi
}
\newcommand{\hatcurCCesoHKmag}[1]{\ifnum#1=22 %
\hatcurCCesoHKmagxxxxxA
\else
\ifnum#1=23 %
\hatcurCCesoHKmagxxxxxB
\else
\ifnum#1=24 %
\hatcurCCesoHKmagxxxxxC
\else
??????\fi
\fi
\fi
}
\newcommand{\hatcurCCesoHmag}[1]{\ifnum#1=22 %
\hatcurCCesoHmagxxxxxA
\else
\ifnum#1=23 %
\hatcurCCesoHmagxxxxxB
\else
\ifnum#1=24 %
\hatcurCCesoHmagxxxxxC
\else
??????\fi
\fi
\fi
}
\newcommand{\hatcurCCesoJHmag}[1]{\ifnum#1=22 %
\hatcurCCesoJHmagxxxxxA
\else
\ifnum#1=23 %
\hatcurCCesoJHmagxxxxxB
\else
\ifnum#1=24 %
\hatcurCCesoJHmagxxxxxC
\else
??????\fi
\fi
\fi
}
\newcommand{\hatcurCCesoJKmag}[1]{\ifnum#1=22 %
\hatcurCCesoJKmagxxxxxA
\else
\ifnum#1=23 %
\hatcurCCesoJKmagxxxxxB
\else
\ifnum#1=24 %
\hatcurCCesoJKmagxxxxxC
\else
??????\fi
\fi
\fi
}
\newcommand{\hatcurCCesoJmag}[1]{\ifnum#1=22 %
\hatcurCCesoJmagxxxxxA
\else
\ifnum#1=23 %
\hatcurCCesoJmagxxxxxB
\else
\ifnum#1=24 %
\hatcurCCesoJmagxxxxxC
\else
??????\fi
\fi
\fi
}
\newcommand{\hatcurCCesoKmag}[1]{\ifnum#1=22 %
\hatcurCCesoKmagxxxxxA
\else
\ifnum#1=23 %
\hatcurCCesoKmagxxxxxB
\else
\ifnum#1=24 %
\hatcurCCesoKmagxxxxxC
\else
??????\fi
\fi
\fi
}
\newcommand{\hatcurCCgsc}[1]{\ifnum#1=22 %
\hatcurCCgscxxxxxA
\else
\ifnum#1=23 %
\hatcurCCgscxxxxxB
\else
\ifnum#1=24 %
\hatcurCCgscxxxxxC
\else
??????\fi
\fi
\fi
}
\newcommand{\hatcurCCmag}[1]{\ifnum#1=22 %
\hatcurCCmagxxxxxA
\else
\ifnum#1=23 %
\hatcurCCmagxxxxxB
\else
\ifnum#1=24 %
\hatcurCCmagxxxxxC
\else
??????\fi
\fi
\fi
}
\newcommand{\hatcurCCpm}[1]{\ifnum#1=22 %
\hatcurCCpmxxxxxA
\else
\ifnum#1=23 %
\hatcurCCpmxxxxxB
\else
\ifnum#1=24 %
\hatcurCCpmxxxxxC
\else
??????\fi
\fi
\fi
}
\newcommand{\hatcurCCpmdec}[1]{\ifnum#1=22 %
\hatcurCCpmdecxxxxxA
\else
\ifnum#1=23 %
\hatcurCCpmdecxxxxxB
\else
\ifnum#1=24 %
\hatcurCCpmdecxxxxxC
\else
??????\fi
\fi
\fi
}
\newcommand{\hatcurCCpmra}[1]{\ifnum#1=22 %
\hatcurCCpmraxxxxxA
\else
\ifnum#1=23 %
\hatcurCCpmraxxxxxB
\else
\ifnum#1=24 %
\hatcurCCpmraxxxxxC
\else
??????\fi
\fi
\fi
}
\newcommand{\hatcurCCra}[1]{\ifnum#1=22 %
\hatcurCCraxxxxxA
\else
\ifnum#1=23 %
\hatcurCCraxxxxxB
\else
\ifnum#1=24 %
\hatcurCCraxxxxxC
\else
??????\fi
\fi
\fi
}
\newcommand{\hatcurCCtassmB}[1]{\ifnum#1=22 %
\hatcurCCtassmBxxxxxA
\else
\ifnum#1=23 %
\hatcurCCtassmBxxxxxB
\else
\ifnum#1=24 %
\hatcurCCtassmBxxxxxC
\else
??????\fi
\fi
\fi
}
\newcommand{\hatcurCCtassmBshort}[1]{\ifnum#1=22 %
\hatcurCCtassmBshortxxxxxA
\else
\ifnum#1=23 %
\hatcurCCtassmBshortxxxxxB
\else
\ifnum#1=24 %
\hatcurCCtassmBshortxxxxxC
\else
??????\fi
\fi
\fi
}
\newcommand{\hatcurCCtassmg}[1]{\ifnum#1=22 %
\hatcurCCtassmgxxxxxA
\else
\ifnum#1=23 %
\hatcurCCtassmgxxxxxB
\else
\ifnum#1=24 %
\hatcurCCtassmgxxxxxC
\else
??????\fi
\fi
\fi
}
\newcommand{\hatcurCCtassmgshort}[1]{\ifnum#1=22 %
\hatcurCCtassmgshortxxxxxA
\else
\ifnum#1=23 %
\hatcurCCtassmgshortxxxxxB
\else
\ifnum#1=24 %
\hatcurCCtassmgshortxxxxxC
\else
??????\fi
\fi
\fi
}
\newcommand{\hatcurCCtassmi}[1]{\ifnum#1=22 %
\hatcurCCtassmixxxxxA
\else
\ifnum#1=23 %
\hatcurCCtassmixxxxxB
\else
\ifnum#1=24 %
\hatcurCCtassmixxxxxC
\else
??????\fi
\fi
\fi
}
\newcommand{\hatcurCCtassmI}[1]{\ifnum#1=22 %
\hatcurCCtassmIxxxxxA
\else
\ifnum#1=23 %
\hatcurCCtassmIxxxxxB
\else
\ifnum#1=24 %
\hatcurCCtassmIxxxxxC
\else
??????\fi
\fi
\fi
}
\newcommand{\hatcurCCtassmishort}[1]{\ifnum#1=22 %
\hatcurCCtassmishortxxxxxA
\else
\ifnum#1=23 %
\hatcurCCtassmishortxxxxxB
\else
\ifnum#1=24 %
\hatcurCCtassmishortxxxxxC
\else
??????\fi
\fi
\fi
}
\newcommand{\hatcurCCtassmIshort}[1]{\ifnum#1=22 %
\hatcurCCtassmIshortxxxxxA
\else
\ifnum#1=23 %
\hatcurCCtassmIshortxxxxxB
\else
\ifnum#1=24 %
\hatcurCCtassmIshortxxxxxC
\else
??????\fi
\fi
\fi
}
\newcommand{\hatcurCCtassmr}[1]{\ifnum#1=22 %
\hatcurCCtassmrxxxxxA
\else
\ifnum#1=23 %
\hatcurCCtassmrxxxxxB
\else
\ifnum#1=24 %
\hatcurCCtassmrxxxxxC
\else
??????\fi
\fi
\fi
}
\newcommand{\hatcurCCtassmrshort}[1]{\ifnum#1=22 %
\hatcurCCtassmrshortxxxxxA
\else
\ifnum#1=23 %
\hatcurCCtassmrshortxxxxxB
\else
\ifnum#1=24 %
\hatcurCCtassmrshortxxxxxC
\else
??????\fi
\fi
\fi
}
\newcommand{\hatcurCCtassmv}[1]{\ifnum#1=22 %
\hatcurCCtassmvxxxxxA
\else
\ifnum#1=23 %
\hatcurCCtassmvxxxxxB
\else
\ifnum#1=24 %
\hatcurCCtassmvxxxxxC
\else
??????\fi
\fi
\fi
}
\newcommand{\hatcurCCtassmvshort}[1]{\ifnum#1=22 %
\hatcurCCtassmvshortxxxxxA
\else
\ifnum#1=23 %
\hatcurCCtassmvshortxxxxxB
\else
\ifnum#1=24 %
\hatcurCCtassmvshortxxxxxC
\else
??????\fi
\fi
\fi
}
\newcommand{\hatcurCCtwomass}[1]{\ifnum#1=22 %
\hatcurCCtwomassxxxxxA
\else
\ifnum#1=23 %
\hatcurCCtwomassxxxxxB
\else
\ifnum#1=24 %
\hatcurCCtwomassxxxxxC
\else
??????\fi
\fi
\fi
}
\newcommand{\hatcurCCtwomassHmag}[1]{\ifnum#1=22 %
\hatcurCCtwomassHmagxxxxxA
\else
\ifnum#1=23 %
\hatcurCCtwomassHmagxxxxxB
\else
\ifnum#1=24 %
\hatcurCCtwomassHmagxxxxxC
\else
??????\fi
\fi
\fi
}
\newcommand{\hatcurCCtwomassJmag}[1]{\ifnum#1=22 %
\hatcurCCtwomassJmagxxxxxA
\else
\ifnum#1=23 %
\hatcurCCtwomassJmagxxxxxB
\else
\ifnum#1=24 %
\hatcurCCtwomassJmagxxxxxC
\else
??????\fi
\fi
\fi
}
\newcommand{\hatcurCCtwomassKmag}[1]{\ifnum#1=22 %
\hatcurCCtwomassKmagxxxxxA
\else
\ifnum#1=23 %
\hatcurCCtwomassKmagxxxxxB
\else
\ifnum#1=24 %
\hatcurCCtwomassKmagxxxxxC
\else
??????\fi
\fi
\fi
}
\newcommand{\hatcurfield}[1]{\ifnum#1=22 %
\hatcurfieldxxxxxA
\else
\ifnum#1=23 %
\hatcurfieldxxxxxB
\else
\ifnum#1=24 %
\hatcurfieldxxxxxC
\else
??????\fi
\fi
\fi
}
\newcommand{\hatcurhtr}[1]{\ifnum#1=22 %
\hatcurhtrxxxxxA
\else
\ifnum#1=23 %
\hatcurhtrxxxxxB
\else
\ifnum#1=24 %
\hatcurhtrxxxxxC
\else
??????\fi
\fi
\fi
}
\newcommand{\hatcurISOage}[1]{\ifnum#1=22 %
\hatcurISOagexxxxxA
\else
\ifnum#1=23 %
\hatcurISOagexxxxxB
\else
\ifnum#1=24 %
\hatcurISOagexxxxxC
\else
??????\fi
\fi
\fi
}
\newcommand{\hatcurISOJK}[1]{\ifnum#1=22 %
\hatcurISOJKxxxxxA
\else
\ifnum#1=23 %
\hatcurISOJKxxxxxB
\else
\ifnum#1=24 %
\hatcurISOJKxxxxxC
\else
??????\fi
\fi
\fi
}
\newcommand{\hatcurISOlogg}[1]{\ifnum#1=22 %
\hatcurISOloggxxxxxA
\else
\ifnum#1=23 %
\hatcurISOloggxxxxxB
\else
\ifnum#1=24 %
\hatcurISOloggxxxxxC
\else
??????\fi
\fi
\fi
}
\newcommand{\hatcurISOlum}[1]{\ifnum#1=22 %
\hatcurISOlumxxxxxA
\else
\ifnum#1=23 %
\hatcurISOlumxxxxxB
\else
\ifnum#1=24 %
\hatcurISOlumxxxxxC
\else
??????\fi
\fi
\fi
}
\newcommand{\hatcurISOlumshort}[1]{\ifnum#1=22 %
\hatcurISOlumshortxxxxxA
\else
\ifnum#1=23 %
\hatcurISOlumshortxxxxxB
\else
\ifnum#1=24 %
\hatcurISOlumshortxxxxxC
\else
??????\fi
\fi
\fi
}
\newcommand{\hatcurISOm}[1]{\ifnum#1=22 %
\hatcurISOmxxxxxA
\else
\ifnum#1=23 %
\hatcurISOmxxxxxB
\else
\ifnum#1=24 %
\hatcurISOmxxxxxC
\else
??????\fi
\fi
\fi
}
\newcommand{\hatcurISOMH}[1]{\ifnum#1=22 %
\hatcurISOMHxxxxxA
\else
\ifnum#1=23 %
\hatcurISOMHxxxxxB
\else
\ifnum#1=24 %
\hatcurISOMHxxxxxC
\else
??????\fi
\fi
\fi
}
\newcommand{\hatcurISOMJ}[1]{\ifnum#1=22 %
\hatcurISOMJxxxxxA
\else
\ifnum#1=23 %
\hatcurISOMJxxxxxB
\else
\ifnum#1=24 %
\hatcurISOMJxxxxxC
\else
??????\fi
\fi
\fi
}
\newcommand{\hatcurISOMK}[1]{\ifnum#1=22 %
\hatcurISOMKxxxxxA
\else
\ifnum#1=23 %
\hatcurISOMKxxxxxB
\else
\ifnum#1=24 %
\hatcurISOMKxxxxxC
\else
??????\fi
\fi
\fi
}
\newcommand{\hatcurISOmlong}[1]{\ifnum#1=22 %
\hatcurISOmlongxxxxxA
\else
\ifnum#1=23 %
\hatcurISOmlongxxxxxB
\else
\ifnum#1=24 %
\hatcurISOmlongxxxxxC
\else
??????\fi
\fi
\fi
}
\newcommand{\hatcurISOmshort}[1]{\ifnum#1=22 %
\hatcurISOmshortxxxxxA
\else
\ifnum#1=23 %
\hatcurISOmshortxxxxxB
\else
\ifnum#1=24 %
\hatcurISOmshortxxxxxC
\else
??????\fi
\fi
\fi
}
\newcommand{\hatcurISOmv}[1]{\ifnum#1=22 %
\hatcurISOmvxxxxxA
\else
\ifnum#1=23 %
\hatcurISOmvxxxxxB
\else
\ifnum#1=24 %
\hatcurISOmvxxxxxC
\else
??????\fi
\fi
\fi
}
\newcommand{\hatcurISOr}[1]{\ifnum#1=22 %
\hatcurISOrxxxxxA
\else
\ifnum#1=23 %
\hatcurISOrxxxxxB
\else
\ifnum#1=24 %
\hatcurISOrxxxxxC
\else
??????\fi
\fi
\fi
}
\newcommand{\hatcurISOrho}[1]{\ifnum#1=22 %
\hatcurISOrhoxxxxxA
\else
\ifnum#1=23 %
\hatcurISOrhoxxxxxB
\else
\ifnum#1=24 %
\hatcurISOrhoxxxxxC
\else
??????\fi
\fi
\fi
}
\newcommand{\hatcurISOrholong}[1]{\ifnum#1=22 %
\hatcurISOrholongxxxxxA
\else
\ifnum#1=23 %
\hatcurISOrholongxxxxxB
\else
\ifnum#1=24 %
\hatcurISOrholongxxxxxC
\else
??????\fi
\fi
\fi
}
\newcommand{\hatcurISOrlong}[1]{\ifnum#1=22 %
\hatcurISOrlongxxxxxA
\else
\ifnum#1=23 %
\hatcurISOrlongxxxxxB
\else
\ifnum#1=24 %
\hatcurISOrlongxxxxxC
\else
??????\fi
\fi
\fi
}
\newcommand{\hatcurISOrshort}[1]{\ifnum#1=22 %
\hatcurISOrshortxxxxxA
\else
\ifnum#1=23 %
\hatcurISOrshortxxxxxB
\else
\ifnum#1=24 %
\hatcurISOrshortxxxxxC
\else
??????\fi
\fi
\fi
}
\newcommand{\hatcurISOsigma}[1]{\ifnum#1=22 %
\hatcurISOsigmaxxxxxA
\else
\ifnum#1=23 %
\hatcurISOsigmaxxxxxB
\else
\ifnum#1=24 %
\hatcurISOsigmaxxxxxC
\else
??????\fi
\fi
\fi
}
\newcommand{\hatcurISOspec}[1]{\ifnum#1=22 %
\hatcurISOspecxxxxxA
\else
\ifnum#1=23 %
\hatcurISOspecxxxxxB
\else
\ifnum#1=24 %
\hatcurISOspecxxxxxC
\else
??????\fi
\fi
\fi
}
\newcommand{\hatcurISOvi}[1]{\ifnum#1=22 %
\hatcurISOvixxxxxA
\else
\ifnum#1=23 %
\hatcurISOvixxxxxB
\else
\ifnum#1=24 %
\hatcurISOvixxxxxC
\else
??????\fi
\fi
\fi
}
\newcommand{\hatcurLBig}[1]{\ifnum#1=22 %
\hatcurLBigxxxxxA
\else
\ifnum#1=23 %
\hatcurLBigxxxxxB
\else
\ifnum#1=24 %
\hatcurLBigxxxxxC
\else
??????\fi
\fi
\fi
}
\newcommand{\hatcurLBii}[1]{\ifnum#1=22 %
\hatcurLBiixxxxxA
\else
\ifnum#1=23 %
\hatcurLBiixxxxxB
\else
\ifnum#1=24 %
\hatcurLBiixxxxxC
\else
??????\fi
\fi
\fi
}
\newcommand{\hatcurLBiI}[1]{\ifnum#1=22 %
\hatcurLBiIxxxxxA
\else
\ifnum#1=23 %
\hatcurLBiIxxxxxB
\else
\ifnum#1=24 %
\hatcurLBiIxxxxxC
\else
??????\fi
\fi
\fi
}
\newcommand{\hatcurLBiig}[1]{\ifnum#1=22 %
\hatcurLBiigxxxxxA
\else
\ifnum#1=23 %
\hatcurLBiigxxxxxB
\else
\ifnum#1=24 %
\hatcurLBiigxxxxxC
\else
??????\fi
\fi
\fi
}
\newcommand{\hatcurLBiii}[1]{\ifnum#1=22 %
\hatcurLBiiixxxxxA
\else
\ifnum#1=23 %
\hatcurLBiiixxxxxB
\else
\ifnum#1=24 %
\hatcurLBiiixxxxxC
\else
??????\fi
\fi
\fi
}
\newcommand{\hatcurLBiiI}[1]{\ifnum#1=22 %
\hatcurLBiiIxxxxxA
\else
\ifnum#1=23 %
\hatcurLBiiIxxxxxB
\else
\ifnum#1=24 %
\hatcurLBiiIxxxxxC
\else
??????\fi
\fi
\fi
}
\newcommand{\hatcurLBiikep}[1]{\ifnum#1=22 %
\hatcurLBiikepxxxxxA
\else
\ifnum#1=23 %
\hatcurLBiikepxxxxxB
\else
\ifnum#1=24 %
\hatcurLBiikepxxxxxC
\else
??????\fi
\fi
\fi
}
\newcommand{\hatcurLBiir}[1]{\ifnum#1=22 %
\hatcurLBiirxxxxxA
\else
\ifnum#1=23 %
\hatcurLBiirxxxxxB
\else
\ifnum#1=24 %
\hatcurLBiirxxxxxC
\else
??????\fi
\fi
\fi
}
\newcommand{\hatcurLBiiR}[1]{\ifnum#1=22 %
\hatcurLBiiRxxxxxA
\else
\ifnum#1=23 %
\hatcurLBiiRxxxxxB
\else
\ifnum#1=24 %
\hatcurLBiiRxxxxxC
\else
??????\fi
\fi
\fi
}
\newcommand{\hatcurLBiiz}[1]{\ifnum#1=22 %
\hatcurLBiizxxxxxA
\else
\ifnum#1=23 %
\hatcurLBiizxxxxxB
\else
\ifnum#1=24 %
\hatcurLBiizxxxxxC
\else
??????\fi
\fi
\fi
}
\newcommand{\hatcurLBikep}[1]{\ifnum#1=22 %
\hatcurLBikepxxxxxA
\else
\ifnum#1=23 %
\hatcurLBikepxxxxxB
\else
\ifnum#1=24 %
\hatcurLBikepxxxxxC
\else
??????\fi
\fi
\fi
}
\newcommand{\hatcurLBir}[1]{\ifnum#1=22 %
\hatcurLBirxxxxxA
\else
\ifnum#1=23 %
\hatcurLBirxxxxxB
\else
\ifnum#1=24 %
\hatcurLBirxxxxxC
\else
??????\fi
\fi
\fi
}
\newcommand{\hatcurLBiR}[1]{\ifnum#1=22 %
\hatcurLBiRxxxxxA
\else
\ifnum#1=23 %
\hatcurLBiRxxxxxB
\else
\ifnum#1=24 %
\hatcurLBiRxxxxxC
\else
??????\fi
\fi
\fi
}
\newcommand{\hatcurLBiz}[1]{\ifnum#1=22 %
\hatcurLBizxxxxxA
\else
\ifnum#1=23 %
\hatcurLBizxxxxxB
\else
\ifnum#1=24 %
\hatcurLBizxxxxxC
\else
??????\fi
\fi
\fi
}
\newcommand{\hatcurLCbsq}[1]{\ifnum#1=22 %
\hatcurLCbsqxxxxxA
\else
\ifnum#1=23 %
\hatcurLCbsqxxxxxB
\else
\ifnum#1=24 %
\hatcurLCbsqxxxxxC
\else
??????\fi
\fi
\fi
}
\newcommand{\hatcurLCbsqsiglowerlim}[1]{\ifnum#1=23 %
\hatcurLCbsqsiglowerlimxxxxxB
\else
??????\fi
}
\newcommand{\hatcurLCdip}[1]{\ifnum#1=22 %
\hatcurLCdipxxxxxA
\else
\ifnum#1=23 %
\hatcurLCdipxxxxxB
\else
\ifnum#1=24 %
\hatcurLCdipxxxxxC
\else
??????\fi
\fi
\fi
}
\newcommand{\hatcurLCdur}[1]{\ifnum#1=22 %
\hatcurLCdurxxxxxA
\else
\ifnum#1=23 %
\hatcurLCdurxxxxxB
\else
\ifnum#1=24 %
\hatcurLCdurxxxxxC
\else
??????\fi
\fi
\fi
}
\newcommand{\hatcurLCdurhr}[1]{\ifnum#1=22 %
\hatcurLCdurhrxxxxxA
\else
\ifnum#1=23 %
\hatcurLCdurhrxxxxxB
\else
\ifnum#1=24 %
\hatcurLCdurhrxxxxxC
\else
??????\fi
\fi
\fi
}
\newcommand{\hatcurLCdurhrshort}[1]{\ifnum#1=22 %
\hatcurLCdurhrshortxxxxxA
\else
\ifnum#1=23 %
\hatcurLCdurhrshortxxxxxB
\else
\ifnum#1=24 %
\hatcurLCdurhrshortxxxxxC
\else
??????\fi
\fi
\fi
}
\newcommand{\hatcurLCdurshort}[1]{\ifnum#1=22 %
\hatcurLCdurshortxxxxxA
\else
\ifnum#1=23 %
\hatcurLCdurshortxxxxxB
\else
\ifnum#1=24 %
\hatcurLCdurshortxxxxxC
\else
??????\fi
\fi
\fi
}
\newcommand{\hatcurLChatnetm}[1]{\ifnum#1=22 %
\hatcurLChatnetmxxxxxA
\else
\ifnum#1=23 %
\hatcurLChatnetmxxxxxB
\else
\ifnum#1=24 %
\hatcurLChatnetmxxxxxC
\else
??????\fi
\fi
\fi
}
\newcommand{\hatcurLCiblend}[1]{\ifnum#1=22 %
\hatcurLCiblendxxxxxA
\else
\ifnum#1=23 %
\hatcurLCiblendxxxxxB
\else
\ifnum#1=24 %
\hatcurLCiblendxxxxxC
\else
??????\fi
\fi
\fi
}
\newcommand{\hatcurLCimp}[1]{\ifnum#1=22 %
\hatcurLCimpxxxxxA
\else
\ifnum#1=23 %
\hatcurLCimpxxxxxB
\else
\ifnum#1=24 %
\hatcurLCimpxxxxxC
\else
??????\fi
\fi
\fi
}
\newcommand{\hatcurLCimpsiglowerlim}[1]{\ifnum#1=23 %
\hatcurLCimpsiglowerlimxxxxxB
\else
??????\fi
}
\newcommand{\hatcurLCingdur}[1]{\ifnum#1=22 %
\hatcurLCingdurxxxxxA
\else
\ifnum#1=23 %
\hatcurLCingdurxxxxxB
\else
\ifnum#1=24 %
\hatcurLCingdurxxxxxC
\else
??????\fi
\fi
\fi
}
\newcommand{\hatcurLCP}[1]{\ifnum#1=22 %
\hatcurLCPxxxxxA
\else
\ifnum#1=23 %
\hatcurLCPxxxxxB
\else
\ifnum#1=24 %
\hatcurLCPxxxxxC
\else
??????\fi
\fi
\fi
}
\newcommand{\hatcurLCPprec}[1]{\ifnum#1=22 %
\hatcurLCPprecxxxxxA
\else
\ifnum#1=23 %
\hatcurLCPprecxxxxxB
\else
\ifnum#1=24 %
\hatcurLCPprecxxxxxC
\else
??????\fi
\fi
\fi
}
\newcommand{\hatcurLCPshort}[1]{\ifnum#1=22 %
\hatcurLCPshortxxxxxA
\else
\ifnum#1=23 %
\hatcurLCPshortxxxxxB
\else
\ifnum#1=24 %
\hatcurLCPshortxxxxxC
\else
??????\fi
\fi
\fi
}
\newcommand{\hatcurLCq}[1]{\ifnum#1=22 %
\hatcurLCqxxxxxA
\else
\ifnum#1=23 %
\hatcurLCqxxxxxB
\else
\ifnum#1=24 %
\hatcurLCqxxxxxC
\else
??????\fi
\fi
\fi
}
\newcommand{\hatcurLCqshort}[1]{\ifnum#1=22 %
\hatcurLCqshortxxxxxA
\else
\ifnum#1=23 %
\hatcurLCqshortxxxxxB
\else
\ifnum#1=24 %
\hatcurLCqshortxxxxxC
\else
??????\fi
\fi
\fi
}
\newcommand{\hatcurLCrho}[1]{\ifnum#1=22 %
\hatcurLCrhoxxxxxA
\else
\ifnum#1=23 %
\hatcurLCrhoxxxxxB
\else
\ifnum#1=24 %
\hatcurLCrhoxxxxxC
\else
??????\fi
\fi
\fi
}
\newcommand{\hatcurLCrprstar}[1]{\ifnum#1=22 %
\hatcurLCrprstarxxxxxA
\else
\ifnum#1=23 %
\hatcurLCrprstarxxxxxB
\else
\ifnum#1=24 %
\hatcurLCrprstarxxxxxC
\else
??????\fi
\fi
\fi
}
\newcommand{\hatcurLCT}[1]{\ifnum#1=22 %
\hatcurLCTxxxxxA
\else
\ifnum#1=23 %
\hatcurLCTxxxxxB
\else
\ifnum#1=24 %
\hatcurLCTxxxxxC
\else
??????\fi
\fi
\fi
}
\newcommand{\hatcurLCTA}[1]{\ifnum#1=22 %
\hatcurLCTAxxxxxA
\else
\ifnum#1=23 %
\hatcurLCTAxxxxxB
\else
\ifnum#1=24 %
\hatcurLCTAxxxxxC
\else
??????\fi
\fi
\fi
}
\newcommand{\hatcurLCTB}[1]{\ifnum#1=22 %
\hatcurLCTBxxxxxA
\else
\ifnum#1=23 %
\hatcurLCTBxxxxxB
\else
\ifnum#1=24 %
\hatcurLCTBxxxxxC
\else
??????\fi
\fi
\fi
}
\newcommand{\hatcurLCzeta}[1]{\ifnum#1=22 %
\hatcurLCzetaxxxxxA
\else
\ifnum#1=23 %
\hatcurLCzetaxxxxxB
\else
\ifnum#1=24 %
\hatcurLCzetaxxxxxC
\else
??????\fi
\fi
\fi
}
\newcommand{\hatcurPPaequiv}[1]{\ifnum#1=22 %
\hatcurPPaequivxxxxxA
\else
\ifnum#1=23 %
\hatcurPPaequivxxxxxB
\else
\ifnum#1=24 %
\hatcurPPaequivxxxxxC
\else
??????\fi
\fi
\fi
}
\newcommand{\hatcurPPar}[1]{\ifnum#1=22 %
\hatcurPParxxxxxA
\else
\ifnum#1=23 %
\hatcurPParxxxxxB
\else
\ifnum#1=24 %
\hatcurPParxxxxxC
\else
??????\fi
\fi
\fi
}
\newcommand{\hatcurPParel}[1]{\ifnum#1=22 %
\hatcurPParelxxxxxA
\else
\ifnum#1=23 %
\hatcurPParelxxxxxB
\else
\ifnum#1=24 %
\hatcurPParelxxxxxC
\else
??????\fi
\fi
\fi
}
\newcommand{\hatcurPPfluxap}[1]{\ifnum#1=22 %
\hatcurPPfluxapxxxxxA
\else
\ifnum#1=23 %
\hatcurPPfluxapxxxxxB
\else
\ifnum#1=24 %
\hatcurPPfluxapxxxxxC
\else
??????\fi
\fi
\fi
}
\newcommand{\hatcurPPfluxapdim}[1]{\ifnum#1=22 %
\hatcurPPfluxapdimxxxxxA
\else
\ifnum#1=23 %
\hatcurPPfluxapdimxxxxxB
\else
\ifnum#1=24 %
\hatcurPPfluxapdimxxxxxC
\else
??????\fi
\fi
\fi
}
\newcommand{\hatcurPPfluxavg}[1]{\ifnum#1=22 %
\hatcurPPfluxavgxxxxxA
\else
\ifnum#1=23 %
\hatcurPPfluxavgxxxxxB
\else
\ifnum#1=24 %
\hatcurPPfluxavgxxxxxC
\else
??????\fi
\fi
\fi
}
\newcommand{\hatcurPPfluxavgdim}[1]{\ifnum#1=22 %
\hatcurPPfluxavgdimxxxxxA
\else
\ifnum#1=23 %
\hatcurPPfluxavgdimxxxxxB
\else
\ifnum#1=24 %
\hatcurPPfluxavgdimxxxxxC
\else
??????\fi
\fi
\fi
}
\newcommand{\hatcurPPfluxavglog}[1]{\ifnum#1=22 %
\hatcurPPfluxavglogxxxxxA
\else
\ifnum#1=23 %
\hatcurPPfluxavglogxxxxxB
\else
\ifnum#1=24 %
\hatcurPPfluxavglogxxxxxC
\else
??????\fi
\fi
\fi
}
\newcommand{\hatcurPPfluxperi}[1]{\ifnum#1=22 %
\hatcurPPfluxperixxxxxA
\else
\ifnum#1=23 %
\hatcurPPfluxperixxxxxB
\else
\ifnum#1=24 %
\hatcurPPfluxperixxxxxC
\else
??????\fi
\fi
\fi
}
\newcommand{\hatcurPPfluxperidim}[1]{\ifnum#1=22 %
\hatcurPPfluxperidimxxxxxA
\else
\ifnum#1=23 %
\hatcurPPfluxperidimxxxxxB
\else
\ifnum#1=24 %
\hatcurPPfluxperidimxxxxxC
\else
??????\fi
\fi
\fi
}
\newcommand{\hatcurPPg}[1]{\ifnum#1=22 %
\hatcurPPgxxxxxA
\else
\ifnum#1=23 %
\hatcurPPgxxxxxB
\else
\ifnum#1=24 %
\hatcurPPgxxxxxC
\else
??????\fi
\fi
\fi
}
\newcommand{\hatcurPPi}[1]{\ifnum#1=22 %
\hatcurPPixxxxxA
\else
\ifnum#1=23 %
\hatcurPPixxxxxB
\else
\ifnum#1=24 %
\hatcurPPixxxxxC
\else
??????\fi
\fi
\fi
}
\newcommand{\hatcurPPitwosigupperlim}[1]{\ifnum#1=23 %
\hatcurPPitwosigupperlimxxxxxB
\else
??????\fi
}
\newcommand{\hatcurPPlogg}[1]{\ifnum#1=22 %
\hatcurPPloggxxxxxA
\else
\ifnum#1=23 %
\hatcurPPloggxxxxxB
\else
\ifnum#1=24 %
\hatcurPPloggxxxxxC
\else
??????\fi
\fi
\fi
}
\newcommand{\hatcurPPm}[1]{\ifnum#1=22 %
\hatcurPPmxxxxxA
\else
\ifnum#1=23 %
\hatcurPPmxxxxxB
\else
\ifnum#1=24 %
\hatcurPPmxxxxxC
\else
??????\fi
\fi
\fi
}
\newcommand{\hatcurPPme}[1]{\ifnum#1=22 %
\hatcurPPmexxxxxA
\else
\ifnum#1=23 %
\hatcurPPmexxxxxB
\else
\ifnum#1=24 %
\hatcurPPmexxxxxC
\else
??????\fi
\fi
\fi
}
\newcommand{\hatcurPPmelong}[1]{\ifnum#1=22 %
\hatcurPPmelongxxxxxA
\else
\ifnum#1=23 %
\hatcurPPmelongxxxxxB
\else
\ifnum#1=24 %
\hatcurPPmelongxxxxxC
\else
??????\fi
\fi
\fi
}
\newcommand{\hatcurPPmeshort}[1]{\ifnum#1=22 %
\hatcurPPmeshortxxxxxA
\else
\ifnum#1=23 %
\hatcurPPmeshortxxxxxB
\else
\ifnum#1=24 %
\hatcurPPmeshortxxxxxC
\else
??????\fi
\fi
\fi
}
\newcommand{\hatcurPPmlong}[1]{\ifnum#1=22 %
\hatcurPPmlongxxxxxA
\else
\ifnum#1=23 %
\hatcurPPmlongxxxxxB
\else
\ifnum#1=24 %
\hatcurPPmlongxxxxxC
\else
??????\fi
\fi
\fi
}
\newcommand{\hatcurPPmrcorr}[1]{\ifnum#1=22 %
\hatcurPPmrcorrxxxxxA
\else
\ifnum#1=23 %
\hatcurPPmrcorrxxxxxB
\else
\ifnum#1=24 %
\hatcurPPmrcorrxxxxxC
\else
??????\fi
\fi
\fi
}
\newcommand{\hatcurPPmshort}[1]{\ifnum#1=22 %
\hatcurPPmshortxxxxxA
\else
\ifnum#1=23 %
\hatcurPPmshortxxxxxB
\else
\ifnum#1=24 %
\hatcurPPmshortxxxxxC
\else
??????\fi
\fi
\fi
}
\newcommand{\hatcurPPperi}[1]{\ifnum#1=22 %
\hatcurPPperixxxxxA
\else
\ifnum#1=23 %
\hatcurPPperixxxxxB
\else
\ifnum#1=24 %
\hatcurPPperixxxxxC
\else
??????\fi
\fi
\fi
}
\newcommand{\hatcurPPphiconj}[1]{\ifnum#1=22 %
\hatcurPPphiconjxxxxxA
\else
\ifnum#1=23 %
\hatcurPPphiconjxxxxxB
\else
\ifnum#1=24 %
\hatcurPPphiconjxxxxxC
\else
??????\fi
\fi
\fi
}
\newcommand{\hatcurPPr}[1]{\ifnum#1=22 %
\hatcurPPrxxxxxA
\else
\ifnum#1=23 %
\hatcurPPrxxxxxB
\else
\ifnum#1=24 %
\hatcurPPrxxxxxC
\else
??????\fi
\fi
\fi
}
\newcommand{\hatcurPPre}[1]{\ifnum#1=22 %
\hatcurPPrexxxxxA
\else
\ifnum#1=23 %
\hatcurPPrexxxxxB
\else
\ifnum#1=24 %
\hatcurPPrexxxxxC
\else
??????\fi
\fi
\fi
}
\newcommand{\hatcurPPrelong}[1]{\ifnum#1=22 %
\hatcurPPrelongxxxxxA
\else
\ifnum#1=23 %
\hatcurPPrelongxxxxxB
\else
\ifnum#1=24 %
\hatcurPPrelongxxxxxC
\else
??????\fi
\fi
\fi
}
\newcommand{\hatcurPPreshort}[1]{\ifnum#1=22 %
\hatcurPPreshortxxxxxA
\else
\ifnum#1=23 %
\hatcurPPreshortxxxxxB
\else
\ifnum#1=24 %
\hatcurPPreshortxxxxxC
\else
??????\fi
\fi
\fi
}
\newcommand{\hatcurPPrho}[1]{\ifnum#1=22 %
\hatcurPPrhoxxxxxA
\else
\ifnum#1=23 %
\hatcurPPrhoxxxxxB
\else
\ifnum#1=24 %
\hatcurPPrhoxxxxxC
\else
??????\fi
\fi
\fi
}
\newcommand{\hatcurPPrlong}[1]{\ifnum#1=22 %
\hatcurPPrlongxxxxxA
\else
\ifnum#1=23 %
\hatcurPPrlongxxxxxB
\else
\ifnum#1=24 %
\hatcurPPrlongxxxxxC
\else
??????\fi
\fi
\fi
}
\newcommand{\hatcurPPrshort}[1]{\ifnum#1=22 %
\hatcurPPrshortxxxxxA
\else
\ifnum#1=23 %
\hatcurPPrshortxxxxxB
\else
\ifnum#1=24 %
\hatcurPPrshortxxxxxC
\else
??????\fi
\fi
\fi
}
\newcommand{\hatcurPPrtwosiglowerlim}[1]{\ifnum#1=23 %
\hatcurPPrtwosiglowerlimxxxxxB
\else
??????\fi
}
\newcommand{\hatcurPPtcirc}[1]{\ifnum#1=22 %
\hatcurPPtcircxxxxxA
\else
\ifnum#1=23 %
\hatcurPPtcircxxxxxB
\else
\ifnum#1=24 %
\hatcurPPtcircxxxxxC
\else
??????\fi
\fi
\fi
}
\newcommand{\hatcurPPteff}[1]{\ifnum#1=22 %
\hatcurPPteffxxxxxA
\else
\ifnum#1=23 %
\hatcurPPteffxxxxxB
\else
\ifnum#1=24 %
\hatcurPPteffxxxxxC
\else
??????\fi
\fi
\fi
}
\newcommand{\hatcurPPtheta}[1]{\ifnum#1=22 %
\hatcurPPthetaxxxxxA
\else
\ifnum#1=23 %
\hatcurPPthetaxxxxxB
\else
\ifnum#1=24 %
\hatcurPPthetaxxxxxC
\else
??????\fi
\fi
\fi
}
\newcommand{\hatcurPPtinfall}[1]{\ifnum#1=22 %
\hatcurPPtinfallxxxxxA
\else
\ifnum#1=23 %
\hatcurPPtinfallxxxxxB
\else
\ifnum#1=24 %
\hatcurPPtinfallxxxxxC
\else
??????\fi
\fi
\fi
}
\newcommand{\hatcurRVeccen}[1]{\ifnum#1=22 %
\hatcurRVeccenxxxxxA
\else
\ifnum#1=23 %
\hatcurRVeccenxxxxxB
\else
\ifnum#1=24 %
\hatcurRVeccenxxxxxC
\else
??????\fi
\fi
\fi
}
\newcommand{\hatcurRVeccentwosiglim}[1]{\ifnum#1=22 %
\hatcurRVeccentwosiglimxxxxxA
\else
\ifnum#1=23 %
\hatcurRVeccentwosiglimxxxxxB
\else
\ifnum#1=24 %
\hatcurRVeccentwosiglimxxxxxC
\else
??????\fi
\fi
\fi
}
\newcommand{\hatcurRVfitrms}[1]{\ifnum#1=23 %
\hatcurRVfitrmsxxxxxB
\else
??????\fi
}
\newcommand{\hatcurRVfitrmsA}[1]{\ifnum#1=22 %
\hatcurRVfitrmsAxxxxxA
\else
\ifnum#1=24 %
\hatcurRVfitrmsAxxxxxC
\else
??????\fi
\fi
}
\newcommand{\hatcurRVfitrmsB}[1]{\ifnum#1=22 %
\hatcurRVfitrmsBxxxxxA
\else
\ifnum#1=24 %
\hatcurRVfitrmsBxxxxxC
\else
??????\fi
\fi
}
\newcommand{\hatcurRVfitrmsC}[1]{\ifnum#1=22 %
\hatcurRVfitrmsCxxxxxA
\else
\ifnum#1=24 %
\hatcurRVfitrmsCxxxxxC
\else
??????\fi
\fi
}
\newcommand{\hatcurRVfitrmsD}[1]{\ifnum#1=24 %
\hatcurRVfitrmsDxxxxxC
\else
??????\fi
}
\newcommand{\hatcurRVgamma}[1]{\ifnum#1=23 %
\hatcurRVgammaxxxxxB
\else
??????\fi
}
\newcommand{\hatcurRVgammaA}[1]{\ifnum#1=22 %
\hatcurRVgammaAxxxxxA
\else
\ifnum#1=24 %
\hatcurRVgammaAxxxxxC
\else
??????\fi
\fi
}
\newcommand{\hatcurRVgammaB}[1]{\ifnum#1=22 %
\hatcurRVgammaBxxxxxA
\else
\ifnum#1=24 %
\hatcurRVgammaBxxxxxC
\else
??????\fi
\fi
}
\newcommand{\hatcurRVgammaC}[1]{\ifnum#1=22 %
\hatcurRVgammaCxxxxxA
\else
\ifnum#1=24 %
\hatcurRVgammaCxxxxxC
\else
??????\fi
\fi
}
\newcommand{\hatcurRVgammaD}[1]{\ifnum#1=24 %
\hatcurRVgammaDxxxxxC
\else
??????\fi
}
\newcommand{\hatcurRVh}[1]{\ifnum#1=22 %
\hatcurRVhxxxxxA
\else
\ifnum#1=23 %
\hatcurRVhxxxxxB
\else
\ifnum#1=24 %
\hatcurRVhxxxxxC
\else
??????\fi
\fi
\fi
}
\newcommand{\hatcurRVjitter}[1]{\ifnum#1=23 %
\hatcurRVjitterxxxxxB
\else
??????\fi
}
\newcommand{\hatcurRVjitterA}[1]{\ifnum#1=22 %
\hatcurRVjitterAxxxxxA
\else
\ifnum#1=24 %
\hatcurRVjitterAxxxxxC
\else
??????\fi
\fi
}
\newcommand{\hatcurRVjitterB}[1]{\ifnum#1=22 %
\hatcurRVjitterBxxxxxA
\else
\ifnum#1=24 %
\hatcurRVjitterBxxxxxC
\else
??????\fi
\fi
}
\newcommand{\hatcurRVjitterC}[1]{\ifnum#1=22 %
\hatcurRVjitterCxxxxxA
\else
\ifnum#1=24 %
\hatcurRVjitterCxxxxxC
\else
??????\fi
\fi
}
\newcommand{\hatcurRVjitterD}[1]{\ifnum#1=24 %
\hatcurRVjitterDxxxxxC
\else
??????\fi
}
\newcommand{\hatcurRVjittertwosiglim}[1]{\ifnum#1=23 %
\hatcurRVjittertwosiglimxxxxxB
\else
??????\fi
}
\newcommand{\hatcurRVjittertwosiglimA}[1]{\ifnum#1=22 %
\hatcurRVjittertwosiglimAxxxxxA
\else
\ifnum#1=24 %
\hatcurRVjittertwosiglimAxxxxxC
\else
??????\fi
\fi
}
\newcommand{\hatcurRVjittertwosiglimB}[1]{\ifnum#1=22 %
\hatcurRVjittertwosiglimBxxxxxA
\else
\ifnum#1=24 %
\hatcurRVjittertwosiglimBxxxxxC
\else
??????\fi
\fi
}
\newcommand{\hatcurRVjittertwosiglimC}[1]{\ifnum#1=22 %
\hatcurRVjittertwosiglimCxxxxxA
\else
\ifnum#1=24 %
\hatcurRVjittertwosiglimCxxxxxC
\else
??????\fi
\fi
}
\newcommand{\hatcurRVjittertwosiglimD}[1]{\ifnum#1=24 %
\hatcurRVjittertwosiglimDxxxxxC
\else
??????\fi
}
\newcommand{\hatcurRVk}[1]{\ifnum#1=22 %
\hatcurRVkxxxxxA
\else
\ifnum#1=23 %
\hatcurRVkxxxxxB
\else
\ifnum#1=24 %
\hatcurRVkxxxxxC
\else
??????\fi
\fi
\fi
}
\newcommand{\hatcurRVK}[1]{\ifnum#1=22 %
\hatcurRVKxxxxxA
\else
\ifnum#1=23 %
\hatcurRVKxxxxxB
\else
\ifnum#1=24 %
\hatcurRVKxxxxxC
\else
??????\fi
\fi
\fi
}
\newcommand{\hatcurRVomega}[1]{\ifnum#1=22 %
\hatcurRVomegaxxxxxA
\else
\ifnum#1=23 %
\hatcurRVomegaxxxxxB
\else
\ifnum#1=24 %
\hatcurRVomegaxxxxxC
\else
??????\fi
\fi
\fi
}
\newcommand{\hatcurRVrh}[1]{\ifnum#1=22 %
\hatcurRVrhxxxxxA
\else
\ifnum#1=23 %
\hatcurRVrhxxxxxB
\else
\ifnum#1=24 %
\hatcurRVrhxxxxxC
\else
??????\fi
\fi
\fi
}
\newcommand{\hatcurRVrk}[1]{\ifnum#1=22 %
\hatcurRVrkxxxxxA
\else
\ifnum#1=23 %
\hatcurRVrkxxxxxB
\else
\ifnum#1=24 %
\hatcurRVrkxxxxxC
\else
??????\fi
\fi
\fi
}
\newcommand{\hatcurRVtrone}[1]{\ifnum#1=22 %
\hatcurRVtronexxxxxA
\else
\ifnum#1=23 %
\hatcurRVtronexxxxxB
\else
\ifnum#1=24 %
\hatcurRVtronexxxxxC
\else
??????\fi
\fi
\fi
}
\newcommand{\hatcurRVtrtwo}[1]{\ifnum#1=22 %
\hatcurRVtrtwoxxxxxA
\else
\ifnum#1=23 %
\hatcurRVtrtwoxxxxxB
\else
\ifnum#1=24 %
\hatcurRVtrtwoxxxxxC
\else
??????\fi
\fi
\fi
}
\newcommand{\hatcurSMEiilogg}[1]{\ifnum#1=22 %
\hatcurSMEiiloggxxxxxA
\else
\ifnum#1=23 %
\hatcurSMEiiloggxxxxxB
\else
\ifnum#1=24 %
\hatcurSMEiiloggxxxxxC
\else
??????\fi
\fi
\fi
}
\newcommand{\hatcurSMEiiteff}[1]{\ifnum#1=22 %
\hatcurSMEiiteffxxxxxA
\else
\ifnum#1=23 %
\hatcurSMEiiteffxxxxxB
\else
\ifnum#1=24 %
\hatcurSMEiiteffxxxxxC
\else
??????\fi
\fi
\fi
}
\newcommand{\hatcurSMEiivsin}[1]{\ifnum#1=22 %
\hatcurSMEiivsinxxxxxA
\else
\ifnum#1=23 %
\hatcurSMEiivsinxxxxxB
\else
\ifnum#1=24 %
\hatcurSMEiivsinxxxxxC
\else
??????\fi
\fi
\fi
}
\newcommand{\hatcurSMEiizfeh}[1]{\ifnum#1=22 %
\hatcurSMEiizfehxxxxxA
\else
\ifnum#1=23 %
\hatcurSMEiizfehxxxxxB
\else
\ifnum#1=24 %
\hatcurSMEiizfehxxxxxC
\else
??????\fi
\fi
\fi
}
\newcommand{\hatcurSMEiizfehshort}[1]{\ifnum#1=22 %
\hatcurSMEiizfehshortxxxxxA
\else
\ifnum#1=23 %
\hatcurSMEiizfehshortxxxxxB
\else
\ifnum#1=24 %
\hatcurSMEiizfehshortxxxxxC
\else
??????\fi
\fi
\fi
}
\newcommand{\hatcurSMEilogg}[1]{\ifnum#1=22 %
\hatcurSMEiloggxxxxxA
\else
\ifnum#1=23 %
\hatcurSMEiloggxxxxxB
\else
\ifnum#1=24 %
\hatcurSMEiloggxxxxxC
\else
??????\fi
\fi
\fi
}
\newcommand{\hatcurSMEiteff}[1]{\ifnum#1=22 %
\hatcurSMEiteffxxxxxA
\else
\ifnum#1=23 %
\hatcurSMEiteffxxxxxB
\else
\ifnum#1=24 %
\hatcurSMEiteffxxxxxC
\else
??????\fi
\fi
\fi
}
\newcommand{\hatcurSMEivmac}[1]{\ifnum#1=22 %
\hatcurSMEivmacxxxxxA
\else
\ifnum#1=23 %
\hatcurSMEivmacxxxxxB
\else
\ifnum#1=24 %
\hatcurSMEivmacxxxxxC
\else
??????\fi
\fi
\fi
}
\newcommand{\hatcurSMEivmic}[1]{\ifnum#1=22 %
\hatcurSMEivmicxxxxxA
\else
\ifnum#1=23 %
\hatcurSMEivmicxxxxxB
\else
\ifnum#1=24 %
\hatcurSMEivmicxxxxxC
\else
??????\fi
\fi
\fi
}
\newcommand{\hatcurSMEivsin}[1]{\ifnum#1=22 %
\hatcurSMEivsinxxxxxA
\else
\ifnum#1=23 %
\hatcurSMEivsinxxxxxB
\else
\ifnum#1=24 %
\hatcurSMEivsinxxxxxC
\else
??????\fi
\fi
\fi
}
\newcommand{\hatcurSMEizfeh}[1]{\ifnum#1=22 %
\hatcurSMEizfehxxxxxA
\else
\ifnum#1=23 %
\hatcurSMEizfehxxxxxB
\else
\ifnum#1=24 %
\hatcurSMEizfehxxxxxC
\else
??????\fi
\fi
\fi
}
\newcommand{\hatcurSMEizfehshort}[1]{\ifnum#1=22 %
\hatcurSMEizfehshortxxxxxA
\else
\ifnum#1=23 %
\hatcurSMEizfehshortxxxxxB
\else
\ifnum#1=24 %
\hatcurSMEizfehshortxxxxxC
\else
??????\fi
\fi
\fi
}
\newcommand{\hatcurXAv}[1]{\ifnum#1=22 %
\hatcurXAvxxxxxA
\else
\ifnum#1=23 %
\hatcurXAvxxxxxB
\else
\ifnum#1=24 %
\hatcurXAvxxxxxC
\else
??????\fi
\fi
\fi
}
\newcommand{\hatcurXdist}[1]{\ifnum#1=22 %
\hatcurXdistxxxxxA
\else
\ifnum#1=23 %
\hatcurXdistxxxxxB
\else
\ifnum#1=24 %
\hatcurXdistxxxxxC
\else
??????\fi
\fi
\fi
}
\newcommand{\hatcurXdistred}[1]{\ifnum#1=22 %
\hatcurXdistredxxxxxA
\else
\ifnum#1=23 %
\hatcurXdistredxxxxxB
\else
\ifnum#1=24 %
\hatcurXdistredxxxxxC
\else
??????\fi
\fi
\fi
}
\newcommand{\hatcurXEBV}[1]{\ifnum#1=22 %
\hatcurXEBVxxxxxA
\else
\ifnum#1=23 %
\hatcurXEBVxxxxxB
\else
\ifnum#1=24 %
\hatcurXEBVxxxxxC
\else
??????\fi
\fi
\fi
}
\newcommand{\hatcurXjhisored}[1]{\ifnum#1=22 %
\hatcurXjhisoredxxxxxA
\else
\ifnum#1=23 %
\hatcurXjhisoredxxxxxB
\else
\ifnum#1=24 %
\hatcurXjhisoredxxxxxC
\else
??????\fi
\fi
\fi
}
\newcommand{\hatcurXjkisored}[1]{\ifnum#1=22 %
\hatcurXjkisoredxxxxxA
\else
\ifnum#1=23 %
\hatcurXjkisoredxxxxxB
\else
\ifnum#1=24 %
\hatcurXjkisoredxxxxxC
\else
??????\fi
\fi
\fi
}
\newcommand{\hatcurXmhisored}[1]{\ifnum#1=22 %
\hatcurXmhisoredxxxxxA
\else
\ifnum#1=23 %
\hatcurXmhisoredxxxxxB
\else
\ifnum#1=24 %
\hatcurXmhisoredxxxxxC
\else
??????\fi
\fi
\fi
}
\newcommand{\hatcurXmiisored}[1]{\ifnum#1=22 %
\hatcurXmiisoredxxxxxA
\else
\ifnum#1=23 %
\hatcurXmiisoredxxxxxB
\else
\ifnum#1=24 %
\hatcurXmiisoredxxxxxC
\else
??????\fi
\fi
\fi
}
\newcommand{\hatcurXmjisored}[1]{\ifnum#1=22 %
\hatcurXmjisoredxxxxxA
\else
\ifnum#1=23 %
\hatcurXmjisoredxxxxxB
\else
\ifnum#1=24 %
\hatcurXmjisoredxxxxxC
\else
??????\fi
\fi
\fi
}
\newcommand{\hatcurXmkisored}[1]{\ifnum#1=22 %
\hatcurXmkisoredxxxxxA
\else
\ifnum#1=23 %
\hatcurXmkisoredxxxxxB
\else
\ifnum#1=24 %
\hatcurXmkisoredxxxxxC
\else
??????\fi
\fi
\fi
}
\newcommand{\hatcurXmvisored}[1]{\ifnum#1=22 %
\hatcurXmvisoredxxxxxA
\else
\ifnum#1=23 %
\hatcurXmvisoredxxxxxB
\else
\ifnum#1=24 %
\hatcurXmvisoredxxxxxC
\else
??????\fi
\fi
\fi
}
\newcommand{\hatcurXsecdur}[1]{\ifnum#1=22 %
\hatcurXsecdurxxxxxA
\else
\ifnum#1=23 %
\hatcurXsecdurxxxxxB
\else
\ifnum#1=24 %
\hatcurXsecdurxxxxxC
\else
??????\fi
\fi
\fi
}
\newcommand{\hatcurXsecingdur}[1]{\ifnum#1=22 %
\hatcurXsecingdurxxxxxA
\else
\ifnum#1=23 %
\hatcurXsecingdurxxxxxB
\else
\ifnum#1=24 %
\hatcurXsecingdurxxxxxC
\else
??????\fi
\fi
\fi
}
\newcommand{\hatcurXsecondary}[1]{\ifnum#1=22 %
\hatcurXsecondaryxxxxxA
\else
\ifnum#1=23 %
\hatcurXsecondaryxxxxxB
\else
\ifnum#1=24 %
\hatcurXsecondaryxxxxxC
\else
??????\fi
\fi
\fi
}
\newcommand{\hatcurXsecphase}[1]{\ifnum#1=22 %
\hatcurXsecphasexxxxxA
\else
\ifnum#1=23 %
\hatcurXsecphasexxxxxB
\else
\ifnum#1=24 %
\hatcurXsecphasexxxxxC
\else
??????\fi
\fi
\fi
}
\newcommand{\hatcurXviisored}[1]{\ifnum#1=22 %
\hatcurXviisoredxxxxxA
\else
\ifnum#1=23 %
\hatcurXviisoredxxxxxB
\else
\ifnum#1=24 %
\hatcurXviisoredxxxxxC
\else
??????\fi
\fi
\fi
}
\newcommand{\hatcurXvkisored}[1]{\ifnum#1=22 %
\hatcurXvkisoredxxxxxA
\else
\ifnum#1=23 %
\hatcurXvkisoredxxxxxB
\else
\ifnum#1=24 %
\hatcurXvkisoredxxxxxC
\else
??????\fi
\fi
\fi
}

\newcommand{\hatcurhtreccenxxxxxA}{HATS610-015}                        
\newcommand{\hatcurfieldeccenxxxxxA}{\ensuremath{string}}              
\newcommand{\hatcurCCraeccenxxxxxA}{\ensuremath{11^{\mathrm h}36^{\mathrm m}02.16{\mathrm s}}}                       
\newcommand{\hatcurCCdececcenxxxxxA}{\ensuremath{-29{\arcdeg}32{\arcmin}35.9{\arcsec}}}                      
\newcommand{\hatcurCCmageccenxxxxxA}{13.455}                           
\newcommand{\hatcurCCtwomasseccenxxxxxA}{2MASS~11360233-2932359}       
\newcommand{\hatcurCCgsceccenxxxxxA}{GSC~6664-00373}                   
\newcommand{\hatcurCCtassmveccenxxxxxA}{\ensuremath{13.455\pm0.040}}   
\newcommand{\hatcurCCtassmvshorteccenxxxxxA}{\ensuremath{13.5}}        
\newcommand{\hatcurCCtassmBeccenxxxxxA}{\ensuremath{14.496\pm0.040}}   
\newcommand{\hatcurCCtassmBshorteccenxxxxxA}{\ensuremath{14.5}}        
\newcommand{\hatcurCCtassmIeccenxxxxxA}{\ensuremath{100\pm1000}}       
\newcommand{\hatcurCCtassmIshorteccenxxxxxA}{\ensuremath{100.0}}       
\newcommand{\hatcurCCtassmgeccenxxxxxA}{\ensuremath{13.943\pm0.030}}   
\newcommand{\hatcurCCtassmgshorteccenxxxxxA}{\ensuremath{13.9}}        
\newcommand{\hatcurCCtassmreccenxxxxxA}{\ensuremath{13.056\pm0.030}}   
\newcommand{\hatcurCCtassmrshorteccenxxxxxA}{\ensuremath{13.1}}        
\newcommand{\hatcurCCtassmieccenxxxxxA}{\ensuremath{12.82\pm0.14}}     
\newcommand{\hatcurCCtassmishorteccenxxxxxA}{\ensuremath{12.8}}        
\newcommand{\hatcurCCtwomassJmageccenxxxxxA}{\ensuremath{11.556\pm0.023}} 
\newcommand{\hatcurCCtwomassHmageccenxxxxxA}{\ensuremath{11.006\pm0.022}} 
\newcommand{\hatcurCCtwomassKmageccenxxxxxA}{\ensuremath{10.942\pm0.019}} 
\newcommand{\hatcurCCcitJmageccenxxxxxA}{\ensuremath{11.560\pm0.024}}  
\newcommand{\hatcurCCcitHmageccenxxxxxA}{\ensuremath{11.001\pm0.023}}  
\newcommand{\hatcurCCcitKmageccenxxxxxA}{\ensuremath{10.966\pm0.020}}  
\newcommand{\hatcurCCbbJmageccenxxxxxA}{\ensuremath{11.629\pm0.026}}   
\newcommand{\hatcurCCbbHmageccenxxxxxA}{\ensuremath{11.023\pm0.024}}   
\newcommand{\hatcurCCbbKmageccenxxxxxA}{\ensuremath{10.986\pm0.020}}   
\newcommand{\hatcurCCesoJmageccenxxxxxA}{\ensuremath{11.634\pm0.028}}  
\newcommand{\hatcurCCesoHmageccenxxxxxA}{\ensuremath{11.016\pm0.026}}  
\newcommand{\hatcurCCesoKmageccenxxxxxA}{\ensuremath{10.984\pm0.021}}  
\newcommand{\hatcurCCesoJHmageccenxxxxxA}{\ensuremath{0.618\pm0.036}}  
\newcommand{\hatcurCCesoJKmageccenxxxxxA}{\ensuremath{0.651\pm0.034}}  
\newcommand{\hatcurCCesoHKmageccenxxxxxA}{\ensuremath{0.032\pm0.033}}  
\newcommand{\hatcurLCdipeccenxxxxxA}{\ensuremath{23.8}}                
\newcommand{\hatcurLCrprstareccenxxxxxA}{\ensuremath{0.1426\pm0.0025}} 
\newcommand{\hatcurLCbsqeccenxxxxxA}{\ensuremath{0.287_{-0.064}^{+0.040}}} 
\newcommand{\hatcurLCimpeccenxxxxxA}{\ensuremath{0.536_{-0.064}^{+0.036}}} 
\newcommand{\hatcurLCzetaeccenxxxxxA}{\ensuremath{26.09_{-0.19}^{+0.28}}} 
\newcommand{\hatcurLCdureccenxxxxxA}{\ensuremath{0.0913\pm0.0015}}     
\newcommand{\hatcurLCdurshorteccenxxxxxA}{\ensuremath{0.0913}}         
\newcommand{\hatcurLCdurhreccenxxxxxA}{\ensuremath{2.191\pm0.035}}     
\newcommand{\hatcurLCdurhrshorteccenxxxxxA}{\ensuremath{2.191}}        
\newcommand{\hatcurLCqeccenxxxxxA}{\ensuremath{0.01930\pm0.00031}}     
\newcommand{\hatcurLCqshorteccenxxxxxA}{\ensuremath{0.019}}            
\newcommand{\hatcurLCingdureccenxxxxxA}{\ensuremath{0.0154\pm0.0014}}  
\newcommand{\hatcurLCPeccenxxxxxA}{\ensuremath{4.7228124\pm0.0000052}} 
\newcommand{\hatcurLCPprececcenxxxxxA}{\ensuremath{4.7228124}}         
\newcommand{\hatcurLCPshorteccenxxxxxA}{\ensuremath{4.7228}}           
\newcommand{\hatcurLCTeccenxxxxxA}{\ensuremath{2457078.58030\pm0.00022}} 
\newcommand{\hatcurLCTAeccenxxxxxA}{\ensuremath{2455671.1822\pm0.0015}} 
\newcommand{\hatcurLCTBeccenxxxxxA}{\ensuremath{2457111.63997\pm0.00023}} 
\newcommand{\hatcurLChatnetmeccenxxxxxA}{\ensuremath{13.130850\pm0.000056}} 
\newcommand{\hatcurLCiblendeccenxxxxxA}{\ensuremath{0.865\pm0.027}}    
\newcommand{\hatcurLCrhoeccenxxxxxA}{\ensuremath{3.26\pm0.68}}         
\newcommand{\hatcurSMEiteffeccenxxxxxA}{\ensuremath{4771\pm80}}        
\newcommand{\hatcurSMEizfeheccenxxxxxA}{\ensuremath{0.000\pm0.060}}    
\newcommand{\hatcurSMEizfehshorteccenxxxxxA}{\ensuremath{0.00}}        
\newcommand{\hatcurSMEiloggeccenxxxxxA}{\ensuremath{4.58\pm0.16}}      
\newcommand{\hatcurSMEivsineccenxxxxxA}{\ensuremath{0.5\pm1.7}}        
\newcommand{\hatcurSMEivmaceccenxxxxxA}{\ensuremath{0.0}}              
\newcommand{\hatcurSMEivmiceccenxxxxxA}{\ensuremath{0.0}}              
\newcommand{\hatcurSMEiiteffeccenxxxxxA}{\ensuremath{4803\pm55}}       
\newcommand{\hatcurSMEiizfeheccenxxxxxA}{\ensuremath{0.000\pm0.040}}   
\newcommand{\hatcurSMEiizfehshorteccenxxxxxA}{\ensuremath{0.0}}        
\newcommand{\hatcurSMEiiloggeccenxxxxxA}{\ensuremath{4.664\pm0.021}}   
\newcommand{\hatcurSMEiivsineccenxxxxxA}{\ensuremath{0.50\pm0.50}}     
\newcommand{\hatcurLBizeccenxxxxxA}{\ensuremath{0.3472}}               
\newcommand{\hatcurLBiizeccenxxxxxA}{\ensuremath{0.2502}}              
\newcommand{\hatcurLBiieccenxxxxxA}{\ensuremath{0.4355}}               
\newcommand{\hatcurLBiiieccenxxxxxA}{\ensuremath{0.2255}}              
\newcommand{\hatcurLBiIeccenxxxxxA}{\ensuremath{0.4055}}               
\newcommand{\hatcurLBiiIeccenxxxxxA}{\ensuremath{0.2342}}              
\newcommand{\hatcurLBigeccenxxxxxA}{\ensuremath{0.8412}}               
\newcommand{\hatcurLBiigeccenxxxxxA}{\ensuremath{-0.0025}}             
\newcommand{\hatcurLBireccenxxxxxA}{\ensuremath{0.5808}}               
\newcommand{\hatcurLBiireccenxxxxxA}{\ensuremath{0.1721}}              
\newcommand{\hatcurLBiReccenxxxxxA}{\ensuremath{0.5404}}               
\newcommand{\hatcurLBiiReccenxxxxxA}{\ensuremath{0.1879}}              
\newcommand{\hatcurLBikepeccenxxxxxA}{\ensuremath{0.1000}}             
\newcommand{\hatcurLBiikepeccenxxxxxA}{\ensuremath{0.1000}}            
\newcommand{\hatcurISOmeccenxxxxxA}{\ensuremath{0.759\pm0.019}}        
\newcommand{\hatcurISOmshorteccenxxxxxA}{\ensuremath{0.76}}            
\newcommand{\hatcurISOmlongeccenxxxxxA}{\ensuremath{0.759\pm0.019}}    
\newcommand{\hatcurISOreccenxxxxxA}{\ensuremath{0.689_{-0.018}^{+0.028}}} 
\newcommand{\hatcurISOrshorteccenxxxxxA}{\ensuremath{0.69}}            
\newcommand{\hatcurISOrlongeccenxxxxxA}{\ensuremath{0.689_{-0.018}^{+0.028}}} 
\newcommand{\hatcurISOrhoeccenxxxxxA}{\ensuremath{3.28_{-0.39}^{+0.25}}} 
\newcommand{\hatcurISOrholongeccenxxxxxA}{\ensuremath{3.28_{-0.39}^{+0.25}}} 
\newcommand{\hatcurISOloggeccenxxxxxA}{\ensuremath{4.644\pm0.028}}     
\newcommand{\hatcurISOlumeccenxxxxxA}{\ensuremath{0.226_{-0.020}^{+0.026}}} 
\newcommand{\hatcurISOlumshorteccenxxxxxA}{\ensuremath{0.23}}          
\newcommand{\hatcurISOmveccenxxxxxA}{\ensuremath{6.71\pm0.13}}         
\newcommand{\hatcurISOvieccenxxxxxA}{\ensuremath{0.986\pm0.028}}       
\newcommand{\hatcurISOageeccenxxxxxA}{\ensuremath{4.6_{-4.0}^{+5.8}}}  
\newcommand{\hatcurISOsigmaeccenxxxxxA}{\ensuremath{0.00770\pm0.00093}} 
\newcommand{\hatcurISOMJeccenxxxxxA}{\ensuremath{4.980\pm0.092}}       
\newcommand{\hatcurISOMHeccenxxxxxA}{\ensuremath{4.449\pm0.083}}       
\newcommand{\hatcurISOMKeccenxxxxxA}{\ensuremath{4.359\pm0.081}}       
\newcommand{\hatcurISOJKeccenxxxxxA}{\ensuremath{0.620\pm0.010}}       
\newcommand{\hatcurISOspececcenxxxxxA}{K}                              
\newcommand{\hatcurRVKeccenxxxxxA}{\ensuremath{399\pm15}}              
\newcommand{\hatcurRVrkeccenxxxxxA}{\ensuremath{0.168_{-0.062}^{+0.046}}} 
\newcommand{\hatcurRVrheccenxxxxxA}{\ensuremath{0.225_{-0.126}^{+0.084}}} 
\newcommand{\hatcurRVkeccenxxxxxA}{\ensuremath{0.043_{-0.011}^{+0.016}}} 
\newcommand{\hatcurRVheccenxxxxxA}{\ensuremath{0.061\pm0.040}}         
\newcommand{\hatcurRVtroneeccenxxxxxA}{\ensuremath{0\pm0}}             
\newcommand{\hatcurRVtrtwoeccenxxxxxA}{\ensuremath{0\pm0}}             
\newcommand{\hatcurRVgammaAeccenxxxxxA}{\ensuremath{-7438\pm32}}       
\newcommand{\hatcurRVjitterAeccenxxxxxA}{\ensuremath{35\pm28}}         
\newcommand{\hatcurRVjittertwosiglimAeccenxxxxxA}{\ensuremath{<98.5}}  
\newcommand{\hatcurRVfitrmsAeccenxxxxxA}{\ensuremath{0.0}}             
\newcommand{\hatcurRVgammaBeccenxxxxxA}{\ensuremath{-7413.9\pm9.4}}    
\newcommand{\hatcurRVjitterBeccenxxxxxA}{\ensuremath{1\pm16}}          
\newcommand{\hatcurRVjittertwosiglimBeccenxxxxxA}{\ensuremath{<36.5}}  
\newcommand{\hatcurRVfitrmsBeccenxxxxxA}{\ensuremath{0.0}}             
\newcommand{\hatcurRVgammaCeccenxxxxxA}{\ensuremath{-7370\pm23}}       
\newcommand{\hatcurRVjitterCeccenxxxxxA}{\ensuremath{1\pm41}}          
\newcommand{\hatcurRVjittertwosiglimCeccenxxxxxA}{\ensuremath{<68.1}}  
\newcommand{\hatcurRVfitrmsCeccenxxxxxA}{\ensuremath{0.0}}             
\newcommand{\hatcurRVecceneccenxxxxxA}{\ensuremath{0.079\pm0.026}}     
\newcommand{\hatcurRVeccentwosiglimeccenxxxxxA}{\ensuremath{<0.132}}   
\newcommand{\hatcurRVomegaeccenxxxxxA}{\ensuremath{56\pm73}}           
\newcommand{\hatcurPPieccenxxxxxA}{\ensuremath{87.96\pm0.21}}          
\newcommand{\hatcurPPgeccenxxxxxA}{\ensuremath{73.9\pm6.1}}            
\newcommand{\hatcurPPloggeccenxxxxxA}{\ensuremath{3.868\pm0.036}}      
\newcommand{\hatcurPPareccenxxxxxA}{\ensuremath{15.70_{-0.65}^{+0.39}}} 
\newcommand{\hatcurPPareleccenxxxxxA}{\ensuremath{0.05025\pm0.00042}}  
\newcommand{\hatcurPPrhoeccenxxxxxA}{\ensuremath{3.89\pm0.45}}         
\newcommand{\hatcurPPmeccenxxxxxA}{\ensuremath{2.74\pm0.11}}           
\newcommand{\hatcurPPmshorteccenxxxxxA}{\ensuremath{2.74}}             
\newcommand{\hatcurPPmlongeccenxxxxxA}{\ensuremath{2.74\pm0.11}}       
\newcommand{\hatcurPPmeeccenxxxxxA}{\ensuremath{872\pm35}}             
\newcommand{\hatcurPPmeshorteccenxxxxxA}{\ensuremath{871.7}}           
\newcommand{\hatcurPPmelongeccenxxxxxA}{\ensuremath{872\pm35}}         
\newcommand{\hatcurPPreccenxxxxxA}{\ensuremath{0.953_{-0.029}^{+0.048}}} 
\newcommand{\hatcurPPrshorteccenxxxxxA}{\ensuremath{0.95}}             
\newcommand{\hatcurPPrlongeccenxxxxxA}{\ensuremath{0.953_{-0.029}^{+0.048}}} 
\newcommand{\hatcurPPreeccenxxxxxA}{\ensuremath{10.68_{-0.33}^{+0.54}}} 
\newcommand{\hatcurPPreshorteccenxxxxxA}{\ensuremath{10.7}}            
\newcommand{\hatcurPPrelongeccenxxxxxA}{\ensuremath{10.68_{-0.33}^{+0.54}}} 
\newcommand{\hatcurPPmrcorreccenxxxxxA}{\ensuremath{0.20}}             
\newcommand{\hatcurPPteffeccenxxxxxA}{\ensuremath{858_{-17}^{+24}}}    
\newcommand{\hatcurPPthetaeccenxxxxxA}{\ensuremath{0.378\pm0.020}}     
\newcommand{\hatcurPPfluxperieccenxxxxxA}{\ensuremath{1.44_{-0.15}^{+0.26}}} 
\newcommand{\hatcurPPfluxperidimeccenxxxxxA}{\ensuremath{8}}           
\newcommand{\hatcurPPfluxapeccenxxxxxA}{\ensuremath{1.050\pm0.088}}    
\newcommand{\hatcurPPfluxapdimeccenxxxxxA}{\ensuremath{8}}             
\newcommand{\hatcurPPfluxavgeccenxxxxxA}{\ensuremath{1.224_{-0.097}^{+0.142}}} 
\newcommand{\hatcurPPfluxavgdimeccenxxxxxA}{\ensuremath{8}}            
\newcommand{\hatcurPPfluxavglogeccenxxxxxA}{\ensuremath{8.088_{-0.036}^{+0.048}}} 
\newcommand{\hatcurXsecphaseeccenxxxxxA}{\ensuremath{0.5274\pm0.0097}} 
\newcommand{\hatcurXsecondaryeccenxxxxxA}{\ensuremath{2457081.071\pm0.046}} 
\newcommand{\hatcurXsecdureccenxxxxxA}{\ensuremath{0.0990\pm0.0048}}   
\newcommand{\hatcurXsecingdureccenxxxxxA}{\ensuremath{0.0179\pm0.0020}} 
\newcommand{\hatcurPPphiconjeccenxxxxxA}{\ensuremath{0.083_{-0.039}^{+0.082}}} 
\newcommand{\hatcurPPperieccenxxxxxA}{\ensuremath{2457078.19\pm0.37}}  
\newcommand{\hatcurPPaequiveccenxxxxxA}{\ensuremath{0.1058\pm0.0048}}  
\newcommand{\hatcurPPtcirceccenxxxxxA}{\ensuremath{7000_{-1600}^{+1200}}} 
\newcommand{\hatcurPPtinfalleccenxxxxxA}{\ensuremath{11900_{-2600}^{+1600}}} 
\newcommand{\hatcurXdisteccenxxxxxA}{\ensuremath{211.4_{-7.3}^{+9.7}}} 
\newcommand{\hatcurXAveccenxxxxxA}{\ensuremath{0.151\pm0.084}}         
\newcommand{\hatcurXdistredeccenxxxxxA}{\ensuremath{207.9_{-6.6}^{+8.9}}} 
\newcommand{\hatcurXEBVeccenxxxxxA}{\ensuremath{0.049\pm0.027}}        
\newcommand{\hatcurXmvisoredeccenxxxxxA}{\ensuremath{13.455\pm0.038}}  
\newcommand{\hatcurXmiisoredeccenxxxxxA}{\ensuremath{12.389\pm0.021}}  
\newcommand{\hatcurXmjisoredeccenxxxxxA}{\ensuremath{11.611\pm0.014}}  
\newcommand{\hatcurXmhisoredeccenxxxxxA}{\ensuremath{11.066\pm0.014}}  
\newcommand{\hatcurXmkisoredeccenxxxxxA}{\ensuremath{10.964\pm0.015}}  
\newcommand{\hatcurXviisoredeccenxxxxxA}{\ensuremath{1.066\pm0.028}}   
\newcommand{\hatcurXvkisoredeccenxxxxxA}{\ensuremath{2.490\pm0.043}}   
\newcommand{\hatcurXjhisoredeccenxxxxxA}{\ensuremath{0.5450\pm0.0072}} 
\newcommand{\hatcurXjkisoredeccenxxxxxA}{\ensuremath{0.6460\pm0.0087}} 
\newcommand{\hatcurCCpmraeccenxxxxxA}{\ensuremath{27.1\pm1.1}}         
\newcommand{\hatcurCCpmdececcenxxxxxA}{\ensuremath{-8.7\pm1.4}}        
\newcommand{\hatcurCCpmeccenxxxxxA}{\ensuremath{28.5\pm1.8}}           

\newcommand{\hatcurhtreccenxxxxxB}{HATS747-017}                      
\newcommand{\hatcurfieldeccenxxxxxB}{\ensuremath{string}}            
\newcommand{\hatcurCCraeccenxxxxxB}{\ensuremath{19^{\mathrm h}05^{\mathrm m}27.96{\mathrm s}}}                     
\newcommand{\hatcurCCdececcenxxxxxB}{\ensuremath{-50{\arcdeg}04{\arcmin}02.5{\arcsec}}}                    
\newcommand{\hatcurCCmageccenxxxxxB}{13.901}                         
\newcommand{\hatcurCCtwomasseccenxxxxxB}{2MASS~19052800-5004024}     
\newcommand{\hatcurCCgsceccenxxxxxB}{GSC~8382-01464}                 
\newcommand{\hatcurCCtassmveccenxxxxxB}{\ensuremath{13.901\pm0.010}} 
\newcommand{\hatcurCCtassmvshorteccenxxxxxB}{\ensuremath{13.9}}      
\newcommand{\hatcurCCtassmBeccenxxxxxB}{\ensuremath{14.625\pm0.010}} 
\newcommand{\hatcurCCtassmBshorteccenxxxxxB}{\ensuremath{14.6}}      
\newcommand{\hatcurCCtassmIeccenxxxxxB}{\ensuremath{nff\pmnff}}      
\newcommand{\hatcurCCtassmIshorteccenxxxxxB}{\ensuremath{0.0}}       
\newcommand{\hatcurCCtassmgeccenxxxxxB}{\ensuremath{14.246\pm0.010}} 
\newcommand{\hatcurCCtassmgshorteccenxxxxxB}{\ensuremath{14.2}}      
\newcommand{\hatcurCCtassmreccenxxxxxB}{\ensuremath{13.735\pm0.010}} 
\newcommand{\hatcurCCtassmrshorteccenxxxxxB}{\ensuremath{13.7}}      
\newcommand{\hatcurCCtassmieccenxxxxxB}{\ensuremath{13.427\pm0.010}} 
\newcommand{\hatcurCCtassmishorteccenxxxxxB}{\ensuremath{13.4}}      
\newcommand{\hatcurCCtwomassJmageccenxxxxxB}{\ensuremath{12.636\pm0.025}} 
\newcommand{\hatcurCCtwomassHmageccenxxxxxB}{\ensuremath{12.293\pm0.025}} 
\newcommand{\hatcurCCtwomassKmageccenxxxxxB}{\ensuremath{12.262\pm0.030}} 
\newcommand{\hatcurCCcitJmageccenxxxxxB}{\ensuremath{12.652\pm0.025}} 
\newcommand{\hatcurCCcitHmageccenxxxxxB}{\ensuremath{12.289\pm0.025}} 
\newcommand{\hatcurCCcitKmageccenxxxxxB}{\ensuremath{12.286\pm0.030}} 
\newcommand{\hatcurCCbbJmageccenxxxxxB}{\ensuremath{12.702\pm0.027}} 
\newcommand{\hatcurCCbbHmageccenxxxxxB}{\ensuremath{12.309\pm0.026}} 
\newcommand{\hatcurCCbbKmageccenxxxxxB}{\ensuremath{12.306\pm0.030}} 
\newcommand{\hatcurCCesoJmageccenxxxxxB}{\ensuremath{12.704\pm0.028}} 
\newcommand{\hatcurCCesoHmageccenxxxxxB}{\ensuremath{12.302\pm0.029}} 
\newcommand{\hatcurCCesoKmageccenxxxxxB}{\ensuremath{12.305\pm0.030}} 
\newcommand{\hatcurCCesoJHmageccenxxxxxB}{\ensuremath{0.4030\pm0.0090}} 
\newcommand{\hatcurCCesoJKmageccenxxxxxB}{\ensuremath{0.399\pm0.042}} 
\newcommand{\hatcurCCesoHKmageccenxxxxxB}{\ensuremath{-0.003\pm0.042}} 
\newcommand{\hatcurLCdipeccenxxxxxB}{\ensuremath{12.8}}              
\newcommand{\hatcurLCrprstareccenxxxxxB}{\ensuremath{0.152\pm0.015}} 
\newcommand{\hatcurLCbsqeccenxxxxxB}{\ensuremath{0.880_{-0.073}^{+0.047}}} 
\newcommand{\hatcurLCimpeccenxxxxxB}{\ensuremath{0.938_{-0.040}^{+0.025}}} 
\newcommand{\hatcurLCbsqsiglowerlimeccenxxxxxB}{\ensuremath{>0.773}} 
\newcommand{\hatcurLCimpsiglowerlimeccenxxxxxB}{\ensuremath{>0.879}} 
\newcommand{\hatcurLCzetaeccenxxxxxB}{\ensuremath{51.3_{-9.0}^{+12.8}}} 
\newcommand{\hatcurLCdureccenxxxxxB}{\ensuremath{0.0752\pm0.0027}}   
\newcommand{\hatcurLCdurshorteccenxxxxxB}{\ensuremath{0.0752}}       
\newcommand{\hatcurLCdurhreccenxxxxxB}{\ensuremath{1.805\pm0.064}}   
\newcommand{\hatcurLCdurhrshorteccenxxxxxB}{\ensuremath{1.805}}      
\newcommand{\hatcurLCqeccenxxxxxB}{\ensuremath{0.0348\pm0.0012}}     
\newcommand{\hatcurLCqshorteccenxxxxxB}{\ensuremath{0.035}}          
\newcommand{\hatcurLCingdureccenxxxxxB}{\ensuremath{0.093\pm0.012}}  
\newcommand{\hatcurLCPeccenxxxxxB}{\ensuremath{2.1605159\pm0.0000049}} 
\newcommand{\hatcurLCPprececcenxxxxxB}{\ensuremath{2.1605159}}       
\newcommand{\hatcurLCPshorteccenxxxxxB}{\ensuremath{2.1605}}         
\newcommand{\hatcurLCTeccenxxxxxB}{\ensuremath{2457042.60544\pm0.00073}} 
\newcommand{\hatcurLCTAeccenxxxxxB}{\ensuremath{2456366.3640\pm0.0016}} 
\newcommand{\hatcurLCTBeccenxxxxxB}{\ensuremath{2457282.42278\pm0.00095}} 
\newcommand{\hatcurLChatnetmeccenxxxxxB}{\ensuremath{13.730550\pm0.000079}} 
\newcommand{\hatcurLCiblendeccenxxxxxB}{\ensuremath{0.750\pm0.059}}  
\newcommand{\hatcurLCrhoeccenxxxxxB}{\ensuremath{1.07_{-0.17}^{+0.23}}} 
\newcommand{\hatcurSMEiteffeccenxxxxxB}{\ensuremath{5950\pm130}}     
\newcommand{\hatcurSMEizfeheccenxxxxxB}{\ensuremath{0.360\pm0.060}}  
\newcommand{\hatcurSMEizfehshorteccenxxxxxB}{\ensuremath{0.36}}      
\newcommand{\hatcurSMEiloggeccenxxxxxB}{\ensuremath{4.61\pm0.19}}    
\newcommand{\hatcurSMEivsineccenxxxxxB}{\ensuremath{4.22\pm0.84}}    
\newcommand{\hatcurSMEivmaceccenxxxxxB}{\ensuremath{0.0}}            
\newcommand{\hatcurSMEivmiceccenxxxxxB}{\ensuremath{0.0}}            
\newcommand{\hatcurSMEiiteffeccenxxxxxB}{\ensuremath{5790\pm140}}    
\newcommand{\hatcurSMEiizfeheccenxxxxxB}{\ensuremath{0.300\pm0.090}} 
\newcommand{\hatcurSMEiizfehshorteccenxxxxxB}{\ensuremath{0.3}}      
\newcommand{\hatcurSMEiiloggeccenxxxxxB}{\ensuremath{4.328\pm0.044}} 
\newcommand{\hatcurSMEiivsineccenxxxxxB}{\ensuremath{4.72\pm0.53}}   
\newcommand{\hatcurLBizeccenxxxxxB}{\ensuremath{0.2118}}             
\newcommand{\hatcurLBiizeccenxxxxxB}{\ensuremath{0.3346}}            
\newcommand{\hatcurLBiieccenxxxxxB}{\ensuremath{0.2780}}             
\newcommand{\hatcurLBiiieccenxxxxxB}{\ensuremath{0.3335}}            
\newcommand{\hatcurLBiIeccenxxxxxB}{\ensuremath{0.2555}}             
\newcommand{\hatcurLBiiIeccenxxxxxB}{\ensuremath{0.3349}}            
\newcommand{\hatcurLBigeccenxxxxxB}{\ensuremath{0.5792}}             
\newcommand{\hatcurLBiigeccenxxxxxB}{\ensuremath{0.2193}}            
\newcommand{\hatcurLBireccenxxxxxB}{\ensuremath{0.3736}}             
\newcommand{\hatcurLBiireccenxxxxxB}{\ensuremath{0.3214}}            
\newcommand{\hatcurLBiReccenxxxxxB}{\ensuremath{0.3470}}             
\newcommand{\hatcurLBiiReccenxxxxxB}{\ensuremath{0.3259}}            
\newcommand{\hatcurLBikepeccenxxxxxB}{\ensuremath{0.1000}}           
\newcommand{\hatcurLBiikepeccenxxxxxB}{\ensuremath{0.1000}}          
\newcommand{\hatcurISOmeccenxxxxxB}{\ensuremath{1.115\pm0.054}}      
\newcommand{\hatcurISOmshorteccenxxxxxB}{\ensuremath{1.12}}          
\newcommand{\hatcurISOmlongeccenxxxxxB}{\ensuremath{1.115\pm0.054}}  
\newcommand{\hatcurISOreccenxxxxxB}{\ensuremath{1.145\pm0.070}}      
\newcommand{\hatcurISOrshorteccenxxxxxB}{\ensuremath{1.15}}          
\newcommand{\hatcurISOrlongeccenxxxxxB}{\ensuremath{1.145\pm0.070}}  
\newcommand{\hatcurISOrhoeccenxxxxxB}{\ensuremath{1.04_{-0.15}^{+0.20}}} 
\newcommand{\hatcurISOrholongeccenxxxxxB}{\ensuremath{1.04_{-0.15}^{+0.20}}} 
\newcommand{\hatcurISOloggeccenxxxxxB}{\ensuremath{4.366\pm0.046}}   
\newcommand{\hatcurISOlumeccenxxxxxB}{\ensuremath{1.31\pm0.24}}      
\newcommand{\hatcurISOlumshorteccenxxxxxB}{\ensuremath{1.31}}        
\newcommand{\hatcurISOmveccenxxxxxB}{\ensuremath{4.52\pm0.21}}       
\newcommand{\hatcurISOvieccenxxxxxB}{\ensuremath{0.702\pm0.042}}     
\newcommand{\hatcurISOageeccenxxxxxB}{\ensuremath{3.6\pm1.8}}        
\newcommand{\hatcurISOsigmaeccenxxxxxB}{\ensuremath{0.00160\pm0.00029}} 
\newcommand{\hatcurISOMJeccenxxxxxB}{\ensuremath{3.38\pm0.16}}       
\newcommand{\hatcurISOMHeccenxxxxxB}{\ensuremath{3.05\pm0.15}}       
\newcommand{\hatcurISOMKeccenxxxxxB}{\ensuremath{2.99\pm0.15}}       
\newcommand{\hatcurISOJKeccenxxxxxB}{\ensuremath{0.20\pm0.19}}       
\newcommand{\hatcurISOspececcenxxxxxB}{G}                            
\newcommand{\hatcurRVKeccenxxxxxB}{\ensuremath{214.6\pm9.4}}         
\newcommand{\hatcurRVrkeccenxxxxxB}{\ensuremath{0.007\pm0.093}}      
\newcommand{\hatcurRVrheccenxxxxxB}{\ensuremath{-0.13\pm0.14}}       
\newcommand{\hatcurRVkeccenxxxxxB}{\ensuremath{0.002_{-0.016}^{+0.021}}} 
\newcommand{\hatcurRVheccenxxxxxB}{\ensuremath{-0.023_{-0.057}^{+0.026}}} 
\newcommand{\hatcurRVtroneeccenxxxxxB}{\ensuremath{0\pm0}}           
\newcommand{\hatcurRVtrtwoeccenxxxxxB}{\ensuremath{0\pm0}}           
\newcommand{\hatcurRVgammaeccenxxxxxB}{\ensuremath{-13370.6\pm6.7}}  
\newcommand{\hatcurRVjittereccenxxxxxB}{\ensuremath{0.0\pm5.7}}      
\newcommand{\hatcurRVjittertwosiglimeccenxxxxxB}{\ensuremath{<11.6}} 
\newcommand{\hatcurRVfitrmseccenxxxxxB}{\ensuremath{.1fym}}          %
\newcommand{\hatcurRVecceneccenxxxxxB}{\ensuremath{0.034\pm0.037}}   
\newcommand{\hatcurRVeccentwosiglimeccenxxxxxB}{\ensuremath{<0.114}} 
\newcommand{\hatcurRVomegaeccenxxxxxB}{\ensuremath{265\pm90}}        
\newcommand{\hatcurPPieccenxxxxxB}{\ensuremath{81.79\pm0.79}}        
\newcommand{\hatcurPPitwosigupperlimeccenxxxxxB}{\ensuremath{<84.5}} 
\newcommand{\hatcurPPgeccenxxxxxB}{\ensuremath{12.8_{-3.0}^{+5.0}}}  
\newcommand{\hatcurPPloggeccenxxxxxB}{\ensuremath{3.11\pm0.12}}      
\newcommand{\hatcurPPareccenxxxxxB}{\ensuremath{6.36\pm0.35}}        
\newcommand{\hatcurPPareleccenxxxxxB}{\ensuremath{0.03392\pm0.00055}} 
\newcommand{\hatcurPPrhoeccenxxxxxB}{\ensuremath{0.38_{-0.12}^{+0.24}}} 
\newcommand{\hatcurPPmeccenxxxxxB}{\ensuremath{1.478\pm0.080}}       
\newcommand{\hatcurPPmshorteccenxxxxxB}{\ensuremath{1.48}}           
\newcommand{\hatcurPPmlongeccenxxxxxB}{\ensuremath{1.478\pm0.080}}   
\newcommand{\hatcurPPmeeccenxxxxxB}{\ensuremath{470\pm25}}           
\newcommand{\hatcurPPmeshorteccenxxxxxB}{\ensuremath{469.8}}         
\newcommand{\hatcurPPmelongeccenxxxxxB}{\ensuremath{470\pm25}}       
\newcommand{\hatcurPPreccenxxxxxB}{\ensuremath{1.69\pm0.24}}         
\newcommand{\hatcurPPrshorteccenxxxxxB}{\ensuremath{1.69}}           
\newcommand{\hatcurPPrlongeccenxxxxxB}{\ensuremath{1.69\pm0.24}}     
\newcommand{\hatcurPPrtwosiglowerlimeccenxxxxxB}{\ensuremath{>1.29}} 
\newcommand{\hatcurPPreeccenxxxxxB}{\ensuremath{19.0\pm2.6}}         
\newcommand{\hatcurPPreshorteccenxxxxxB}{\ensuremath{19.0}}          
\newcommand{\hatcurPPrelongeccenxxxxxB}{\ensuremath{19.0\pm2.6}}     
\newcommand{\hatcurPPmrcorreccenxxxxxB}{\ensuremath{0.22}}           
\newcommand{\hatcurPPteffeccenxxxxxB}{\ensuremath{1622\pm61}}        
\newcommand{\hatcurPPthetaeccenxxxxxB}{\ensuremath{0.0532\pm0.0080}} 
\newcommand{\hatcurPPfluxperieccenxxxxxB}{\ensuremath{1.71\pm0.24}}  
\newcommand{\hatcurPPfluxperidimeccenxxxxxB}{\ensuremath{9}}         
\newcommand{\hatcurPPfluxapeccenxxxxxB}{\ensuremath{1.45\pm0.27}}    
\newcommand{\hatcurPPfluxapdimeccenxxxxxB}{\ensuremath{9}}           
\newcommand{\hatcurPPfluxavgeccenxxxxxB}{\ensuremath{1.57\pm0.24}}   
\newcommand{\hatcurPPfluxavgdimeccenxxxxxB}{\ensuremath{9}}          
\newcommand{\hatcurPPfluxavglogeccenxxxxxB}{\ensuremath{9.195\pm0.065}} 
\newcommand{\hatcurXsecphaseeccenxxxxxB}{\ensuremath{0.501\pm0.013}} 
\newcommand{\hatcurXsecondaryeccenxxxxxB}{\ensuremath{2457043.687\pm0.029}} 
\newcommand{\hatcurXsecdureccenxxxxxB}{\ensuremath{0.0820\pm0.0071}} 
\newcommand{\hatcurXsecingdureccenxxxxxB}{\ensuremath{0.0406\pm0.0099}} 
\newcommand{\hatcurPPphiconjeccenxxxxxB}{\ensuremath{0.14_{-0.57}^{+0.32}}} 
\newcommand{\hatcurPPperieccenxxxxxB}{\ensuremath{2457042.31\pm0.82}} 
\newcommand{\hatcurPPaequiveccenxxxxxB}{\ensuremath{0.0296\pm0.0023}} 
\newcommand{\hatcurPPtcirceccenxxxxxB}{\ensuremath{9.5_{-4.5}^{+12.7}}} 
\newcommand{\hatcurPPtinfalleccenxxxxxB}{\ensuremath{161_{-38}^{+53}}} 
\newcommand{\hatcurXdisteccenxxxxxB}{\ensuremath{729\pm50}}          
\newcommand{\hatcurXAveccenxxxxxB}{\ensuremath{0.105\pm0.085}}       
\newcommand{\hatcurXdistredeccenxxxxxB}{\ensuremath{715\pm50}}       
\newcommand{\hatcurXEBVeccenxxxxxB}{\ensuremath{0.034\pm0.027}}      
\newcommand{\hatcurXmvisoredeccenxxxxxB}{\ensuremath{13.903\pm0.012}} 
\newcommand{\hatcurXmiisoredeccenxxxxxB}{\ensuremath{13.144\pm0.013}} 
\newcommand{\hatcurXmjisoredeccenxxxxxB}{\ensuremath{12.681\pm0.019}} 
\newcommand{\hatcurXmhisoredeccenxxxxxB}{\ensuremath{12.341\pm0.029}} 
\newcommand{\hatcurXmkisoredeccenxxxxxB}{\ensuremath{12.279_{-0.025}^{+0.019}}} 
\newcommand{\hatcurXviisoredeccenxxxxxB}{\ensuremath{0.759\pm0.015}} 
\newcommand{\hatcurXvkisoredeccenxxxxxB}{\ensuremath{1.623\pm0.037}} 
\newcommand{\hatcurXjhisoredeccenxxxxxB}{\ensuremath{0.340_{-0.012}^{+0.016}}} 
\newcommand{\hatcurXjkisoredeccenxxxxxB}{\ensuremath{0.4020_{-0.0090}^{+0.0140}}} 
\newcommand{\hatcurCCpmraeccenxxxxxB}{\ensuremath{3.3\pm1.4}}        
\newcommand{\hatcurCCpmdececcenxxxxxB}{\ensuremath{-1.1\pm1.5}}      
\newcommand{\hatcurCCpmeccenxxxxxB}{\ensuremath{3.5\pm2.1}}          

\newcommand{\hatcurhtreccenxxxxxC}{HATS776-001}                      
\newcommand{\hatcurfieldeccenxxxxxC}{\ensuremath{string}}            
\newcommand{\hatcurCCraeccenxxxxxC}{\ensuremath{17^{\mathrm h}55^{\mathrm m}33.60{\mathrm s}}}                     
\newcommand{\hatcurCCdececcenxxxxxC}{\ensuremath{-61{\arcdeg}44{\arcmin}50.3{\arcsec}}}                    
\newcommand{\hatcurCCmageccenxxxxxC}{12.830}                         
\newcommand{\hatcurCCtwomasseccenxxxxxC}{2MASS~17553376-6144503}     
\newcommand{\hatcurCCgsceccenxxxxxC}{GSC~9054-00129}                 
\newcommand{\hatcurCCtassmveccenxxxxxC}{\ensuremath{12.830\pm0.010}} 
\newcommand{\hatcurCCtassmvshorteccenxxxxxC}{\ensuremath{12.8}}      
\newcommand{\hatcurCCtassmBeccenxxxxxC}{\ensuremath{13.404\pm0.010}} 
\newcommand{\hatcurCCtassmBshorteccenxxxxxC}{\ensuremath{13.4}}      
\newcommand{\hatcurCCtassmIeccenxxxxxC}{\ensuremath{100\pm1000}}     
\newcommand{\hatcurCCtassmIshorteccenxxxxxC}{\ensuremath{100.0}}     
\newcommand{\hatcurCCtassmgeccenxxxxxC}{\ensuremath{13.071\pm0.010}} 
\newcommand{\hatcurCCtassmgshorteccenxxxxxC}{\ensuremath{13.1}}      
\newcommand{\hatcurCCtassmreccenxxxxxC}{\ensuremath{12.643\pm0.010}} 
\newcommand{\hatcurCCtassmrshorteccenxxxxxC}{\ensuremath{12.6}}      
\newcommand{\hatcurCCtassmieccenxxxxxC}{\ensuremath{12.518\pm0.090}} 
\newcommand{\hatcurCCtassmishorteccenxxxxxC}{\ensuremath{12.5}}      
\newcommand{\hatcurCCtwomassJmageccenxxxxxC}{\ensuremath{11.678\pm0.022}} 
\newcommand{\hatcurCCtwomassHmageccenxxxxxC}{\ensuremath{11.447\pm0.025}} 
\newcommand{\hatcurCCtwomassKmageccenxxxxxC}{\ensuremath{11.382\pm0.023}} 
\newcommand{\hatcurCCcitJmageccenxxxxxC}{\ensuremath{11.699\pm0.022}} 
\newcommand{\hatcurCCcitHmageccenxxxxxC}{\ensuremath{11.442\pm0.025}} 
\newcommand{\hatcurCCcitKmageccenxxxxxC}{\ensuremath{11.406\pm0.023}} 
\newcommand{\hatcurCCbbJmageccenxxxxxC}{\ensuremath{11.742\pm0.024}} 
\newcommand{\hatcurCCbbHmageccenxxxxxC}{\ensuremath{11.463\pm0.026}} 
\newcommand{\hatcurCCbbKmageccenxxxxxC}{\ensuremath{11.426\pm0.023}} 
\newcommand{\hatcurCCesoJmageccenxxxxxC}{\ensuremath{11.743\pm0.025}} 
\newcommand{\hatcurCCesoHmageccenxxxxxC}{\ensuremath{11.458\pm0.029}} 
\newcommand{\hatcurCCesoKmageccenxxxxxC}{\ensuremath{11.425\pm0.024}} 
\newcommand{\hatcurCCesoJHmageccenxxxxxC}{\ensuremath{0.285\pm0.036}} 
\newcommand{\hatcurCCesoJKmageccenxxxxxC}{\ensuremath{0.319\pm0.034}} 
\newcommand{\hatcurCCesoHKmageccenxxxxxC}{\ensuremath{0.034\pm0.038}} 
\newcommand{\hatcurLCdipeccenxxxxxC}{\ensuremath{17.3}}              
\newcommand{\hatcurLCrprstareccenxxxxxC}{\ensuremath{0.1302\pm0.0028}} 
\newcommand{\hatcurLCbsqeccenxxxxxC}{\ensuremath{0.046_{-0.029}^{+0.036}}} 
\newcommand{\hatcurLCimpeccenxxxxxC}{\ensuremath{0.214_{-0.082}^{+0.073}}} 
\newcommand{\hatcurLCzetaeccenxxxxxC}{\ensuremath{22.74_{-0.27}^{+0.18}}} 
\newcommand{\hatcurLCdureccenxxxxxC}{\ensuremath{0.1002\pm0.0010}}   
\newcommand{\hatcurLCdurshorteccenxxxxxC}{\ensuremath{0.1002}}       
\newcommand{\hatcurLCdurhreccenxxxxxC}{\ensuremath{2.405\pm0.024}}   
\newcommand{\hatcurLCdurhrshorteccenxxxxxC}{\ensuremath{2.405}}      
\newcommand{\hatcurLCqeccenxxxxxC}{\ensuremath{0.07430\pm0.00076}}   
\newcommand{\hatcurLCqshorteccenxxxxxC}{\ensuremath{0.074}}          
\newcommand{\hatcurLCingdureccenxxxxxC}{\ensuremath{0.01208\pm0.00056}} 
\newcommand{\hatcurLCPeccenxxxxxC}{\ensuremath{1.3484956\pm0.0000011}} 
\newcommand{\hatcurLCPprececcenxxxxxC}{\ensuremath{1.3484956}}       
\newcommand{\hatcurLCPshorteccenxxxxxC}{\ensuremath{1.3485}}         
\newcommand{\hatcurLCTeccenxxxxxC}{\ensuremath{2457045.21571\pm0.00040}} 
\newcommand{\hatcurLCTAeccenxxxxxC}{\ensuremath{2455696.7201\pm0.0012}} 
\newcommand{\hatcurLCTBeccenxxxxxC}{\ensuremath{2457181.41374\pm0.00040}} 
\newcommand{\hatcurLChatnetmeccenxxxxxC}{\ensuremath{12.60993\pm0.00015}} 
\newcommand{\hatcurLCiblendeccenxxxxxC}{\ensuremath{0.787\pm0.048}}  
\newcommand{\hatcurLCrhoeccenxxxxxC}{\ensuremath{1.09_{-0.15}^{+0.23}}} 
\newcommand{\hatcurSMEiteffeccenxxxxxC}{\ensuremath{6300\pm100}}     
\newcommand{\hatcurSMEizfeheccenxxxxxC}{\ensuremath{-0.040\pm0.060}} 
\newcommand{\hatcurSMEizfehshorteccenxxxxxC}{\ensuremath{-0.04}}     
\newcommand{\hatcurSMEiloggeccenxxxxxC}{\ensuremath{4.20\pm0.15}}    
\newcommand{\hatcurSMEivsineccenxxxxxC}{\ensuremath{9.52\pm0.28}}    
\newcommand{\hatcurSMEivmaceccenxxxxxC}{\ensuremath{0.0}}            
\newcommand{\hatcurSMEivmiceccenxxxxxC}{\ensuremath{0.0}}            
\newcommand{\hatcurSMEiiteffeccenxxxxxC}{\ensuremath{6346\pm81}}     
\newcommand{\hatcurSMEiizfeheccenxxxxxC}{\ensuremath{0.000\pm0.050}} 
\newcommand{\hatcurSMEiizfehshorteccenxxxxxC}{\ensuremath{0.00}}     
\newcommand{\hatcurSMEiiloggeccenxxxxxC}{\ensuremath{4\pm0}}         
\newcommand{\hatcurSMEiivsineccenxxxxxC}{\ensuremath{9.44\pm0.21}}   
\newcommand{\hatcurLBizeccenxxxxxC}{\ensuremath{0.1405}}             
\newcommand{\hatcurLBiizeccenxxxxxC}{\ensuremath{0.3577}}            
\newcommand{\hatcurLBiieccenxxxxxC}{\ensuremath{0.1919}}             
\newcommand{\hatcurLBiiieccenxxxxxC}{\ensuremath{0.3654}}            
\newcommand{\hatcurLBiIeccenxxxxxC}{\ensuremath{0.1738}}             
\newcommand{\hatcurLBiiIeccenxxxxxC}{\ensuremath{0.3639}}            
\newcommand{\hatcurLBigeccenxxxxxC}{\ensuremath{0.4251}}             
\newcommand{\hatcurLBiigeccenxxxxxC}{\ensuremath{0.3251}}            
\newcommand{\hatcurLBireccenxxxxxC}{\ensuremath{0.2638}}             
\newcommand{\hatcurLBiireccenxxxxxC}{\ensuremath{0.3753}}            
\newcommand{\hatcurLBiReccenxxxxxC}{\ensuremath{0.2434}}             
\newcommand{\hatcurLBiiReccenxxxxxC}{\ensuremath{0.3740}}            
\newcommand{\hatcurLBikepeccenxxxxxC}{\ensuremath{0.1000}}           
\newcommand{\hatcurLBiikepeccenxxxxxC}{\ensuremath{0.1000}}          
\newcommand{\hatcurISOmeccenxxxxxC}{\ensuremath{1.218\pm0.036}}      
\newcommand{\hatcurISOmshorteccenxxxxxC}{\ensuremath{1.22}}          
\newcommand{\hatcurISOmlongeccenxxxxxC}{\ensuremath{1.218\pm0.036}}  
\newcommand{\hatcurISOreccenxxxxxC}{\ensuremath{1.194_{-0.041}^{+0.066}}} 
\newcommand{\hatcurISOrshorteccenxxxxxC}{\ensuremath{1.19}}          
\newcommand{\hatcurISOrlongeccenxxxxxC}{\ensuremath{1.194_{-0.041}^{+0.066}}} 
\newcommand{\hatcurISOrhoeccenxxxxxC}{\ensuremath{1.004_{-0.140}^{+0.096}}} 
\newcommand{\hatcurISOrholongeccenxxxxxC}{\ensuremath{1.004_{-0.140}^{+0.096}}} 
\newcommand{\hatcurISOloggeccenxxxxxC}{\ensuremath{4.370\pm0.043}}   
\newcommand{\hatcurISOlumeccenxxxxxC}{\ensuremath{2.05_{-0.21}^{+0.28}}} 
\newcommand{\hatcurISOlumshorteccenxxxxxC}{\ensuremath{2.05}}        
\newcommand{\hatcurISOmveccenxxxxxC}{\ensuremath{3.98\pm0.16}}       
\newcommand{\hatcurISOvieccenxxxxxC}{\ensuremath{0.544\pm0.020}}     
\newcommand{\hatcurISOageeccenxxxxxC}{\ensuremath{1.13_{-0.59}^{+0.89}}} 
\newcommand{\hatcurISOsigmaeccenxxxxxC}{\ensuremath{0.00240\pm0.00078}} 
\newcommand{\hatcurISOMJeccenxxxxxC}{\ensuremath{3.10\pm0.13}}       
\newcommand{\hatcurISOMHeccenxxxxxC}{\ensuremath{2.85\pm0.12}}       
\newcommand{\hatcurISOMKeccenxxxxxC}{\ensuremath{2.80\pm0.12}}       
\newcommand{\hatcurISOJKeccenxxxxxC}{\ensuremath{0.300\pm0.020}}     
\newcommand{\hatcurISOspececcenxxxxxC}{F}                            
\newcommand{\hatcurRVKeccenxxxxxC}{\ensuremath{388\pm34}}            
\newcommand{\hatcurRVrkeccenxxxxxC}{\ensuremath{-0.32\pm0.14}}       
\newcommand{\hatcurRVrheccenxxxxxC}{\ensuremath{0.066_{-0.087}^{+0.138}}} 
\newcommand{\hatcurRVkeccenxxxxxC}{\ensuremath{-0.115_{-0.096}^{+0.071}}} 
\newcommand{\hatcurRVheccenxxxxxC}{\ensuremath{0.021_{-0.029}^{+0.054}}} 
\newcommand{\hatcurRVtroneeccenxxxxxC}{\ensuremath{0\pm0}}           
\newcommand{\hatcurRVtrtwoeccenxxxxxC}{\ensuremath{0\pm0}}           
\newcommand{\hatcurRVgammaAeccenxxxxxC}{\ensuremath{-3219\pm25}}     
\newcommand{\hatcurRVjitterAeccenxxxxxC}{\ensuremath{0\pm23}}        
\newcommand{\hatcurRVjittertwosiglimAeccenxxxxxC}{\ensuremath{<45.0}} 
\newcommand{\hatcurRVfitrmsAeccenxxxxxC}{\ensuremath{0.0}}           
\newcommand{\hatcurRVgammaBeccenxxxxxC}{\ensuremath{-3234\pm51}}     
\newcommand{\hatcurRVjitterBeccenxxxxxC}{\ensuremath{1\pm50}}        
\newcommand{\hatcurRVjittertwosiglimBeccenxxxxxC}{\ensuremath{<119.4}} 
\newcommand{\hatcurRVfitrmsBeccenxxxxxC}{\ensuremath{0.0}}           
\newcommand{\hatcurRVgammaCeccenxxxxxC}{\ensuremath{-3440\pm220}}    
\newcommand{\hatcurRVjitterCeccenxxxxxC}{\ensuremath{130\pm220}}     
\newcommand{\hatcurRVjittertwosiglimCeccenxxxxxC}{\ensuremath{<611.0}} 
\newcommand{\hatcurRVfitrmsCeccenxxxxxC}{\ensuremath{0.0}}           
\newcommand{\hatcurRVgammaDeccenxxxxxC}{\ensuremath{-3262\pm75}}     
\newcommand{\hatcurRVjitterDeccenxxxxxC}{\ensuremath{117\pm79}}      
\newcommand{\hatcurRVjittertwosiglimDeccenxxxxxC}{\ensuremath{<319.2}} 
\newcommand{\hatcurRVfitrmsDeccenxxxxxC}{\ensuremath{.1fym}}         %
\newcommand{\hatcurRVecceneccenxxxxxC}{\ensuremath{0.128\pm0.074}}   
\newcommand{\hatcurRVeccentwosiglimeccenxxxxxC}{\ensuremath{<0.242}} 
\newcommand{\hatcurRVomegaeccenxxxxxC}{\ensuremath{168\pm33}}        
\newcommand{\hatcurPPieccenxxxxxC}{\ensuremath{87.1\pm1.1}}          
\newcommand{\hatcurPPgeccenxxxxxC}{\ensuremath{25.5\pm3.1}}          
\newcommand{\hatcurPPloggeccenxxxxxC}{\ensuremath{3.407\pm0.051}}    
\newcommand{\hatcurPPareccenxxxxxC}{\ensuremath{4.59_{-0.22}^{+0.14}}} 
\newcommand{\hatcurPPareleccenxxxxxC}{\ensuremath{0.02551\pm0.00025}} 
\newcommand{\hatcurPPrhoeccenxxxxxC}{\ensuremath{0.84\pm0.14}}       
\newcommand{\hatcurPPmeccenxxxxxC}{\ensuremath{2.39_{-0.12}^{+0.21}}} 
\newcommand{\hatcurPPmshorteccenxxxxxC}{\ensuremath{2.39}}           
\newcommand{\hatcurPPmlongeccenxxxxxC}{\ensuremath{2.39_{-0.12}^{+0.21}}} 
\newcommand{\hatcurPPmeeccenxxxxxC}{\ensuremath{758_{-39}^{+65}}}    
\newcommand{\hatcurPPmeshorteccenxxxxxC}{\ensuremath{758.4}}         
\newcommand{\hatcurPPmelongeccenxxxxxC}{\ensuremath{758_{-39}^{+65}}} 
\newcommand{\hatcurPPreccenxxxxxC}{\ensuremath{1.516_{-0.065}^{+0.085}}} 
\newcommand{\hatcurPPrshorteccenxxxxxC}{\ensuremath{1.52}}           
\newcommand{\hatcurPPrlongeccenxxxxxC}{\ensuremath{1.516_{-0.065}^{+0.085}}} 
\newcommand{\hatcurPPreeccenxxxxxC}{\ensuremath{16.99_{-0.73}^{+0.95}}} 
\newcommand{\hatcurPPreshorteccenxxxxxC}{\ensuremath{17.0}}          
\newcommand{\hatcurPPrelongeccenxxxxxC}{\ensuremath{16.99_{-0.73}^{+0.95}}} 
\newcommand{\hatcurPPmrcorreccenxxxxxC}{\ensuremath{0.34}}           
\newcommand{\hatcurPPteffeccenxxxxxC}{\ensuremath{2095\pm70}}        
\newcommand{\hatcurPPthetaeccenxxxxxC}{\ensuremath{0.0654_{-0.0040}^{+0.0054}}} 
\newcommand{\hatcurPPfluxperieccenxxxxxC}{\ensuremath{5.7_{-1.0}^{+1.5}}} 
\newcommand{\hatcurPPfluxperidimeccenxxxxxC}{\ensuremath{9}}         
\newcommand{\hatcurPPfluxapeccenxxxxxC}{\ensuremath{3.45\pm0.44}}    
\newcommand{\hatcurPPfluxapdimeccenxxxxxC}{\ensuremath{9}}           
\newcommand{\hatcurPPfluxavgeccenxxxxxC}{\ensuremath{4.35_{-0.40}^{+0.56}}} 
\newcommand{\hatcurPPfluxavgdimeccenxxxxxC}{\ensuremath{9}}          
\newcommand{\hatcurPPfluxavglogeccenxxxxxC}{\ensuremath{9.638\pm0.057}} 
\newcommand{\hatcurXsecphaseeccenxxxxxC}{\ensuremath{0.427\pm0.047}} 
\newcommand{\hatcurXsecondaryeccenxxxxxC}{\ensuremath{2457045.791\pm0.064}} 
\newcommand{\hatcurXsecdureccenxxxxxC}{\ensuremath{0.104\pm0.011}}   
\newcommand{\hatcurXsecingdureccenxxxxxC}{\ensuremath{0.0128\pm0.0015}} 
\newcommand{\hatcurPPphiconjeccenxxxxxC}{\ensuremath{-0.171_{-0.046}^{+0.079}}} 
\newcommand{\hatcurPPperieccenxxxxxC}{\ensuremath{2457045.45\pm0.13}} 
\newcommand{\hatcurPPaequiveccenxxxxxC}{\ensuremath{0.0178\pm0.0011}} 
\newcommand{\hatcurPPtcirceccenxxxxxC}{\ensuremath{3.3\pm1.0}}       
\newcommand{\hatcurPPtinfalleccenxxxxxC}{\ensuremath{13.1\pm2.9}}    
\newcommand{\hatcurXdisteccenxxxxxC}{\ensuremath{530_{-20}^{+30}}}   
\newcommand{\hatcurXAveccenxxxxxC}{\ensuremath{0.269\pm0.065}}       
\newcommand{\hatcurXdistredeccenxxxxxC}{\ensuremath{519_{-19}^{+29}}} 
\newcommand{\hatcurXEBVeccenxxxxxC}{\ensuremath{0.087\pm0.021}}      
\newcommand{\hatcurXmvisoredeccenxxxxxC}{\ensuremath{12.8300\pm0.0100}} 
\newcommand{\hatcurXmiisoredeccenxxxxxC}{\ensuremath{12.145\pm0.016}} 
\newcommand{\hatcurXmjisoredeccenxxxxxC}{\ensuremath{11.751\pm0.014}} 
\newcommand{\hatcurXmhisoredeccenxxxxxC}{\ensuremath{11.473\pm0.015}} 
\newcommand{\hatcurXmkisoredeccenxxxxxC}{\ensuremath{11.410\pm0.016}} 
\newcommand{\hatcurXviisoredeccenxxxxxC}{\ensuremath{0.685\pm0.017}} 
\newcommand{\hatcurXvkisoredeccenxxxxxC}{\ensuremath{1.420\pm0.020}} 
\newcommand{\hatcurXjhisoredeccenxxxxxC}{\ensuremath{0.2780\pm0.0089}} 
\newcommand{\hatcurXjkisoredeccenxxxxxC}{\ensuremath{0.3410\pm0.0070}} 
\newcommand{\hatcurCCpmraeccenxxxxxC}{\ensuremath{2.9\pm2.6}}        
\newcommand{\hatcurCCpmdececcenxxxxxC}{\ensuremath{-11.0\pm2.6}}     
\newcommand{\hatcurCCpmeccenxxxxxC}{\ensuremath{11.4\pm3.7}}         

\newcommand{\hatcurCCbbHmageccen}[1]{\ifnum#1=22 %
\hatcurCCbbHmageccenxxxxxA
\else
\ifnum#1=23 %
\hatcurCCbbHmageccenxxxxxB
\else
\ifnum#1=24 %
\hatcurCCbbHmageccenxxxxxC
\else
??????\fi
\fi
\fi
}
\newcommand{\hatcurCCbbJmageccen}[1]{\ifnum#1=22 %
\hatcurCCbbJmageccenxxxxxA
\else
\ifnum#1=23 %
\hatcurCCbbJmageccenxxxxxB
\else
\ifnum#1=24 %
\hatcurCCbbJmageccenxxxxxC
\else
??????\fi
\fi
\fi
}
\newcommand{\hatcurCCbbKmageccen}[1]{\ifnum#1=22 %
\hatcurCCbbKmageccenxxxxxA
\else
\ifnum#1=23 %
\hatcurCCbbKmageccenxxxxxB
\else
\ifnum#1=24 %
\hatcurCCbbKmageccenxxxxxC
\else
??????\fi
\fi
\fi
}
\newcommand{\hatcurCCcitHmageccen}[1]{\ifnum#1=22 %
\hatcurCCcitHmageccenxxxxxA
\else
\ifnum#1=23 %
\hatcurCCcitHmageccenxxxxxB
\else
\ifnum#1=24 %
\hatcurCCcitHmageccenxxxxxC
\else
??????\fi
\fi
\fi
}
\newcommand{\hatcurCCcitJmageccen}[1]{\ifnum#1=22 %
\hatcurCCcitJmageccenxxxxxA
\else
\ifnum#1=23 %
\hatcurCCcitJmageccenxxxxxB
\else
\ifnum#1=24 %
\hatcurCCcitJmageccenxxxxxC
\else
??????\fi
\fi
\fi
}
\newcommand{\hatcurCCcitKmageccen}[1]{\ifnum#1=22 %
\hatcurCCcitKmageccenxxxxxA
\else
\ifnum#1=23 %
\hatcurCCcitKmageccenxxxxxB
\else
\ifnum#1=24 %
\hatcurCCcitKmageccenxxxxxC
\else
??????\fi
\fi
\fi
}
\newcommand{\hatcurCCdececcen}[1]{\ifnum#1=22 %
\hatcurCCdececcenxxxxxA
\else
\ifnum#1=23 %
\hatcurCCdececcenxxxxxB
\else
\ifnum#1=24 %
\hatcurCCdececcenxxxxxC
\else
??????\fi
\fi
\fi
}
\newcommand{\hatcurCCesoHKmageccen}[1]{\ifnum#1=22 %
\hatcurCCesoHKmageccenxxxxxA
\else
\ifnum#1=23 %
\hatcurCCesoHKmageccenxxxxxB
\else
\ifnum#1=24 %
\hatcurCCesoHKmageccenxxxxxC
\else
??????\fi
\fi
\fi
}
\newcommand{\hatcurCCesoHmageccen}[1]{\ifnum#1=22 %
\hatcurCCesoHmageccenxxxxxA
\else
\ifnum#1=23 %
\hatcurCCesoHmageccenxxxxxB
\else
\ifnum#1=24 %
\hatcurCCesoHmageccenxxxxxC
\else
??????\fi
\fi
\fi
}
\newcommand{\hatcurCCesoJHmageccen}[1]{\ifnum#1=22 %
\hatcurCCesoJHmageccenxxxxxA
\else
\ifnum#1=23 %
\hatcurCCesoJHmageccenxxxxxB
\else
\ifnum#1=24 %
\hatcurCCesoJHmageccenxxxxxC
\else
??????\fi
\fi
\fi
}
\newcommand{\hatcurCCesoJKmageccen}[1]{\ifnum#1=22 %
\hatcurCCesoJKmageccenxxxxxA
\else
\ifnum#1=23 %
\hatcurCCesoJKmageccenxxxxxB
\else
\ifnum#1=24 %
\hatcurCCesoJKmageccenxxxxxC
\else
??????\fi
\fi
\fi
}
\newcommand{\hatcurCCesoJmageccen}[1]{\ifnum#1=22 %
\hatcurCCesoJmageccenxxxxxA
\else
\ifnum#1=23 %
\hatcurCCesoJmageccenxxxxxB
\else
\ifnum#1=24 %
\hatcurCCesoJmageccenxxxxxC
\else
??????\fi
\fi
\fi
}
\newcommand{\hatcurCCesoKmageccen}[1]{\ifnum#1=22 %
\hatcurCCesoKmageccenxxxxxA
\else
\ifnum#1=23 %
\hatcurCCesoKmageccenxxxxxB
\else
\ifnum#1=24 %
\hatcurCCesoKmageccenxxxxxC
\else
??????\fi
\fi
\fi
}
\newcommand{\hatcurCCgsceccen}[1]{\ifnum#1=22 %
\hatcurCCgsceccenxxxxxA
\else
\ifnum#1=23 %
\hatcurCCgsceccenxxxxxB
\else
\ifnum#1=24 %
\hatcurCCgsceccenxxxxxC
\else
??????\fi
\fi
\fi
}
\newcommand{\hatcurCCmageccen}[1]{\ifnum#1=22 %
\hatcurCCmageccenxxxxxA
\else
\ifnum#1=23 %
\hatcurCCmageccenxxxxxB
\else
\ifnum#1=24 %
\hatcurCCmageccenxxxxxC
\else
??????\fi
\fi
\fi
}
\newcommand{\hatcurCCpmdececcen}[1]{\ifnum#1=22 %
\hatcurCCpmdececcenxxxxxA
\else
\ifnum#1=23 %
\hatcurCCpmdececcenxxxxxB
\else
\ifnum#1=24 %
\hatcurCCpmdececcenxxxxxC
\else
??????\fi
\fi
\fi
}
\newcommand{\hatcurCCpmeccen}[1]{\ifnum#1=22 %
\hatcurCCpmeccenxxxxxA
\else
\ifnum#1=23 %
\hatcurCCpmeccenxxxxxB
\else
\ifnum#1=24 %
\hatcurCCpmeccenxxxxxC
\else
??????\fi
\fi
\fi
}
\newcommand{\hatcurCCpmraeccen}[1]{\ifnum#1=22 %
\hatcurCCpmraeccenxxxxxA
\else
\ifnum#1=23 %
\hatcurCCpmraeccenxxxxxB
\else
\ifnum#1=24 %
\hatcurCCpmraeccenxxxxxC
\else
??????\fi
\fi
\fi
}
\newcommand{\hatcurCCraeccen}[1]{\ifnum#1=22 %
\hatcurCCraeccenxxxxxA
\else
\ifnum#1=23 %
\hatcurCCraeccenxxxxxB
\else
\ifnum#1=24 %
\hatcurCCraeccenxxxxxC
\else
??????\fi
\fi
\fi
}
\newcommand{\hatcurCCtassmBeccen}[1]{\ifnum#1=22 %
\hatcurCCtassmBeccenxxxxxA
\else
\ifnum#1=23 %
\hatcurCCtassmBeccenxxxxxB
\else
\ifnum#1=24 %
\hatcurCCtassmBeccenxxxxxC
\else
??????\fi
\fi
\fi
}
\newcommand{\hatcurCCtassmBshorteccen}[1]{\ifnum#1=22 %
\hatcurCCtassmBshorteccenxxxxxA
\else
\ifnum#1=23 %
\hatcurCCtassmBshorteccenxxxxxB
\else
\ifnum#1=24 %
\hatcurCCtassmBshorteccenxxxxxC
\else
??????\fi
\fi
\fi
}
\newcommand{\hatcurCCtassmgeccen}[1]{\ifnum#1=22 %
\hatcurCCtassmgeccenxxxxxA
\else
\ifnum#1=23 %
\hatcurCCtassmgeccenxxxxxB
\else
\ifnum#1=24 %
\hatcurCCtassmgeccenxxxxxC
\else
??????\fi
\fi
\fi
}
\newcommand{\hatcurCCtassmgshorteccen}[1]{\ifnum#1=22 %
\hatcurCCtassmgshorteccenxxxxxA
\else
\ifnum#1=23 %
\hatcurCCtassmgshorteccenxxxxxB
\else
\ifnum#1=24 %
\hatcurCCtassmgshorteccenxxxxxC
\else
??????\fi
\fi
\fi
}
\newcommand{\hatcurCCtassmieccen}[1]{\ifnum#1=22 %
\hatcurCCtassmieccenxxxxxA
\else
\ifnum#1=23 %
\hatcurCCtassmieccenxxxxxB
\else
\ifnum#1=24 %
\hatcurCCtassmieccenxxxxxC
\else
??????\fi
\fi
\fi
}
\newcommand{\hatcurCCtassmIeccen}[1]{\ifnum#1=22 %
\hatcurCCtassmIeccenxxxxxA
\else
\ifnum#1=23 %
\hatcurCCtassmIeccenxxxxxB
\else
\ifnum#1=24 %
\hatcurCCtassmIeccenxxxxxC
\else
??????\fi
\fi
\fi
}
\newcommand{\hatcurCCtassmishorteccen}[1]{\ifnum#1=22 %
\hatcurCCtassmishorteccenxxxxxA
\else
\ifnum#1=23 %
\hatcurCCtassmishorteccenxxxxxB
\else
\ifnum#1=24 %
\hatcurCCtassmishorteccenxxxxxC
\else
??????\fi
\fi
\fi
}
\newcommand{\hatcurCCtassmIshorteccen}[1]{\ifnum#1=22 %
\hatcurCCtassmIshorteccenxxxxxA
\else
\ifnum#1=23 %
\hatcurCCtassmIshorteccenxxxxxB
\else
\ifnum#1=24 %
\hatcurCCtassmIshorteccenxxxxxC
\else
??????\fi
\fi
\fi
}
\newcommand{\hatcurCCtassmreccen}[1]{\ifnum#1=22 %
\hatcurCCtassmreccenxxxxxA
\else
\ifnum#1=23 %
\hatcurCCtassmreccenxxxxxB
\else
\ifnum#1=24 %
\hatcurCCtassmreccenxxxxxC
\else
??????\fi
\fi
\fi
}
\newcommand{\hatcurCCtassmrshorteccen}[1]{\ifnum#1=22 %
\hatcurCCtassmrshorteccenxxxxxA
\else
\ifnum#1=23 %
\hatcurCCtassmrshorteccenxxxxxB
\else
\ifnum#1=24 %
\hatcurCCtassmrshorteccenxxxxxC
\else
??????\fi
\fi
\fi
}
\newcommand{\hatcurCCtassmveccen}[1]{\ifnum#1=22 %
\hatcurCCtassmveccenxxxxxA
\else
\ifnum#1=23 %
\hatcurCCtassmveccenxxxxxB
\else
\ifnum#1=24 %
\hatcurCCtassmveccenxxxxxC
\else
??????\fi
\fi
\fi
}
\newcommand{\hatcurCCtassmvshorteccen}[1]{\ifnum#1=22 %
\hatcurCCtassmvshorteccenxxxxxA
\else
\ifnum#1=23 %
\hatcurCCtassmvshorteccenxxxxxB
\else
\ifnum#1=24 %
\hatcurCCtassmvshorteccenxxxxxC
\else
??????\fi
\fi
\fi
}
\newcommand{\hatcurCCtwomasseccen}[1]{\ifnum#1=22 %
\hatcurCCtwomasseccenxxxxxA
\else
\ifnum#1=23 %
\hatcurCCtwomasseccenxxxxxB
\else
\ifnum#1=24 %
\hatcurCCtwomasseccenxxxxxC
\else
??????\fi
\fi
\fi
}
\newcommand{\hatcurCCtwomassHmageccen}[1]{\ifnum#1=22 %
\hatcurCCtwomassHmageccenxxxxxA
\else
\ifnum#1=23 %
\hatcurCCtwomassHmageccenxxxxxB
\else
\ifnum#1=24 %
\hatcurCCtwomassHmageccenxxxxxC
\else
??????\fi
\fi
\fi
}
\newcommand{\hatcurCCtwomassJmageccen}[1]{\ifnum#1=22 %
\hatcurCCtwomassJmageccenxxxxxA
\else
\ifnum#1=23 %
\hatcurCCtwomassJmageccenxxxxxB
\else
\ifnum#1=24 %
\hatcurCCtwomassJmageccenxxxxxC
\else
??????\fi
\fi
\fi
}
\newcommand{\hatcurCCtwomassKmageccen}[1]{\ifnum#1=22 %
\hatcurCCtwomassKmageccenxxxxxA
\else
\ifnum#1=23 %
\hatcurCCtwomassKmageccenxxxxxB
\else
\ifnum#1=24 %
\hatcurCCtwomassKmageccenxxxxxC
\else
??????\fi
\fi
\fi
}
\newcommand{\hatcurfieldeccen}[1]{\ifnum#1=22 %
\hatcurfieldeccenxxxxxA
\else
\ifnum#1=23 %
\hatcurfieldeccenxxxxxB
\else
\ifnum#1=24 %
\hatcurfieldeccenxxxxxC
\else
??????\fi
\fi
\fi
}
\newcommand{\hatcurhtreccen}[1]{\ifnum#1=22 %
\hatcurhtreccenxxxxxA
\else
\ifnum#1=23 %
\hatcurhtreccenxxxxxB
\else
\ifnum#1=24 %
\hatcurhtreccenxxxxxC
\else
??????\fi
\fi
\fi
}
\newcommand{\hatcurISOageeccen}[1]{\ifnum#1=22 %
\hatcurISOageeccenxxxxxA
\else
\ifnum#1=23 %
\hatcurISOageeccenxxxxxB
\else
\ifnum#1=24 %
\hatcurISOageeccenxxxxxC
\else
??????\fi
\fi
\fi
}
\newcommand{\hatcurISOJKeccen}[1]{\ifnum#1=22 %
\hatcurISOJKeccenxxxxxA
\else
\ifnum#1=23 %
\hatcurISOJKeccenxxxxxB
\else
\ifnum#1=24 %
\hatcurISOJKeccenxxxxxC
\else
??????\fi
\fi
\fi
}
\newcommand{\hatcurISOloggeccen}[1]{\ifnum#1=22 %
\hatcurISOloggeccenxxxxxA
\else
\ifnum#1=23 %
\hatcurISOloggeccenxxxxxB
\else
\ifnum#1=24 %
\hatcurISOloggeccenxxxxxC
\else
??????\fi
\fi
\fi
}
\newcommand{\hatcurISOlumeccen}[1]{\ifnum#1=22 %
\hatcurISOlumeccenxxxxxA
\else
\ifnum#1=23 %
\hatcurISOlumeccenxxxxxB
\else
\ifnum#1=24 %
\hatcurISOlumeccenxxxxxC
\else
??????\fi
\fi
\fi
}
\newcommand{\hatcurISOlumshorteccen}[1]{\ifnum#1=22 %
\hatcurISOlumshorteccenxxxxxA
\else
\ifnum#1=23 %
\hatcurISOlumshorteccenxxxxxB
\else
\ifnum#1=24 %
\hatcurISOlumshorteccenxxxxxC
\else
??????\fi
\fi
\fi
}
\newcommand{\hatcurISOmeccen}[1]{\ifnum#1=22 %
\hatcurISOmeccenxxxxxA
\else
\ifnum#1=23 %
\hatcurISOmeccenxxxxxB
\else
\ifnum#1=24 %
\hatcurISOmeccenxxxxxC
\else
??????\fi
\fi
\fi
}
\newcommand{\hatcurISOMHeccen}[1]{\ifnum#1=22 %
\hatcurISOMHeccenxxxxxA
\else
\ifnum#1=23 %
\hatcurISOMHeccenxxxxxB
\else
\ifnum#1=24 %
\hatcurISOMHeccenxxxxxC
\else
??????\fi
\fi
\fi
}
\newcommand{\hatcurISOMJeccen}[1]{\ifnum#1=22 %
\hatcurISOMJeccenxxxxxA
\else
\ifnum#1=23 %
\hatcurISOMJeccenxxxxxB
\else
\ifnum#1=24 %
\hatcurISOMJeccenxxxxxC
\else
??????\fi
\fi
\fi
}
\newcommand{\hatcurISOMKeccen}[1]{\ifnum#1=22 %
\hatcurISOMKeccenxxxxxA
\else
\ifnum#1=23 %
\hatcurISOMKeccenxxxxxB
\else
\ifnum#1=24 %
\hatcurISOMKeccenxxxxxC
\else
??????\fi
\fi
\fi
}
\newcommand{\hatcurISOmlongeccen}[1]{\ifnum#1=22 %
\hatcurISOmlongeccenxxxxxA
\else
\ifnum#1=23 %
\hatcurISOmlongeccenxxxxxB
\else
\ifnum#1=24 %
\hatcurISOmlongeccenxxxxxC
\else
??????\fi
\fi
\fi
}
\newcommand{\hatcurISOmshorteccen}[1]{\ifnum#1=22 %
\hatcurISOmshorteccenxxxxxA
\else
\ifnum#1=23 %
\hatcurISOmshorteccenxxxxxB
\else
\ifnum#1=24 %
\hatcurISOmshorteccenxxxxxC
\else
??????\fi
\fi
\fi
}
\newcommand{\hatcurISOmveccen}[1]{\ifnum#1=22 %
\hatcurISOmveccenxxxxxA
\else
\ifnum#1=23 %
\hatcurISOmveccenxxxxxB
\else
\ifnum#1=24 %
\hatcurISOmveccenxxxxxC
\else
??????\fi
\fi
\fi
}
\newcommand{\hatcurISOreccen}[1]{\ifnum#1=22 %
\hatcurISOreccenxxxxxA
\else
\ifnum#1=23 %
\hatcurISOreccenxxxxxB
\else
\ifnum#1=24 %
\hatcurISOreccenxxxxxC
\else
??????\fi
\fi
\fi
}
\newcommand{\hatcurISOrhoeccen}[1]{\ifnum#1=22 %
\hatcurISOrhoeccenxxxxxA
\else
\ifnum#1=23 %
\hatcurISOrhoeccenxxxxxB
\else
\ifnum#1=24 %
\hatcurISOrhoeccenxxxxxC
\else
??????\fi
\fi
\fi
}
\newcommand{\hatcurISOrholongeccen}[1]{\ifnum#1=22 %
\hatcurISOrholongeccenxxxxxA
\else
\ifnum#1=23 %
\hatcurISOrholongeccenxxxxxB
\else
\ifnum#1=24 %
\hatcurISOrholongeccenxxxxxC
\else
??????\fi
\fi
\fi
}
\newcommand{\hatcurISOrlongeccen}[1]{\ifnum#1=22 %
\hatcurISOrlongeccenxxxxxA
\else
\ifnum#1=23 %
\hatcurISOrlongeccenxxxxxB
\else
\ifnum#1=24 %
\hatcurISOrlongeccenxxxxxC
\else
??????\fi
\fi
\fi
}
\newcommand{\hatcurISOrshorteccen}[1]{\ifnum#1=22 %
\hatcurISOrshorteccenxxxxxA
\else
\ifnum#1=23 %
\hatcurISOrshorteccenxxxxxB
\else
\ifnum#1=24 %
\hatcurISOrshorteccenxxxxxC
\else
??????\fi
\fi
\fi
}
\newcommand{\hatcurISOsigmaeccen}[1]{\ifnum#1=22 %
\hatcurISOsigmaeccenxxxxxA
\else
\ifnum#1=23 %
\hatcurISOsigmaeccenxxxxxB
\else
\ifnum#1=24 %
\hatcurISOsigmaeccenxxxxxC
\else
??????\fi
\fi
\fi
}
\newcommand{\hatcurISOspececcen}[1]{\ifnum#1=22 %
\hatcurISOspececcenxxxxxA
\else
\ifnum#1=23 %
\hatcurISOspececcenxxxxxB
\else
\ifnum#1=24 %
\hatcurISOspececcenxxxxxC
\else
??????\fi
\fi
\fi
}
\newcommand{\hatcurISOvieccen}[1]{\ifnum#1=22 %
\hatcurISOvieccenxxxxxA
\else
\ifnum#1=23 %
\hatcurISOvieccenxxxxxB
\else
\ifnum#1=24 %
\hatcurISOvieccenxxxxxC
\else
??????\fi
\fi
\fi
}
\newcommand{\hatcurLBigeccen}[1]{\ifnum#1=22 %
\hatcurLBigeccenxxxxxA
\else
\ifnum#1=23 %
\hatcurLBigeccenxxxxxB
\else
\ifnum#1=24 %
\hatcurLBigeccenxxxxxC
\else
??????\fi
\fi
\fi
}
\newcommand{\hatcurLBiieccen}[1]{\ifnum#1=22 %
\hatcurLBiieccenxxxxxA
\else
\ifnum#1=23 %
\hatcurLBiieccenxxxxxB
\else
\ifnum#1=24 %
\hatcurLBiieccenxxxxxC
\else
??????\fi
\fi
\fi
}
\newcommand{\hatcurLBiIeccen}[1]{\ifnum#1=22 %
\hatcurLBiIeccenxxxxxA
\else
\ifnum#1=23 %
\hatcurLBiIeccenxxxxxB
\else
\ifnum#1=24 %
\hatcurLBiIeccenxxxxxC
\else
??????\fi
\fi
\fi
}
\newcommand{\hatcurLBiigeccen}[1]{\ifnum#1=22 %
\hatcurLBiigeccenxxxxxA
\else
\ifnum#1=23 %
\hatcurLBiigeccenxxxxxB
\else
\ifnum#1=24 %
\hatcurLBiigeccenxxxxxC
\else
??????\fi
\fi
\fi
}
\newcommand{\hatcurLBiiieccen}[1]{\ifnum#1=22 %
\hatcurLBiiieccenxxxxxA
\else
\ifnum#1=23 %
\hatcurLBiiieccenxxxxxB
\else
\ifnum#1=24 %
\hatcurLBiiieccenxxxxxC
\else
??????\fi
\fi
\fi
}
\newcommand{\hatcurLBiiIeccen}[1]{\ifnum#1=22 %
\hatcurLBiiIeccenxxxxxA
\else
\ifnum#1=23 %
\hatcurLBiiIeccenxxxxxB
\else
\ifnum#1=24 %
\hatcurLBiiIeccenxxxxxC
\else
??????\fi
\fi
\fi
}
\newcommand{\hatcurLBiikepeccen}[1]{\ifnum#1=22 %
\hatcurLBiikepeccenxxxxxA
\else
\ifnum#1=23 %
\hatcurLBiikepeccenxxxxxB
\else
\ifnum#1=24 %
\hatcurLBiikepeccenxxxxxC
\else
??????\fi
\fi
\fi
}
\newcommand{\hatcurLBiireccen}[1]{\ifnum#1=22 %
\hatcurLBiireccenxxxxxA
\else
\ifnum#1=23 %
\hatcurLBiireccenxxxxxB
\else
\ifnum#1=24 %
\hatcurLBiireccenxxxxxC
\else
??????\fi
\fi
\fi
}
\newcommand{\hatcurLBiiReccen}[1]{\ifnum#1=22 %
\hatcurLBiiReccenxxxxxA
\else
\ifnum#1=23 %
\hatcurLBiiReccenxxxxxB
\else
\ifnum#1=24 %
\hatcurLBiiReccenxxxxxC
\else
??????\fi
\fi
\fi
}
\newcommand{\hatcurLBiizeccen}[1]{\ifnum#1=22 %
\hatcurLBiizeccenxxxxxA
\else
\ifnum#1=23 %
\hatcurLBiizeccenxxxxxB
\else
\ifnum#1=24 %
\hatcurLBiizeccenxxxxxC
\else
??????\fi
\fi
\fi
}
\newcommand{\hatcurLBikepeccen}[1]{\ifnum#1=22 %
\hatcurLBikepeccenxxxxxA
\else
\ifnum#1=23 %
\hatcurLBikepeccenxxxxxB
\else
\ifnum#1=24 %
\hatcurLBikepeccenxxxxxC
\else
??????\fi
\fi
\fi
}
\newcommand{\hatcurLBireccen}[1]{\ifnum#1=22 %
\hatcurLBireccenxxxxxA
\else
\ifnum#1=23 %
\hatcurLBireccenxxxxxB
\else
\ifnum#1=24 %
\hatcurLBireccenxxxxxC
\else
??????\fi
\fi
\fi
}
\newcommand{\hatcurLBiReccen}[1]{\ifnum#1=22 %
\hatcurLBiReccenxxxxxA
\else
\ifnum#1=23 %
\hatcurLBiReccenxxxxxB
\else
\ifnum#1=24 %
\hatcurLBiReccenxxxxxC
\else
??????\fi
\fi
\fi
}
\newcommand{\hatcurLBizeccen}[1]{\ifnum#1=22 %
\hatcurLBizeccenxxxxxA
\else
\ifnum#1=23 %
\hatcurLBizeccenxxxxxB
\else
\ifnum#1=24 %
\hatcurLBizeccenxxxxxC
\else
??????\fi
\fi
\fi
}
\newcommand{\hatcurLCbsqeccen}[1]{\ifnum#1=22 %
\hatcurLCbsqeccenxxxxxA
\else
\ifnum#1=23 %
\hatcurLCbsqeccenxxxxxB
\else
\ifnum#1=24 %
\hatcurLCbsqeccenxxxxxC
\else
??????\fi
\fi
\fi
}
\newcommand{\hatcurLCbsqsiglowerlimeccen}[1]{\ifnum#1=23 %
\hatcurLCbsqsiglowerlimeccenxxxxxB
\else
??????\fi
}
\newcommand{\hatcurLCdipeccen}[1]{\ifnum#1=22 %
\hatcurLCdipeccenxxxxxA
\else
\ifnum#1=23 %
\hatcurLCdipeccenxxxxxB
\else
\ifnum#1=24 %
\hatcurLCdipeccenxxxxxC
\else
??????\fi
\fi
\fi
}
\newcommand{\hatcurLCdureccen}[1]{\ifnum#1=22 %
\hatcurLCdureccenxxxxxA
\else
\ifnum#1=23 %
\hatcurLCdureccenxxxxxB
\else
\ifnum#1=24 %
\hatcurLCdureccenxxxxxC
\else
??????\fi
\fi
\fi
}
\newcommand{\hatcurLCdurhreccen}[1]{\ifnum#1=22 %
\hatcurLCdurhreccenxxxxxA
\else
\ifnum#1=23 %
\hatcurLCdurhreccenxxxxxB
\else
\ifnum#1=24 %
\hatcurLCdurhreccenxxxxxC
\else
??????\fi
\fi
\fi
}
\newcommand{\hatcurLCdurhrshorteccen}[1]{\ifnum#1=22 %
\hatcurLCdurhrshorteccenxxxxxA
\else
\ifnum#1=23 %
\hatcurLCdurhrshorteccenxxxxxB
\else
\ifnum#1=24 %
\hatcurLCdurhrshorteccenxxxxxC
\else
??????\fi
\fi
\fi
}
\newcommand{\hatcurLCdurshorteccen}[1]{\ifnum#1=22 %
\hatcurLCdurshorteccenxxxxxA
\else
\ifnum#1=23 %
\hatcurLCdurshorteccenxxxxxB
\else
\ifnum#1=24 %
\hatcurLCdurshorteccenxxxxxC
\else
??????\fi
\fi
\fi
}
\newcommand{\hatcurLChatnetmeccen}[1]{\ifnum#1=22 %
\hatcurLChatnetmeccenxxxxxA
\else
\ifnum#1=23 %
\hatcurLChatnetmeccenxxxxxB
\else
\ifnum#1=24 %
\hatcurLChatnetmeccenxxxxxC
\else
??????\fi
\fi
\fi
}
\newcommand{\hatcurLCiblendeccen}[1]{\ifnum#1=22 %
\hatcurLCiblendeccenxxxxxA
\else
\ifnum#1=23 %
\hatcurLCiblendeccenxxxxxB
\else
\ifnum#1=24 %
\hatcurLCiblendeccenxxxxxC
\else
??????\fi
\fi
\fi
}
\newcommand{\hatcurLCimpeccen}[1]{\ifnum#1=22 %
\hatcurLCimpeccenxxxxxA
\else
\ifnum#1=23 %
\hatcurLCimpeccenxxxxxB
\else
\ifnum#1=24 %
\hatcurLCimpeccenxxxxxC
\else
??????\fi
\fi
\fi
}
\newcommand{\hatcurLCimpsiglowerlimeccen}[1]{\ifnum#1=23 %
\hatcurLCimpsiglowerlimeccenxxxxxB
\else
??????\fi
}
\newcommand{\hatcurLCingdureccen}[1]{\ifnum#1=22 %
\hatcurLCingdureccenxxxxxA
\else
\ifnum#1=23 %
\hatcurLCingdureccenxxxxxB
\else
\ifnum#1=24 %
\hatcurLCingdureccenxxxxxC
\else
??????\fi
\fi
\fi
}
\newcommand{\hatcurLCPeccen}[1]{\ifnum#1=22 %
\hatcurLCPeccenxxxxxA
\else
\ifnum#1=23 %
\hatcurLCPeccenxxxxxB
\else
\ifnum#1=24 %
\hatcurLCPeccenxxxxxC
\else
??????\fi
\fi
\fi
}
\newcommand{\hatcurLCPprececcen}[1]{\ifnum#1=22 %
\hatcurLCPprececcenxxxxxA
\else
\ifnum#1=23 %
\hatcurLCPprececcenxxxxxB
\else
\ifnum#1=24 %
\hatcurLCPprececcenxxxxxC
\else
??????\fi
\fi
\fi
}
\newcommand{\hatcurLCPshorteccen}[1]{\ifnum#1=22 %
\hatcurLCPshorteccenxxxxxA
\else
\ifnum#1=23 %
\hatcurLCPshorteccenxxxxxB
\else
\ifnum#1=24 %
\hatcurLCPshorteccenxxxxxC
\else
??????\fi
\fi
\fi
}
\newcommand{\hatcurLCqeccen}[1]{\ifnum#1=22 %
\hatcurLCqeccenxxxxxA
\else
\ifnum#1=23 %
\hatcurLCqeccenxxxxxB
\else
\ifnum#1=24 %
\hatcurLCqeccenxxxxxC
\else
??????\fi
\fi
\fi
}
\newcommand{\hatcurLCqshorteccen}[1]{\ifnum#1=22 %
\hatcurLCqshorteccenxxxxxA
\else
\ifnum#1=23 %
\hatcurLCqshorteccenxxxxxB
\else
\ifnum#1=24 %
\hatcurLCqshorteccenxxxxxC
\else
??????\fi
\fi
\fi
}
\newcommand{\hatcurLCrhoeccen}[1]{\ifnum#1=22 %
\hatcurLCrhoeccenxxxxxA
\else
\ifnum#1=23 %
\hatcurLCrhoeccenxxxxxB
\else
\ifnum#1=24 %
\hatcurLCrhoeccenxxxxxC
\else
??????\fi
\fi
\fi
}
\newcommand{\hatcurLCrprstareccen}[1]{\ifnum#1=22 %
\hatcurLCrprstareccenxxxxxA
\else
\ifnum#1=23 %
\hatcurLCrprstareccenxxxxxB
\else
\ifnum#1=24 %
\hatcurLCrprstareccenxxxxxC
\else
??????\fi
\fi
\fi
}
\newcommand{\hatcurLCTAeccen}[1]{\ifnum#1=22 %
\hatcurLCTAeccenxxxxxA
\else
\ifnum#1=23 %
\hatcurLCTAeccenxxxxxB
\else
\ifnum#1=24 %
\hatcurLCTAeccenxxxxxC
\else
??????\fi
\fi
\fi
}
\newcommand{\hatcurLCTBeccen}[1]{\ifnum#1=22 %
\hatcurLCTBeccenxxxxxA
\else
\ifnum#1=23 %
\hatcurLCTBeccenxxxxxB
\else
\ifnum#1=24 %
\hatcurLCTBeccenxxxxxC
\else
??????\fi
\fi
\fi
}
\newcommand{\hatcurLCTeccen}[1]{\ifnum#1=22 %
\hatcurLCTeccenxxxxxA
\else
\ifnum#1=23 %
\hatcurLCTeccenxxxxxB
\else
\ifnum#1=24 %
\hatcurLCTeccenxxxxxC
\else
??????\fi
\fi
\fi
}
\newcommand{\hatcurLCzetaeccen}[1]{\ifnum#1=22 %
\hatcurLCzetaeccenxxxxxA
\else
\ifnum#1=23 %
\hatcurLCzetaeccenxxxxxB
\else
\ifnum#1=24 %
\hatcurLCzetaeccenxxxxxC
\else
??????\fi
\fi
\fi
}
\newcommand{\hatcurPPaequiveccen}[1]{\ifnum#1=22 %
\hatcurPPaequiveccenxxxxxA
\else
\ifnum#1=23 %
\hatcurPPaequiveccenxxxxxB
\else
\ifnum#1=24 %
\hatcurPPaequiveccenxxxxxC
\else
??????\fi
\fi
\fi
}
\newcommand{\hatcurPPareccen}[1]{\ifnum#1=22 %
\hatcurPPareccenxxxxxA
\else
\ifnum#1=23 %
\hatcurPPareccenxxxxxB
\else
\ifnum#1=24 %
\hatcurPPareccenxxxxxC
\else
??????\fi
\fi
\fi
}
\newcommand{\hatcurPPareleccen}[1]{\ifnum#1=22 %
\hatcurPPareleccenxxxxxA
\else
\ifnum#1=23 %
\hatcurPPareleccenxxxxxB
\else
\ifnum#1=24 %
\hatcurPPareleccenxxxxxC
\else
??????\fi
\fi
\fi
}
\newcommand{\hatcurPPfluxapdimeccen}[1]{\ifnum#1=22 %
\hatcurPPfluxapdimeccenxxxxxA
\else
\ifnum#1=23 %
\hatcurPPfluxapdimeccenxxxxxB
\else
\ifnum#1=24 %
\hatcurPPfluxapdimeccenxxxxxC
\else
??????\fi
\fi
\fi
}
\newcommand{\hatcurPPfluxapeccen}[1]{\ifnum#1=22 %
\hatcurPPfluxapeccenxxxxxA
\else
\ifnum#1=23 %
\hatcurPPfluxapeccenxxxxxB
\else
\ifnum#1=24 %
\hatcurPPfluxapeccenxxxxxC
\else
??????\fi
\fi
\fi
}
\newcommand{\hatcurPPfluxavgdimeccen}[1]{\ifnum#1=22 %
\hatcurPPfluxavgdimeccenxxxxxA
\else
\ifnum#1=23 %
\hatcurPPfluxavgdimeccenxxxxxB
\else
\ifnum#1=24 %
\hatcurPPfluxavgdimeccenxxxxxC
\else
??????\fi
\fi
\fi
}
\newcommand{\hatcurPPfluxavgeccen}[1]{\ifnum#1=22 %
\hatcurPPfluxavgeccenxxxxxA
\else
\ifnum#1=23 %
\hatcurPPfluxavgeccenxxxxxB
\else
\ifnum#1=24 %
\hatcurPPfluxavgeccenxxxxxC
\else
??????\fi
\fi
\fi
}
\newcommand{\hatcurPPfluxavglogeccen}[1]{\ifnum#1=22 %
\hatcurPPfluxavglogeccenxxxxxA
\else
\ifnum#1=23 %
\hatcurPPfluxavglogeccenxxxxxB
\else
\ifnum#1=24 %
\hatcurPPfluxavglogeccenxxxxxC
\else
??????\fi
\fi
\fi
}
\newcommand{\hatcurPPfluxperidimeccen}[1]{\ifnum#1=22 %
\hatcurPPfluxperidimeccenxxxxxA
\else
\ifnum#1=23 %
\hatcurPPfluxperidimeccenxxxxxB
\else
\ifnum#1=24 %
\hatcurPPfluxperidimeccenxxxxxC
\else
??????\fi
\fi
\fi
}
\newcommand{\hatcurPPfluxperieccen}[1]{\ifnum#1=22 %
\hatcurPPfluxperieccenxxxxxA
\else
\ifnum#1=23 %
\hatcurPPfluxperieccenxxxxxB
\else
\ifnum#1=24 %
\hatcurPPfluxperieccenxxxxxC
\else
??????\fi
\fi
\fi
}
\newcommand{\hatcurPPgeccen}[1]{\ifnum#1=22 %
\hatcurPPgeccenxxxxxA
\else
\ifnum#1=23 %
\hatcurPPgeccenxxxxxB
\else
\ifnum#1=24 %
\hatcurPPgeccenxxxxxC
\else
??????\fi
\fi
\fi
}
\newcommand{\hatcurPPieccen}[1]{\ifnum#1=22 %
\hatcurPPieccenxxxxxA
\else
\ifnum#1=23 %
\hatcurPPieccenxxxxxB
\else
\ifnum#1=24 %
\hatcurPPieccenxxxxxC
\else
??????\fi
\fi
\fi
}
\newcommand{\hatcurPPitwosigupperlimeccen}[1]{\ifnum#1=23 %
\hatcurPPitwosigupperlimeccenxxxxxB
\else
??????\fi
}
\newcommand{\hatcurPPloggeccen}[1]{\ifnum#1=22 %
\hatcurPPloggeccenxxxxxA
\else
\ifnum#1=23 %
\hatcurPPloggeccenxxxxxB
\else
\ifnum#1=24 %
\hatcurPPloggeccenxxxxxC
\else
??????\fi
\fi
\fi
}
\newcommand{\hatcurPPmeccen}[1]{\ifnum#1=22 %
\hatcurPPmeccenxxxxxA
\else
\ifnum#1=23 %
\hatcurPPmeccenxxxxxB
\else
\ifnum#1=24 %
\hatcurPPmeccenxxxxxC
\else
??????\fi
\fi
\fi
}
\newcommand{\hatcurPPmeeccen}[1]{\ifnum#1=22 %
\hatcurPPmeeccenxxxxxA
\else
\ifnum#1=23 %
\hatcurPPmeeccenxxxxxB
\else
\ifnum#1=24 %
\hatcurPPmeeccenxxxxxC
\else
??????\fi
\fi
\fi
}
\newcommand{\hatcurPPmelongeccen}[1]{\ifnum#1=22 %
\hatcurPPmelongeccenxxxxxA
\else
\ifnum#1=23 %
\hatcurPPmelongeccenxxxxxB
\else
\ifnum#1=24 %
\hatcurPPmelongeccenxxxxxC
\else
??????\fi
\fi
\fi
}
\newcommand{\hatcurPPmeshorteccen}[1]{\ifnum#1=22 %
\hatcurPPmeshorteccenxxxxxA
\else
\ifnum#1=23 %
\hatcurPPmeshorteccenxxxxxB
\else
\ifnum#1=24 %
\hatcurPPmeshorteccenxxxxxC
\else
??????\fi
\fi
\fi
}
\newcommand{\hatcurPPmlongeccen}[1]{\ifnum#1=22 %
\hatcurPPmlongeccenxxxxxA
\else
\ifnum#1=23 %
\hatcurPPmlongeccenxxxxxB
\else
\ifnum#1=24 %
\hatcurPPmlongeccenxxxxxC
\else
??????\fi
\fi
\fi
}
\newcommand{\hatcurPPmrcorreccen}[1]{\ifnum#1=22 %
\hatcurPPmrcorreccenxxxxxA
\else
\ifnum#1=23 %
\hatcurPPmrcorreccenxxxxxB
\else
\ifnum#1=24 %
\hatcurPPmrcorreccenxxxxxC
\else
??????\fi
\fi
\fi
}
\newcommand{\hatcurPPmshorteccen}[1]{\ifnum#1=22 %
\hatcurPPmshorteccenxxxxxA
\else
\ifnum#1=23 %
\hatcurPPmshorteccenxxxxxB
\else
\ifnum#1=24 %
\hatcurPPmshorteccenxxxxxC
\else
??????\fi
\fi
\fi
}
\newcommand{\hatcurPPperieccen}[1]{\ifnum#1=22 %
\hatcurPPperieccenxxxxxA
\else
\ifnum#1=23 %
\hatcurPPperieccenxxxxxB
\else
\ifnum#1=24 %
\hatcurPPperieccenxxxxxC
\else
??????\fi
\fi
\fi
}
\newcommand{\hatcurPPphiconjeccen}[1]{\ifnum#1=22 %
\hatcurPPphiconjeccenxxxxxA
\else
\ifnum#1=23 %
\hatcurPPphiconjeccenxxxxxB
\else
\ifnum#1=24 %
\hatcurPPphiconjeccenxxxxxC
\else
??????\fi
\fi
\fi
}
\newcommand{\hatcurPPreccen}[1]{\ifnum#1=22 %
\hatcurPPreccenxxxxxA
\else
\ifnum#1=23 %
\hatcurPPreccenxxxxxB
\else
\ifnum#1=24 %
\hatcurPPreccenxxxxxC
\else
??????\fi
\fi
\fi
}
\newcommand{\hatcurPPreeccen}[1]{\ifnum#1=22 %
\hatcurPPreeccenxxxxxA
\else
\ifnum#1=23 %
\hatcurPPreeccenxxxxxB
\else
\ifnum#1=24 %
\hatcurPPreeccenxxxxxC
\else
??????\fi
\fi
\fi
}
\newcommand{\hatcurPPrelongeccen}[1]{\ifnum#1=22 %
\hatcurPPrelongeccenxxxxxA
\else
\ifnum#1=23 %
\hatcurPPrelongeccenxxxxxB
\else
\ifnum#1=24 %
\hatcurPPrelongeccenxxxxxC
\else
??????\fi
\fi
\fi
}
\newcommand{\hatcurPPreshorteccen}[1]{\ifnum#1=22 %
\hatcurPPreshorteccenxxxxxA
\else
\ifnum#1=23 %
\hatcurPPreshorteccenxxxxxB
\else
\ifnum#1=24 %
\hatcurPPreshorteccenxxxxxC
\else
??????\fi
\fi
\fi
}
\newcommand{\hatcurPPrhoeccen}[1]{\ifnum#1=22 %
\hatcurPPrhoeccenxxxxxA
\else
\ifnum#1=23 %
\hatcurPPrhoeccenxxxxxB
\else
\ifnum#1=24 %
\hatcurPPrhoeccenxxxxxC
\else
??????\fi
\fi
\fi
}
\newcommand{\hatcurPPrlongeccen}[1]{\ifnum#1=22 %
\hatcurPPrlongeccenxxxxxA
\else
\ifnum#1=23 %
\hatcurPPrlongeccenxxxxxB
\else
\ifnum#1=24 %
\hatcurPPrlongeccenxxxxxC
\else
??????\fi
\fi
\fi
}
\newcommand{\hatcurPPrshorteccen}[1]{\ifnum#1=22 %
\hatcurPPrshorteccenxxxxxA
\else
\ifnum#1=23 %
\hatcurPPrshorteccenxxxxxB
\else
\ifnum#1=24 %
\hatcurPPrshorteccenxxxxxC
\else
??????\fi
\fi
\fi
}
\newcommand{\hatcurPPrtwosiglowerlimeccen}[1]{\ifnum#1=23 %
\hatcurPPrtwosiglowerlimeccenxxxxxB
\else
??????\fi
}
\newcommand{\hatcurPPtcirceccen}[1]{\ifnum#1=22 %
\hatcurPPtcirceccenxxxxxA
\else
\ifnum#1=23 %
\hatcurPPtcirceccenxxxxxB
\else
\ifnum#1=24 %
\hatcurPPtcirceccenxxxxxC
\else
??????\fi
\fi
\fi
}
\newcommand{\hatcurPPteffeccen}[1]{\ifnum#1=22 %
\hatcurPPteffeccenxxxxxA
\else
\ifnum#1=23 %
\hatcurPPteffeccenxxxxxB
\else
\ifnum#1=24 %
\hatcurPPteffeccenxxxxxC
\else
??????\fi
\fi
\fi
}
\newcommand{\hatcurPPthetaeccen}[1]{\ifnum#1=22 %
\hatcurPPthetaeccenxxxxxA
\else
\ifnum#1=23 %
\hatcurPPthetaeccenxxxxxB
\else
\ifnum#1=24 %
\hatcurPPthetaeccenxxxxxC
\else
??????\fi
\fi
\fi
}
\newcommand{\hatcurPPtinfalleccen}[1]{\ifnum#1=22 %
\hatcurPPtinfalleccenxxxxxA
\else
\ifnum#1=23 %
\hatcurPPtinfalleccenxxxxxB
\else
\ifnum#1=24 %
\hatcurPPtinfalleccenxxxxxC
\else
??????\fi
\fi
\fi
}
\newcommand{\hatcurRVecceneccen}[1]{\ifnum#1=22 %
\hatcurRVecceneccenxxxxxA
\else
\ifnum#1=23 %
\hatcurRVecceneccenxxxxxB
\else
\ifnum#1=24 %
\hatcurRVecceneccenxxxxxC
\else
??????\fi
\fi
\fi
}
\newcommand{\hatcurRVeccentwosiglimeccen}[1]{\ifnum#1=22 %
\hatcurRVeccentwosiglimeccenxxxxxA
\else
\ifnum#1=23 %
\hatcurRVeccentwosiglimeccenxxxxxB
\else
\ifnum#1=24 %
\hatcurRVeccentwosiglimeccenxxxxxC
\else
??????\fi
\fi
\fi
}
\newcommand{\hatcurRVfitrmsAeccen}[1]{\ifnum#1=22 %
\hatcurRVfitrmsAeccenxxxxxA
\else
\ifnum#1=24 %
\hatcurRVfitrmsAeccenxxxxxC
\else
??????\fi
\fi
}
\newcommand{\hatcurRVfitrmsBeccen}[1]{\ifnum#1=22 %
\hatcurRVfitrmsBeccenxxxxxA
\else
\ifnum#1=24 %
\hatcurRVfitrmsBeccenxxxxxC
\else
??????\fi
\fi
}
\newcommand{\hatcurRVfitrmsCeccen}[1]{\ifnum#1=22 %
\hatcurRVfitrmsCeccenxxxxxA
\else
\ifnum#1=24 %
\hatcurRVfitrmsCeccenxxxxxC
\else
??????\fi
\fi
}
\newcommand{\hatcurRVfitrmsDeccen}[1]{\ifnum#1=24 %
\hatcurRVfitrmsDeccenxxxxxC
\else
??????\fi
}
\newcommand{\hatcurRVfitrmseccen}[1]{\ifnum#1=23 %
\hatcurRVfitrmseccenxxxxxB
\else
??????\fi
}
\newcommand{\hatcurRVgammaAeccen}[1]{\ifnum#1=22 %
\hatcurRVgammaAeccenxxxxxA
\else
\ifnum#1=24 %
\hatcurRVgammaAeccenxxxxxC
\else
??????\fi
\fi
}
\newcommand{\hatcurRVgammaBeccen}[1]{\ifnum#1=22 %
\hatcurRVgammaBeccenxxxxxA
\else
\ifnum#1=24 %
\hatcurRVgammaBeccenxxxxxC
\else
??????\fi
\fi
}
\newcommand{\hatcurRVgammaCeccen}[1]{\ifnum#1=22 %
\hatcurRVgammaCeccenxxxxxA
\else
\ifnum#1=24 %
\hatcurRVgammaCeccenxxxxxC
\else
??????\fi
\fi
}
\newcommand{\hatcurRVgammaDeccen}[1]{\ifnum#1=24 %
\hatcurRVgammaDeccenxxxxxC
\else
??????\fi
}
\newcommand{\hatcurRVgammaeccen}[1]{\ifnum#1=23 %
\hatcurRVgammaeccenxxxxxB
\else
??????\fi
}
\newcommand{\hatcurRVheccen}[1]{\ifnum#1=22 %
\hatcurRVheccenxxxxxA
\else
\ifnum#1=23 %
\hatcurRVheccenxxxxxB
\else
\ifnum#1=24 %
\hatcurRVheccenxxxxxC
\else
??????\fi
\fi
\fi
}
\newcommand{\hatcurRVjitterAeccen}[1]{\ifnum#1=22 %
\hatcurRVjitterAeccenxxxxxA
\else
\ifnum#1=24 %
\hatcurRVjitterAeccenxxxxxC
\else
??????\fi
\fi
}
\newcommand{\hatcurRVjitterBeccen}[1]{\ifnum#1=22 %
\hatcurRVjitterBeccenxxxxxA
\else
\ifnum#1=24 %
\hatcurRVjitterBeccenxxxxxC
\else
??????\fi
\fi
}
\newcommand{\hatcurRVjitterCeccen}[1]{\ifnum#1=22 %
\hatcurRVjitterCeccenxxxxxA
\else
\ifnum#1=24 %
\hatcurRVjitterCeccenxxxxxC
\else
??????\fi
\fi
}
\newcommand{\hatcurRVjitterDeccen}[1]{\ifnum#1=24 %
\hatcurRVjitterDeccenxxxxxC
\else
??????\fi
}
\newcommand{\hatcurRVjittereccen}[1]{\ifnum#1=23 %
\hatcurRVjittereccenxxxxxB
\else
??????\fi
}
\newcommand{\hatcurRVjittertwosiglimAeccen}[1]{\ifnum#1=22 %
\hatcurRVjittertwosiglimAeccenxxxxxA
\else
\ifnum#1=24 %
\hatcurRVjittertwosiglimAeccenxxxxxC
\else
??????\fi
\fi
}
\newcommand{\hatcurRVjittertwosiglimBeccen}[1]{\ifnum#1=22 %
\hatcurRVjittertwosiglimBeccenxxxxxA
\else
\ifnum#1=24 %
\hatcurRVjittertwosiglimBeccenxxxxxC
\else
??????\fi
\fi
}
\newcommand{\hatcurRVjittertwosiglimCeccen}[1]{\ifnum#1=22 %
\hatcurRVjittertwosiglimCeccenxxxxxA
\else
\ifnum#1=24 %
\hatcurRVjittertwosiglimCeccenxxxxxC
\else
??????\fi
\fi
}
\newcommand{\hatcurRVjittertwosiglimDeccen}[1]{\ifnum#1=24 %
\hatcurRVjittertwosiglimDeccenxxxxxC
\else
??????\fi
}
\newcommand{\hatcurRVjittertwosiglimeccen}[1]{\ifnum#1=23 %
\hatcurRVjittertwosiglimeccenxxxxxB
\else
??????\fi
}
\newcommand{\hatcurRVkeccen}[1]{\ifnum#1=22 %
\hatcurRVkeccenxxxxxA
\else
\ifnum#1=23 %
\hatcurRVkeccenxxxxxB
\else
\ifnum#1=24 %
\hatcurRVkeccenxxxxxC
\else
??????\fi
\fi
\fi
}
\newcommand{\hatcurRVKeccen}[1]{\ifnum#1=22 %
\hatcurRVKeccenxxxxxA
\else
\ifnum#1=23 %
\hatcurRVKeccenxxxxxB
\else
\ifnum#1=24 %
\hatcurRVKeccenxxxxxC
\else
??????\fi
\fi
\fi
}
\newcommand{\hatcurRVomegaeccen}[1]{\ifnum#1=22 %
\hatcurRVomegaeccenxxxxxA
\else
\ifnum#1=23 %
\hatcurRVomegaeccenxxxxxB
\else
\ifnum#1=24 %
\hatcurRVomegaeccenxxxxxC
\else
??????\fi
\fi
\fi
}
\newcommand{\hatcurRVrheccen}[1]{\ifnum#1=22 %
\hatcurRVrheccenxxxxxA
\else
\ifnum#1=23 %
\hatcurRVrheccenxxxxxB
\else
\ifnum#1=24 %
\hatcurRVrheccenxxxxxC
\else
??????\fi
\fi
\fi
}
\newcommand{\hatcurRVrkeccen}[1]{\ifnum#1=22 %
\hatcurRVrkeccenxxxxxA
\else
\ifnum#1=23 %
\hatcurRVrkeccenxxxxxB
\else
\ifnum#1=24 %
\hatcurRVrkeccenxxxxxC
\else
??????\fi
\fi
\fi
}
\newcommand{\hatcurRVtroneeccen}[1]{\ifnum#1=22 %
\hatcurRVtroneeccenxxxxxA
\else
\ifnum#1=23 %
\hatcurRVtroneeccenxxxxxB
\else
\ifnum#1=24 %
\hatcurRVtroneeccenxxxxxC
\else
??????\fi
\fi
\fi
}
\newcommand{\hatcurRVtrtwoeccen}[1]{\ifnum#1=22 %
\hatcurRVtrtwoeccenxxxxxA
\else
\ifnum#1=23 %
\hatcurRVtrtwoeccenxxxxxB
\else
\ifnum#1=24 %
\hatcurRVtrtwoeccenxxxxxC
\else
??????\fi
\fi
\fi
}
\newcommand{\hatcurSMEiiloggeccen}[1]{\ifnum#1=22 %
\hatcurSMEiiloggeccenxxxxxA
\else
\ifnum#1=23 %
\hatcurSMEiiloggeccenxxxxxB
\else
\ifnum#1=24 %
\hatcurSMEiiloggeccenxxxxxC
\else
??????\fi
\fi
\fi
}
\newcommand{\hatcurSMEiiteffeccen}[1]{\ifnum#1=22 %
\hatcurSMEiiteffeccenxxxxxA
\else
\ifnum#1=23 %
\hatcurSMEiiteffeccenxxxxxB
\else
\ifnum#1=24 %
\hatcurSMEiiteffeccenxxxxxC
\else
??????\fi
\fi
\fi
}
\newcommand{\hatcurSMEiivsineccen}[1]{\ifnum#1=22 %
\hatcurSMEiivsineccenxxxxxA
\else
\ifnum#1=23 %
\hatcurSMEiivsineccenxxxxxB
\else
\ifnum#1=24 %
\hatcurSMEiivsineccenxxxxxC
\else
??????\fi
\fi
\fi
}
\newcommand{\hatcurSMEiizfeheccen}[1]{\ifnum#1=22 %
\hatcurSMEiizfeheccenxxxxxA
\else
\ifnum#1=23 %
\hatcurSMEiizfeheccenxxxxxB
\else
\ifnum#1=24 %
\hatcurSMEiizfeheccenxxxxxC
\else
??????\fi
\fi
\fi
}
\newcommand{\hatcurSMEiizfehshorteccen}[1]{\ifnum#1=22 %
\hatcurSMEiizfehshorteccenxxxxxA
\else
\ifnum#1=23 %
\hatcurSMEiizfehshorteccenxxxxxB
\else
\ifnum#1=24 %
\hatcurSMEiizfehshorteccenxxxxxC
\else
??????\fi
\fi
\fi
}
\newcommand{\hatcurSMEiloggeccen}[1]{\ifnum#1=22 %
\hatcurSMEiloggeccenxxxxxA
\else
\ifnum#1=23 %
\hatcurSMEiloggeccenxxxxxB
\else
\ifnum#1=24 %
\hatcurSMEiloggeccenxxxxxC
\else
??????\fi
\fi
\fi
}
\newcommand{\hatcurSMEiteffeccen}[1]{\ifnum#1=22 %
\hatcurSMEiteffeccenxxxxxA
\else
\ifnum#1=23 %
\hatcurSMEiteffeccenxxxxxB
\else
\ifnum#1=24 %
\hatcurSMEiteffeccenxxxxxC
\else
??????\fi
\fi
\fi
}
\newcommand{\hatcurSMEivmaceccen}[1]{\ifnum#1=22 %
\hatcurSMEivmaceccenxxxxxA
\else
\ifnum#1=23 %
\hatcurSMEivmaceccenxxxxxB
\else
\ifnum#1=24 %
\hatcurSMEivmaceccenxxxxxC
\else
??????\fi
\fi
\fi
}
\newcommand{\hatcurSMEivmiceccen}[1]{\ifnum#1=22 %
\hatcurSMEivmiceccenxxxxxA
\else
\ifnum#1=23 %
\hatcurSMEivmiceccenxxxxxB
\else
\ifnum#1=24 %
\hatcurSMEivmiceccenxxxxxC
\else
??????\fi
\fi
\fi
}
\newcommand{\hatcurSMEivsineccen}[1]{\ifnum#1=22 %
\hatcurSMEivsineccenxxxxxA
\else
\ifnum#1=23 %
\hatcurSMEivsineccenxxxxxB
\else
\ifnum#1=24 %
\hatcurSMEivsineccenxxxxxC
\else
??????\fi
\fi
\fi
}
\newcommand{\hatcurSMEizfeheccen}[1]{\ifnum#1=22 %
\hatcurSMEizfeheccenxxxxxA
\else
\ifnum#1=23 %
\hatcurSMEizfeheccenxxxxxB
\else
\ifnum#1=24 %
\hatcurSMEizfeheccenxxxxxC
\else
??????\fi
\fi
\fi
}
\newcommand{\hatcurSMEizfehshorteccen}[1]{\ifnum#1=22 %
\hatcurSMEizfehshorteccenxxxxxA
\else
\ifnum#1=23 %
\hatcurSMEizfehshorteccenxxxxxB
\else
\ifnum#1=24 %
\hatcurSMEizfehshorteccenxxxxxC
\else
??????\fi
\fi
\fi
}
\newcommand{\hatcurXAveccen}[1]{\ifnum#1=22 %
\hatcurXAveccenxxxxxA
\else
\ifnum#1=23 %
\hatcurXAveccenxxxxxB
\else
\ifnum#1=24 %
\hatcurXAveccenxxxxxC
\else
??????\fi
\fi
\fi
}
\newcommand{\hatcurXdisteccen}[1]{\ifnum#1=22 %
\hatcurXdisteccenxxxxxA
\else
\ifnum#1=23 %
\hatcurXdisteccenxxxxxB
\else
\ifnum#1=24 %
\hatcurXdisteccenxxxxxC
\else
??????\fi
\fi
\fi
}
\newcommand{\hatcurXdistredeccen}[1]{\ifnum#1=22 %
\hatcurXdistredeccenxxxxxA
\else
\ifnum#1=23 %
\hatcurXdistredeccenxxxxxB
\else
\ifnum#1=24 %
\hatcurXdistredeccenxxxxxC
\else
??????\fi
\fi
\fi
}
\newcommand{\hatcurXEBVeccen}[1]{\ifnum#1=22 %
\hatcurXEBVeccenxxxxxA
\else
\ifnum#1=23 %
\hatcurXEBVeccenxxxxxB
\else
\ifnum#1=24 %
\hatcurXEBVeccenxxxxxC
\else
??????\fi
\fi
\fi
}
\newcommand{\hatcurXjhisoredeccen}[1]{\ifnum#1=22 %
\hatcurXjhisoredeccenxxxxxA
\else
\ifnum#1=23 %
\hatcurXjhisoredeccenxxxxxB
\else
\ifnum#1=24 %
\hatcurXjhisoredeccenxxxxxC
\else
??????\fi
\fi
\fi
}
\newcommand{\hatcurXjkisoredeccen}[1]{\ifnum#1=22 %
\hatcurXjkisoredeccenxxxxxA
\else
\ifnum#1=23 %
\hatcurXjkisoredeccenxxxxxB
\else
\ifnum#1=24 %
\hatcurXjkisoredeccenxxxxxC
\else
??????\fi
\fi
\fi
}
\newcommand{\hatcurXmhisoredeccen}[1]{\ifnum#1=22 %
\hatcurXmhisoredeccenxxxxxA
\else
\ifnum#1=23 %
\hatcurXmhisoredeccenxxxxxB
\else
\ifnum#1=24 %
\hatcurXmhisoredeccenxxxxxC
\else
??????\fi
\fi
\fi
}
\newcommand{\hatcurXmiisoredeccen}[1]{\ifnum#1=22 %
\hatcurXmiisoredeccenxxxxxA
\else
\ifnum#1=23 %
\hatcurXmiisoredeccenxxxxxB
\else
\ifnum#1=24 %
\hatcurXmiisoredeccenxxxxxC
\else
??????\fi
\fi
\fi
}
\newcommand{\hatcurXmjisoredeccen}[1]{\ifnum#1=22 %
\hatcurXmjisoredeccenxxxxxA
\else
\ifnum#1=23 %
\hatcurXmjisoredeccenxxxxxB
\else
\ifnum#1=24 %
\hatcurXmjisoredeccenxxxxxC
\else
??????\fi
\fi
\fi
}
\newcommand{\hatcurXmkisoredeccen}[1]{\ifnum#1=22 %
\hatcurXmkisoredeccenxxxxxA
\else
\ifnum#1=23 %
\hatcurXmkisoredeccenxxxxxB
\else
\ifnum#1=24 %
\hatcurXmkisoredeccenxxxxxC
\else
??????\fi
\fi
\fi
}
\newcommand{\hatcurXmvisoredeccen}[1]{\ifnum#1=22 %
\hatcurXmvisoredeccenxxxxxA
\else
\ifnum#1=23 %
\hatcurXmvisoredeccenxxxxxB
\else
\ifnum#1=24 %
\hatcurXmvisoredeccenxxxxxC
\else
??????\fi
\fi
\fi
}
\newcommand{\hatcurXsecdureccen}[1]{\ifnum#1=22 %
\hatcurXsecdureccenxxxxxA
\else
\ifnum#1=23 %
\hatcurXsecdureccenxxxxxB
\else
\ifnum#1=24 %
\hatcurXsecdureccenxxxxxC
\else
??????\fi
\fi
\fi
}
\newcommand{\hatcurXsecingdureccen}[1]{\ifnum#1=22 %
\hatcurXsecingdureccenxxxxxA
\else
\ifnum#1=23 %
\hatcurXsecingdureccenxxxxxB
\else
\ifnum#1=24 %
\hatcurXsecingdureccenxxxxxC
\else
??????\fi
\fi
\fi
}
\newcommand{\hatcurXsecondaryeccen}[1]{\ifnum#1=22 %
\hatcurXsecondaryeccenxxxxxA
\else
\ifnum#1=23 %
\hatcurXsecondaryeccenxxxxxB
\else
\ifnum#1=24 %
\hatcurXsecondaryeccenxxxxxC
\else
??????\fi
\fi
\fi
}
\newcommand{\hatcurXsecphaseeccen}[1]{\ifnum#1=22 %
\hatcurXsecphaseeccenxxxxxA
\else
\ifnum#1=23 %
\hatcurXsecphaseeccenxxxxxB
\else
\ifnum#1=24 %
\hatcurXsecphaseeccenxxxxxC
\else
??????\fi
\fi
\fi
}
\newcommand{\hatcurXviisoredeccen}[1]{\ifnum#1=22 %
\hatcurXviisoredeccenxxxxxA
\else
\ifnum#1=23 %
\hatcurXviisoredeccenxxxxxB
\else
\ifnum#1=24 %
\hatcurXviisoredeccenxxxxxC
\else
??????\fi
\fi
\fi
}
\newcommand{\hatcurXvkisoredeccen}[1]{\ifnum#1=22 %
\hatcurXvkisoredeccenxxxxxA
\else
\ifnum#1=23 %
\hatcurXvkisoredeccenxxxxxB
\else
\ifnum#1=24 %
\hatcurXvkisoredeccenxxxxxC
\else
??????\fi
\fi
\fi
}

\newcommand{\hatcurxxxxxA}{HATS-22}
\newcommand{\hatcurbxxxxxA}{HATS-22b}
\newcommand{\hatcurcxxxxxA}{HATS-22c}

\newcommand{\hatcurplanetnumxxxxxA}{22}

\newcommand{\hatcurRVgammaabsxxxxxA}{\hatcurRVgammaCeccen{\hatcurplanetnumxxxxxA}}                           

\newcommand{\hatcurRVgammainstxxxxxA}{CORALIE}

\newcommand{\hatcurRVgammarelxxxxxA}{\hatcurRVgammaCeccen{\hatcurplanetnumxxxxxA}}                           

\newcommand{\hatcurCCtassvixxxxxA}{NULL}                  

\newcommand{\hatcurSMEversionxxxxxA}{ii}                                       

\newcommand{\hatcurisoshortxxxxxA}{YY}
\newcommand{\hatcurisofullxxxxxA}{Yonsei-Yale (YY)}
\newcommand{\hatcurisocitexxxxxA}{yi:2001}

\newcommand{\hatcurlumindxxxxxA}{\arstar}

\newcommand{\hatcurjhkfilsetxxxxxA}{ESO}

\newcommand{\hatcurSMEteffxxxxxA}{\ifthenelse{\equal{\hatcurSMEversionxxxxxA}{i}}{\hatcurSMEiteff{\hatcurplanetnumxxxxxA}}{\hatcurSMEiiteff{\hatcurplanetnumxxxxxA}}}
\newcommand{\hatcurSMEzfehxxxxxA}{\ifthenelse{\equal{\hatcurSMEversionxxxxxA}{i}}{\hatcurSMEizfeh{\hatcurplanetnumxxxxxA}}{\hatcurSMEiizfeh{\hatcurplanetnumxxxxxA}}}
\newcommand{\hatcurSMEzfehshortxxxxxA}{\ifthenelse{\equal{\hatcurSMEversionxxxxxA}{i}}{\hatcurSMEizfehshort{\hatcurplanetnumxxxxxA}}{\hatcurSMEiizfehshort{\hatcurplanetnumxxxxxA}}}
\newcommand{\hatcurSMEloggxxxxxA}{\ifthenelse{\equal{\hatcurSMEversionxxxxxA}{i}}{\hatcurSMEilogg{\hatcurplanetnumxxxxxA}}{\hatcurSMEiilogg{\hatcurplanetnumxxxxxA}}}
\newcommand{\hatcurSMEvsinxxxxxA}{\ifthenelse{\equal{\hatcurSMEversionxxxxxA}{i}}{\hatcurSMEivsin{\hatcurplanetnumxxxxxA}}{\hatcurSMEiivsin{\hatcurplanetnumxxxxxA}}}
\newcommand{\hatcurSMEvmacxxxxxA}{\ifthenelse{\equal{\hatcurSMEversionxxxxxA}{i}}{\hatcurSMEivmac{\hatcurplanetnumxxxxxA}}{\hatcurSMEiivmac{\hatcurplanetnumxxxxxA}}}
\newcommand{\hatcurSMEvmicxxxxxA}{\ifthenelse{\equal{\hatcurSMEversionxxxxxA}{i}}{\hatcurSMEivmic{\hatcurplanetnumxxxxxA}}{\hatcurSMEiivmic{\hatcurplanetnumxxxxxA}}}

\newcommand{\hatcurxxxxxB}{HATS-23}
\newcommand{\hatcurbxxxxxB}{HATS-23b}
\newcommand{\hatcurcxxxxxB}{HATS-23c}

\newcommand{\hatcurplanetnumxxxxxB}{23}

\newcommand{\hatcurRVgammaabsxxxxxB}{\hatcurRVgamma{\hatcurplanetnumxxxxxB}}                           

\newcommand{\hatcurRVgammainstxxxxxB}{FEROS}

\newcommand{\hatcurRVgammarelxxxxxB}{\hatcurRVgamma{\hatcurplanetnumxxxxxB}}                           

\newcommand{\hatcurCCtassvixxxxxB}{NULL}

\newcommand{\hatcurSMEversionxxxxxB}{ii}                                       

\newcommand{\hatcurisoshortxxxxxB}{YY}
\newcommand{\hatcurisofullxxxxxB}{Yonsei-Yale (YY)}
\newcommand{\hatcurisocitexxxxxB}{yi:2001}

\newcommand{\hatcurlumindxxxxxB}{\arstar}

\newcommand{\hatcurjhkfilsetxxxxxB}{ESO}

\newcommand{\hatcurSMEteffxxxxxB}{\ifthenelse{\equal{\hatcurSMEversionxxxxxB}{i}}{\hatcurSMEiteff{\hatcurplanetnumxxxxxB}}{\hatcurSMEiiteff{\hatcurplanetnumxxxxxB}}}
\newcommand{\hatcurSMEzfehxxxxxB}{\ifthenelse{\equal{\hatcurSMEversionxxxxxB}{i}}{\hatcurSMEizfeh{\hatcurplanetnumxxxxxB}}{\hatcurSMEiizfeh{\hatcurplanetnumxxxxxB}}}
\newcommand{\hatcurSMEzfehshortxxxxxB}{\ifthenelse{\equal{\hatcurSMEversionxxxxxB}{i}}{\hatcurSMEizfehshort{\hatcurplanetnumxxxxxB}}{\hatcurSMEiizfehshort{\hatcurplanetnumxxxxxB}}}
\newcommand{\hatcurSMEloggxxxxxB}{\ifthenelse{\equal{\hatcurSMEversionxxxxxB}{i}}{\hatcurSMEilogg{\hatcurplanetnumxxxxxB}}{\hatcurSMEiilogg{\hatcurplanetnumxxxxxB}}}
\newcommand{\hatcurSMEvsinxxxxxB}{\ifthenelse{\equal{\hatcurSMEversionxxxxxB}{i}}{\hatcurSMEivsin{\hatcurplanetnumxxxxxB}}{\hatcurSMEiivsin{\hatcurplanetnumxxxxxB}}}
\newcommand{\hatcurSMEvmacxxxxxB}{\ifthenelse{\equal{\hatcurSMEversionxxxxxB}{i}}{\hatcurSMEivmac{\hatcurplanetnumxxxxxB}}{\hatcurSMEiivmac{\hatcurplanetnumxxxxxB}}}
\newcommand{\hatcurSMEvmicxxxxxB}{\ifthenelse{\equal{\hatcurSMEversionxxxxxB}{i}}{\hatcurSMEivmic{\hatcurplanetnumxxxxxB}}{\hatcurSMEiivmic{\hatcurplanetnumxxxxxB}}}

\newcommand{\hatcurxxxxxC}{HATS-24}
\newcommand{\hatcurbxxxxxC}{HATS-24b}
\newcommand{\hatcurcxxxxxC}{HATS-24c}

\newcommand{\hatcurplanetnumxxxxxC}{24}

\newcommand{\hatcurRVgammaabsxxxxxC}{\hatcurRVgammaA{\hatcurplanetnumxxxxxC}}                           

\newcommand{\hatcurRVgammainstxxxxxC}{FEROS}

\newcommand{\hatcurRVgammarelxxxxxC}{\hatcurRVgammaA{\hatcurplanetnumxxxxxC}}                           

\newcommand{\hatcurCCtassvixxxxxC}{NULL}

\newcommand{\hatcurSMEversionxxxxxC}{ii}                                       

\newcommand{\hatcurisoshortxxxxxC}{YY}
\newcommand{\hatcurisofullxxxxxC}{Yonsei-Yale (YY)}
\newcommand{\hatcurisocitexxxxxC}{yi:2001}

\newcommand{\hatcurlumindxxxxxC}{\arstar}

\newcommand{\hatcurjhkfilsetxxxxxC}{ESO}

\newcommand{\hatcurSMEteffxxxxxC}{\ifthenelse{\equal{\hatcurSMEversionxxxxxC}{i}}{\hatcurSMEiteff{\hatcurplanetnumxxxxxC}}{\hatcurSMEiiteff{\hatcurplanetnumxxxxxC}}}
\newcommand{\hatcurSMEzfehxxxxxC}{\ifthenelse{\equal{\hatcurSMEversionxxxxxC}{i}}{\hatcurSMEizfeh{\hatcurplanetnumxxxxxC}}{\hatcurSMEiizfeh{\hatcurplanetnumxxxxxC}}}
\newcommand{\hatcurSMEzfehshortxxxxxC}{\ifthenelse{\equal{\hatcurSMEversionxxxxxC}{i}}{\hatcurSMEizfehshort{\hatcurplanetnumxxxxxC}}{\hatcurSMEiizfehshort{\hatcurplanetnumxxxxxC}}}
\newcommand{\hatcurSMEloggxxxxxC}{\ifthenelse{\equal{\hatcurSMEversionxxxxxC}{i}}{\hatcurSMEilogg{\hatcurplanetnumxxxxxC}}{\hatcurSMEiilogg{\hatcurplanetnumxxxxxC}}}
\newcommand{\hatcurSMEvsinxxxxxC}{\ifthenelse{\equal{\hatcurSMEversionxxxxxC}{i}}{\hatcurSMEivsin{\hatcurplanetnumxxxxxC}}{\hatcurSMEiivsin{\hatcurplanetnumxxxxxC}}}
\newcommand{\hatcurSMEvmacxxxxxC}{\ifthenelse{\equal{\hatcurSMEversionxxxxxC}{i}}{\hatcurSMEivmac{\hatcurplanetnumxxxxxC}}{\hatcurSMEiivmac{\hatcurplanetnumxxxxxC}}}
\newcommand{\hatcurSMEvmicxxxxxC}{\ifthenelse{\equal{\hatcurSMEversionxxxxxC}{i}}{\hatcurSMEivmic{\hatcurplanetnumxxxxxC}}{\hatcurSMEiivmic{\hatcurplanetnumxxxxxC}}}

\newcommand{\hatcur}[1]{\ifnum#1=22 %
\hatcurxxxxxA
\else
\ifnum#1=23 %
\hatcurxxxxxB
\else
\ifnum#1=24 %
\hatcurxxxxxC
\else
??????\fi
\fi
\fi
}
\newcommand{\hatcurb}[1]{\ifnum#1=22 %
\hatcurbxxxxxA
\else
\ifnum#1=23 %
\hatcurbxxxxxB
\else
\ifnum#1=24 %
\hatcurbxxxxxC
\else
??????\fi
\fi
\fi
}
\newcommand{\hatcurc}[1]{\ifnum#1=22 %
\hatcurcxxxxxA
\else
\ifnum#1=23 %
\hatcurcxxxxxB
\else
\ifnum#1=24 %
\hatcurcxxxxxC
\else
??????\fi
\fi
\fi
}
\newcommand{\hatcurCCtassvi}[1]{\ifnum#1=22 %
\hatcurCCtassvixxxxxA
\else
\ifnum#1=23 %
\hatcurCCtassvixxxxxB
\else
\ifnum#1=24 %
\hatcurCCtassvixxxxxC
\else
??????\fi
\fi
\fi
}
\newcommand{\hatcurisocite}[1]{\ifnum#1=22 %
\hatcurisocitexxxxxA
\else
\ifnum#1=23 %
\hatcurisocitexxxxxB
\else
\ifnum#1=24 %
\hatcurisocitexxxxxC
\else
??????\fi
\fi
\fi
}
\newcommand{\hatcurisofull}[1]{\ifnum#1=22 %
\hatcurisofullxxxxxA
\else
\ifnum#1=23 %
\hatcurisofullxxxxxB
\else
\ifnum#1=24 %
\hatcurisofullxxxxxC
\else
??????\fi
\fi
\fi
}
\newcommand{\hatcurisoshort}[1]{\ifnum#1=22 %
\hatcurisoshortxxxxxA
\else
\ifnum#1=23 %
\hatcurisoshortxxxxxB
\else
\ifnum#1=24 %
\hatcurisoshortxxxxxC
\else
??????\fi
\fi
\fi
}
\newcommand{\hatcurjhkfilset}[1]{\ifnum#1=22 %
\hatcurjhkfilsetxxxxxA
\else
\ifnum#1=23 %
\hatcurjhkfilsetxxxxxB
\else
\ifnum#1=24 %
\hatcurjhkfilsetxxxxxC
\else
??????\fi
\fi
\fi
}
\newcommand{\hatcurlumind}[1]{\ifnum#1=22 %
\hatcurlumindxxxxxA
\else
\ifnum#1=23 %
\hatcurlumindxxxxxB
\else
\ifnum#1=24 %
\hatcurlumindxxxxxC
\else
??????\fi
\fi
\fi
}
\newcommand{\hatcurplanetnum}[1]{\ifnum#1=22 %
\hatcurplanetnumxxxxxA
\else
\ifnum#1=23 %
\hatcurplanetnumxxxxxB
\else
\ifnum#1=24 %
\hatcurplanetnumxxxxxC
\else
??????\fi
\fi
\fi
}
\newcommand{\hatcurRVgammaabs}[1]{\ifnum#1=22 %
\hatcurRVgammaabsxxxxxA
\else
\ifnum#1=23 %
\hatcurRVgammaabsxxxxxB
\else
\ifnum#1=24 %
\hatcurRVgammaabsxxxxxC
\else
??????\fi
\fi
\fi
}
\newcommand{\hatcurRVgammainst}[1]{\ifnum#1=22 %
\hatcurRVgammainstxxxxxA
\else
\ifnum#1=23 %
\hatcurRVgammainstxxxxxB
\else
\ifnum#1=24 %
\hatcurRVgammainstxxxxxC
\else
??????\fi
\fi
\fi
}
\newcommand{\hatcurRVgammarel}[1]{\ifnum#1=22 %
\hatcurRVgammarelxxxxxA
\else
\ifnum#1=23 %
\hatcurRVgammarelxxxxxB
\else
\ifnum#1=24 %
\hatcurRVgammarelxxxxxC
\else
??????\fi
\fi
\fi
}
\newcommand{\hatcurSMElogg}[1]{\ifnum#1=22 %
\hatcurSMEloggxxxxxA
\else
\ifnum#1=23 %
\hatcurSMEloggxxxxxB
\else
\ifnum#1=24 %
\hatcurSMEloggxxxxxC
\else
??????\fi
\fi
\fi
}
\newcommand{\hatcurSMEteff}[1]{\ifnum#1=22 %
\hatcurSMEteffxxxxxA
\else
\ifnum#1=23 %
\hatcurSMEteffxxxxxB
\else
\ifnum#1=24 %
\hatcurSMEteffxxxxxC
\else
??????\fi
\fi
\fi
}
\newcommand{\hatcurSMEversion}[1]{\ifnum#1=22 %
\hatcurSMEversionxxxxxA
\else
\ifnum#1=23 %
\hatcurSMEversionxxxxxB
\else
\ifnum#1=24 %
\hatcurSMEversionxxxxxC
\else
??????\fi
\fi
\fi
}
\newcommand{\hatcurSMEvmac}[1]{\ifnum#1=22 %
\hatcurSMEvmacxxxxxA
\else
\ifnum#1=23 %
\hatcurSMEvmacxxxxxB
\else
\ifnum#1=24 %
\hatcurSMEvmacxxxxxC
\else
??????\fi
\fi
\fi
}
\newcommand{\hatcurSMEvmic}[1]{\ifnum#1=22 %
\hatcurSMEvmicxxxxxA
\else
\ifnum#1=23 %
\hatcurSMEvmicxxxxxB
\else
\ifnum#1=24 %
\hatcurSMEvmicxxxxxC
\else
??????\fi
\fi
\fi
}
\newcommand{\hatcurSMEvsin}[1]{\ifnum#1=22 %
\hatcurSMEvsinxxxxxA
\else
\ifnum#1=23 %
\hatcurSMEvsinxxxxxB
\else
\ifnum#1=24 %
\hatcurSMEvsinxxxxxC
\else
??????\fi
\fi
\fi
}
\newcommand{\hatcurSMEzfeh}[1]{\ifnum#1=22 %
\hatcurSMEzfehxxxxxA
\else
\ifnum#1=23 %
\hatcurSMEzfehxxxxxB
\else
\ifnum#1=24 %
\hatcurSMEzfehxxxxxC
\else
??????\fi
\fi
\fi
}
\newcommand{\hatcurSMEzfehshort}[1]{\ifnum#1=22 %
\hatcurSMEzfehshortxxxxxA
\else
\ifnum#1=23 %
\hatcurSMEzfehshortxxxxxB
\else
\ifnum#1=24 %
\hatcurSMEzfehshortxxxxxC
\else
??????\fi
\fi
\fi
}

\newcounter{planetcounter}

\newboolean{emulateapj}
\setboolean{emulateapj}{true}

\newboolean{rvtablelong}
\setboolean{rvtablelong}{true}

\newboolean{astroph}
\setboolean{astroph}{true}

\newlength{\plotwidthtwo}

\shortauthors{Bento et al.}
\shorttitle{\hatcur{22}\lowercase{b}--\hatcur{24}\lowercase{b}}
\ifthenelse{\boolean{emulateapj}}{
    \newcommand{\titledag}{$\dagger$}
}{
    \newcommand{\titledag}{\dagger}
}

\begin{document}

\title{
\hatcur{22}\lowercase{b}, \hatcur{23}\lowercase{b} and \hatcur{24}\lowercase{b}: Three New Transiting Super-Jupiters from the HATSouth project\altaffilmark{\titledag}
}

\author{J.~Bento$^{1}$, B.~Schmidt$^{1}$, J.~D.~Hartman$^{2}$, G.~\'A. Bakos$^{2}$, S.~Ciceri$^{9,10}$, R.~Brahm$^{3}$, D.~Bayliss$^{5}$, N.~Espinoza$^{3}$, G.~Zhou$^{4}$, M.~Rabus$^{3,9}$, W.~Bhatti$^{2}$, K.~Penev$^{2}$, Z.~Csubry$^{2}$, A.~Jord\'an$^{3,9}$, L.~Mancini$^{9}$, T.~Henning$^{9}$, M.~de~Val-Borro$^{2}$, C.~G.~Tinney$^{6,7}$, 
D.~J.~Wright$^{6,7}$, S.~Durkan$^{11}$, V.~Suc$^{3}$, R.~Noyes$^{4}$, 
J.~L\'az\'ar$^{8}$, I.~Papp$^{8}$, 
P.~S\'ari$^{8}$}

\altaffiltext{$^{1}$}{Research School of Astronomy and Astrophysics, Mount Stromlo Observatory, Australian National University, Cotter Road, Weston, ACT 2611, Australia. joao.bento@anu.edu.au} 
\altaffiltext{$^{2}$}{Department of Astrophysical Sciences, 4 Ivy Ln., Princeton, NJ 08544}
\altaffiltext{$^{3}$}{Instituto de Astrof\'isica, Facultad de F\'isica, Pontificia Universidad
Cat\'olica de Chile, Av. Vicu\~{n}a Mackenna 4860, 7820436 Macul, Santiago, Chile}
\altaffiltext{$^{4}$}{Harvard-Smithsonian Center for Astrophysics, Cambridge, MA 02138, USA}
\altaffiltext{$^{5}$}{Observatoire Astronomique de l'Universit\'e de Gen\'eve, 51 ch. des Maillettes, 1290 Versoix, Switzerland}
\altaffiltext{$^{6}$}{Australian Centre for Astrobiology, School of Physics,
University of New South Wales, NSW 2052, Australia}
\altaffiltext{$^{7}$}{Exoplanetary Science at UNSW, School of Physics,
University of New South Wales, NSW 2052, Australia}
\altaffiltext{$^{8}$}{Hungarian Astronomical Association, Budapest, Hungary}
\altaffiltext{$^{9}$}{Max Plank Institute for Astronomy, K\"onigstuhl 17, 69117 Heidelberg, Germany}
\altaffiltext{$^{10}$}{Department of Astronomy, Stockholm University, SE-106 91 Stockholm, Sweden}
\altaffiltext{$^{11}$}{Astrophysics Research Centre, School of Mathematics \& Physics, Queen's University, Belfast BT7 1NN, UK}

\altaffiltext{$\dagger$}{
 The HATSouth network is operated by a collaboration consisting of
Princeton University (PU), the Max Planck Institute f\"ur Astronomie
(MPIA), the Australian National University (ANU), and the Pontificia
Universidad Cat\'olica de Chile (PUC).  The station at Las Campanas
Observatory (LCO) of the Carnegie Institute is operated by PU in
conjunction with PUC, the station at the High Energy Spectroscopic
Survey (H.E.S.S.) site is operated in conjunction with MPIA, and the
station at Siding Spring Observatory (SSO) is operated jointly with
ANU.
 Based in part on data collected at Subaru Telescope, which is
 operated by the National Astronomical Observatory of Japan. Based in
 part on observations made with the MPG~2.2\,m Telescope at the ESO
 Observatory in La Silla.
}


\begin{abstract}

\setcounter{footnote}{10}
We report the discovery of three moderately high-mass transiting \hjs from the HATSouth survey: \hatcurb{22}, \hatcurb{23} and \hatcurb{24}. These planets add to the number of known planets in the  $\sim 2\mjup$ regime. \hatcurb{22} is a $\hatcurPPmlongeccen{22}\,\mjup$ mass and $\hatcurPPrlongeccen{22}\,\rjup$ radius planet orbiting a $V=\hatcurCCtassmveccen{22}$ sub-solar mass ($\mstar = \hatcurISOmlongeccen{22}\,\msun$;$\rstar = \hatcurISOmlongeccen{22}\,\rsun$) K-dwarf host star on an eccentric ($e = \hatcurRVecceneccen{22}$) orbit. This planet's high planet-to-stellar mass ratio is further evidence that migration mechanisms for \hjs may rely on exciting orbital eccentricities that bring the planets closer to their parent stars followed by tidal circularisation. \hatcurb{23} is a $\hatcurPPmeccen{23}\,\mjup$ mass and $\hatcurPPrlongeccen{23}\,\rjup$ radius planet on a grazing orbit around a $V=\hatcurCCtassmveccen{23}$ G-dwarf with properties very similar to those of the Sun ($\mstar = \hatcurISOmlongeccen{23}$; $\rstar = \hatcurISOrlongeccen{23}$). \hatcurb{24} orbits a moderately bright V=\hatcurCCtassmv{24}\ \hatcurISOspec{24}\ dwarf star ($\mstar = \hatcurISOmlongeccen{24}\,\msun$; $\rstar = \hatcurISOrlongeccen{24}\,\rsun$). This planet has a mass of $\hatcurPPmlongeccen{24}\,\mjup$ and an inflated  radius of $\hatcurPPrlongeccen{24}\,\rjup$.
\setcounter{footnote}{0}
\end{abstract}

\keywords{
    planetary systems ---
    stars: individual (
\setcounter{planetcounter}{1}
\hatcur{22},
\hatcurCCgsc{22}\loopcommanoperiod
\setcounter{planetcounter}{2}
\hatcur{23},
\hatcurCCgsc{23}\loopcommanoperiod
\setcounter{planetcounter}{3}
\hatcur{24},
\hatcurCCgsc{24}\loopcommanoperiod
\setcounter{planetcounter}{4}
) 
    techniques: spectroscopic, photometric
}


\section{Introduction}
\label{sec:introduction}

Transiting planets are the key towards understanding the structure and composition of planetary systems. The breadth of system parameters that can be determined from the discovery data sets and follow-up studies surpasses any other detection method, the most important being the mass and radius, yielding an estimate of the bulk density. Moreover, these planets are amenable to transmission studies during transit \citep[e.g.][]{bento:2014,jordan:2013,marley:2013,seager:2000, pont:2008,sing:2011}, a direct measurement of the planet's day-side emission as an estimate of the surface temperature during secondary eclipse \citep{zhou:2013,zhou:2014:sec,knutson:2008,desert:2011,croll:2011}, and other properties \citep{louden:2015,zhou:2016,collier:2010, hartman:2015, gandolfi:2012}. 

In particular, the hundreds of \hjs (broadly Jupiter mass planets orbiting close to their host stars on less that $\sim 10$ day periods) found to date have challenged planetary formation theories and structure models. Despite an early suggestion of the possibility that such planets may exist by \citet{struve:1952}, explaining their existence is not trivial as they are not generally expected to form in-situ \citep{boss:1995, lissauer:1995,bodenheimer:2000}, with the migration potentially taking place in the very early early stages of formation \citep{donati:2016}. Recent work suggests that there is a potential mechanism that can form such planets in-situ \citep{batygin:2016}, but the general consensus is that these planets are formed at large separations and migrate inwards to their current positions, and several possible mechanisms have been suggested for this process. A disk migration scenario has been proposed in which the orbiting planet exchanges angular momentum with the protoplanetary disk and loses orbital momentum, thereby starting out at large separations and making its way in to close to the host star \citep[e.g.][and references therein]{alibert:2005,chambers:2009,rice:2008}. Alternatively, interactions with other bodies in the system can cause scattering/ejection events and the planet in question is forced into an eccentric orbit that brings it closer to the host star \citep[e.g.][]{rasio:1996,ford:2008}. Tidal interactions are then thought to circularise the orbit resulting in close-in planets. Other processes suggested include Kozai migration, first proposed by \cite{wu:2003}, which states that a highly inclined stellar companion can induce Kozai oscillations \citep{kozai:1962} in the planet and excite it to progressively higher eccentricity. We note, however, that \citet{ngo:2016} suggest that only a small fraction ($<$20\%) of \hj host stars have stellar companions capable of inducing such oscillations. Very recent works by \cite{petrovich:2015} and \cite{wu:2011} suggest that {\em secular migrations} may occur due to interactions between two-or-more well-spaced, eccentric planets, which can cause one of them to become very eccentric on long timescales, leading to both enhanced eccentricity and tidal dissipation over larger timescales \citep{lithwick:2011}. However, recent results show that an understanding of planet formation and migration has not been achieved yet. \citet{antonini:2016} suggest that perhaps \hjs with outer companions are unlikely to have migrated through high-eccentricity processes due to the instability of their orbits, while \citet{schlaufman:2016} find that warm Jupiters are no more likely to have wide orbit planetary companions than those in longer orbits, which is at odds with an eccentric migration scenario. 

The possibility that there is a mass dependence in the question of planet migration and eccentricity is supported by evidence that higher mass planets tend to show higher eccentricity than those less massive than 2$\mjup$ \citep{mazeh:1997,marcy:2005,southworth:2009}. Moreover, it seems that planets at higher orbital separation/period also have a higher tendency to show non-zero eccentricities \citep{pont:2011} versus close-in planets. This raises questions such as: are high-mass planets more susceptible to retain large eccentricities on longer timescales? And, if so, is this an indication that planet-planet scattering, predicted to generate high eccentric orbits, is likely to be the main migration mechanism for planetary systems? Is the structure and evolution of high and low-mass planets fundamentally different? Are \hj structures fundamentally affected by extreme cases of inwards migration and current irradiation levels? The answer lies on a larger sample and better understanding of the composition of these planets. 

In this paper we report the discovery of three new transiting super Jupiters with masses higher than 1.4\mjup\, from the HATSouth survey: \hatcurb{22}, \hatcurb{23} and \hatcurb{24}. These planets add to the list of of well-characterised massive \hjs which collectively pose a challenge to models of planetary formation and migration.

In Section \ref{sec:obs} we describe the photometric and spectroscopic observations undertaken for all three targets in the pursuit of determining their planetary nature. Section \ref{sec:analysis} contains a description of the global data analysis and presents the modelled stellar and planetary parameters. We also describe the methods employed to reject false positive scenarios. Our findings are finally discussed in Section \ref{sec:discussion}.

\section{Observations}
\label{sec:obs}

Periodic planetary transit-like signals in any time-series photometric survey can be created by a range of astrophysical events which include grazing binary stellar eclipses, transits by planet-sized dwarf stars and eclipsing binary systems whose light is blended with a nearby foreground or background star. As such, a substantial follow-up campaign is required using both photometric and spectroscopic observations. In this section we describe the full set of observations that led to the detection and confirmation of the planets presented in this paper. 

\subsection{Photometric detection}
\label{sec:detection}

The HATSouth project is a collaboration between Princeton, the Australian National University, The Max Plank Institute for Astronomy, and the Pontificia Universidad
Cat\'olica de Chile, dedicated to finding transiting planets hosted by bright stars in the southern hemisphere \citep{bakos:2013:hatsouth}. It is the largest ground-based search for transiting extrasolar planets in the world, with a three-site network (Las Campanas Observatory in Chile, the High Energy Spectroscopic Survey (H.E.S.S.) site in Namibia and Siding Spring Observatory, Australia) capable of continuously monitoring 128 sq degree fields in the southern hemisphere. The project has commissioned six enclosures, two per site, each containing four telescopes on a single mount. Discoveries include the notable case of HATS-17b \citep{brahm:2016}, which is the largest period transiting exoplanet found to date from a ground-based survey, thereby demonstrating HATSouth's unique strength in its longitude coverage. A full list of discovered planets along with discovery \lcs\, can be found at http://hatsouth.org/

Table \ref{tab:photobs} shows a summary of the photometric observations for the planetary systems \hatcur{22}, \hatcur{23} and \hatcur{24}. For HATSouth data we list the HATSouth unit, CCD and field name from which the observations were taken. The detection of all targets relied on data from all HATSouth telescopes, HS-1 and HS-2 located in Chile, HS-3 and HS-4 in Namibia and HS-5 and HS-6 in Australia. The data gathered at different time periods between 2011/04 and 2013/11 for different targets, as described in Table \ref{tab:photobs}, resulted in a total of 13,129 data points for \hatcur{22}, 22,937 observations of \hatcur{23} and 4,406 points for \hatcur{24}. 

All HATSouth observations are obtained through a Sloan \emph{r} filter with a typical exposure time of 240 seconds. The data were reduced with the custom pipeline described by \cite{penev:2013:hats1} and the \lcs{} were detrended using an External Parameter Decorrelation method \citep{bakos:2013:hatsouth} followed by the application of the Trend Filtering Algorithm \citep[TFA,][]{kovacs:2005:TFA}. A Box Least-Squares algorithm \citep[BLS; see][]{kovacs:2002:BLS} was then used to search for periodic transit-like signals. The resulting discovery \lcs\ phase-folded to the highest likelihood periods are shown in Figure \ref{fig:hatsouth}. 

\begin{figure*}[t]
{
\centering

\includegraphics[width=\linewidth]{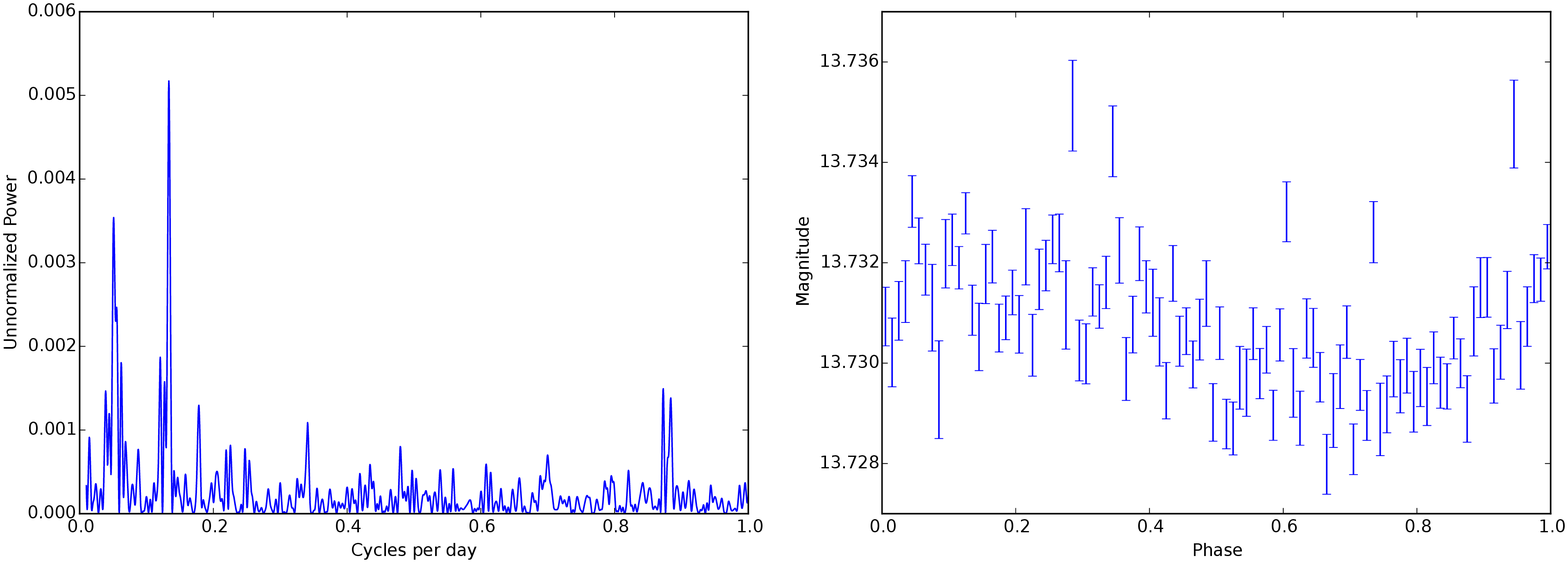}
}
\caption{
    Additional period detected on the discovery \lc\, of \hatcurb{23}. \emph{Left:} We show the Lomb-Scargle periodogram produced using the method of \citet{zechmeister:2009} after removal of the transit signal. \emph{Right:} The phase folded lightcurve on the 7.4965 day period equivalent to the highest peak binned in 0.01 phase intervals. The error bars shown are calculated through post binning error propagation of the original magnitude errors of individual measurements.
}
\label{fig:otherphase}
\end{figure*}

After having removed the best fit Box Least Squares model corresponding to the hot-Jupiter transit signal from the \lcs , we searched for additional periodic signals with the aim to detect potential stellar activity or other transiting planets in each of the three systems. The \lcs\, for all targets did not reveal any other significant signals, defined as those with a formal false alarm probability, assuming Gaussian white noise, of less than 0.1\%, on a second BLS pass of the residuals. We therefore find no evidence for additional transiting planets in the systems. However, a Generalised Lomb Scargle \citep[GLS,][]{zechmeister:2009} routine, looking for sinusoidal patterns that can be related to activity, revealed a significant peak at a period of $7.49 \pm 0.25$ days for \hatcur{23} with a false alarm probability of $10^{-21}$. This is shown in Figure \ref{fig:otherphase} (left-hand panel). Further inspection of this signal reveals a sinusoidal signal (right-hand panel) that can be attributed to activity such as stellar spots on the surface of the host star modulating the \lc. We note that the detected 7.49 day period is not consistent with the stellar rotation period estimated from \vsini\, ($13.1 \pm 1.4$ days) assuming an aligned stellar rotation axis. If the detected sinusoidal signal is indeed real and related to the stellar rotation, this may suggest a moderately high misalignment between the orbital plane and the stellar rotational axis. On the other hand, we can not definitively exclude the possibility that the true rotation period is double this value, which would be consistent with the observed \vsini. However, when the \lc\, is folded at twice the 7.49 day period it reveals a double oscillation and the power of the window function at this period is substantially lower. Nevertheless, more observations of this system are required to address this dichotomy. 

%
%
\ifthenelse{\boolean{emulateapj}}{
    \begin{figure*}[!ht]
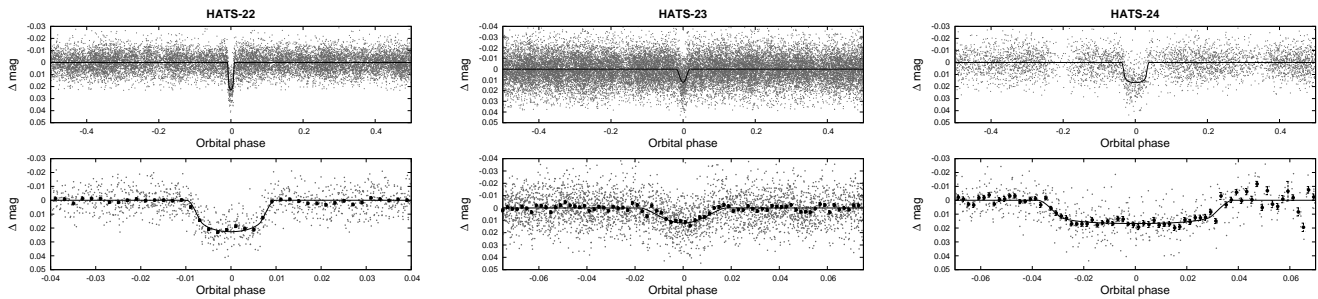

}{
    \begin{figure}[!ht]
}
{
\centering
\setlength{\plotwidthtwo}{0.31\linewidth}
\includegraphics[width={\plotwidthtwo}]{\hatcurhtr{22}-hs-eccen.eps}
\hfil
\includegraphics[width={\plotwidthtwo}]{\hatcurhtr{23}-hs.eps}
\hfil
\includegraphics[width={\plotwidthtwo}]{\hatcurhtr{24}-hs.eps}
}
\caption[]{
    Phase-folded unbinned HATSouth light curves for \hatcur{22} (left), \hatcur{23} (middle) and \hatcur{24} (right). In each case we show two panels. The
    top panel shows the full light curve, while the bottom panel shows
    the light curve zoomed-in on the transit. The solid lines show the
    model fits to the light curves. The dark filled circles in the
    bottom panels show the light curves binned in phase with a bin
    size of 0.002.
\label{fig:hatsouth}}
\ifthenelse{\boolean{emulateapj}}{
    \end{figure*}
}{
    \end{figure}
}

\ifthenelse{\boolean{emulateapj}}{
    \begin{deluxetable*}{llrrrr}
}{
    \begin{deluxetable}{llrrrr}
}
\tablewidth{0pc}
\tabletypesize{\scriptsize}
\tablecaption{
    Summary of photometric observations
    \label{tab:photobs}
}
\tablehead{
    \multicolumn{1}{c}{Instrument/Field\tablenotemark{a}} &
    \multicolumn{1}{c}{Date(s)} &
    \multicolumn{1}{c}{\# Images} &
    \multicolumn{1}{c}{Cadence\tablenotemark{b}} &
    \multicolumn{1}{c}{Filter} &
    \multicolumn{1}{c}{Precision\tablenotemark{c}} \\
    \multicolumn{1}{c}{} &
    \multicolumn{1}{c}{} &
    \multicolumn{1}{c}{} &
    \multicolumn{1}{c}{(sec)} &
    \multicolumn{1}{c}{} &
    \multicolumn{1}{c}{(mmag)}
}
\startdata
\sidehead{\textbf{\hatcur{22}}}
~~~~HS-2.2/G610 & 2011 Apr--2013 Jul & 5368 & 280 & $r$ & 7.9 \\
~~~~HS-4.2/G610 & 2013 Jan--2013 Jul & 3755 & 289 & $r$ & 7.2 \\
~~~~HS-6.2/G610 & 2011 Apr--2013 Jul & 4006 & 282 & $r$ & 7.6 \\
~~~~LCOGT~1\,m+CTIO/sinistro & 2015 Mar 30 & 85 & 226 & $i$ & 1.0 \\
\sidehead{\textbf{\hatcur{23}}}
~~~~HS-1.2/G747 & 2013 Mar--2013 Oct & 4233 & 287 & $r$ & 12.6 \\
~~~~HS-2.2/G747 & 2013 Sep--2013 Oct & 648 & 287 & $r$ & 12.3 \\
~~~~HS-3.2/G747 & 2013 Apr--2013 Nov & 9020 & 297 & $r$ & 12.1 \\
~~~~HS-4.2/G747 & 2013 Sep--2013 Nov & 1460 & 297 & $r$ & 13.6 \\
~~~~HS-5.2/G747 & 2013 Mar--2013 Nov & 6013 & 297 & $r$ & 11.9 \\
~~~~HS-6.2/G747 & 2013 Sep--2013 Nov & 1563 & 290 & $r$ & 14.9 \\
~~~~LCOGT~1\,m+SSO/SBIG & 2015 Jul 07 & 22 & 194 & $i$ & 2.1 \\
~~~~Swope~1\,m/e2v \tablenotemark{d} & 2015 Jul 15 & 51 & 139 & $i$ & 13.7 \\
~~~~LCOGT~1\,m+SSO/SBIG & 2015 Aug 30 & 34 & 192 & $i$ & 3.1 \\
~~~~LCOGT~1\,m+CTIO/sinistro & 2015 Sep 05 & 47 & 223 & $i$ & 1.2 \\
~~~~LCOGT~1\,m+SAAO/SBIG & 2015 Sep 16 & 39 & 201 & $z$ & 4.8 \\
\sidehead{\textbf{\hatcur{24}}}
~~~~HS-1.1/G777 & 2011 May--2012 Sep & 1513 & 298 & $r$ & 9.1 \\
~~~~HS-3.1/G777 & 2011 Jul--2012 Sep & 1688 & 297 & $r$ & 9.4 \\
~~~~HS-5.1/G777 & 2011 May--2012 Sep & 1205 & 296 & $r$ & 9.3 \\
~~~~LCOGT~1\,m+SAAO/SBIG & 2015 Jun 07 & 90 & 151 & $i$ & 1.6 \\
\enddata
\tablenotetext{a}{
    For HATSouth data we list the HATSouth unit, CCD and field name
    from which the observations are taken. HS-1 and -2 are located at
    Las Campanas Observatory in Chile, HS-3 and -4 are located at the
    H.E.S.S. site in Namibia, and HS-5 and -6 are located at Siding
    Spring Observatory in Australia. Each unit has 4 CCDs. Each field
    corresponds to one of 838 fixed pointings used to cover the full
    4$\pi$ celestial sphere. All data from a given HATSouth field and
    CCD number are reduced together, while detrending through External
    Parameter Decorrelation (EPD) is done independently for each
    unique unit+CCD+field combination.
}
\tablenotetext{b}{
    The median time between consecutive images rounded to the nearest
    second. Due to factors such as weather, the day--night cycle,
    guiding and focus corrections the cadence is only approximately
    uniform over short timescales.
}
\tablenotetext{c}{
    The RMS of the residuals from the best-fit model.
}
\tablenotetext{d}{
    The Swope~1\,m observations of \hatcur{23} produced very poor
    quality photometry due to adverse weather conditions, so we excluded them from the analysis of this
    system.  
} \ifthenelse{\boolean{emulateapj}}{
    \end{deluxetable*}
}{
    \end{deluxetable}
}


\subsection{Spectroscopic Observations}
\label{sec:obsspec}

\subsubsection{Reconnaissance spectroscopic observations}
\label{sec:recspec}

The initial follow-up phase for all HATSouth planet candidates is carried out with reconnaissance spectra taken with the WiFeS instrument on the 2.3m ANU telescope at Siding Spring Observatory (SSO) \citep{Dopita:2007}. Observations at $R \equiv \Delta \lambda / \lambda \approx 3000$ were taken to determine the stellar type of the host star, using the blue arm of the spectrograph. We estimate three key stellar properties, the effective temperature $\teff$, $\loggstar$ and $\feh$, by performing a grid search minimizing the $\chi ^2$ between the observed normalised spectrum and synthetic templates from the MARCS model atmospheres \citep{gustafsson:2008}. 2MASS J-K colors are used to restrict the $\teff$ parameter space and extinction correction is applied using the method of \citet{cardelli:1989}. A detailed description of the observing and data reduction procedure is described in \citet{bayliss:2013:hats3}. This type of observation is performed to identify giant host stars, for which the observed dip in its \lc\, could only have been caused by a stellar companion, and to identify stars not suitable for precise radial velocity follow-up due to high \teff\, or high \vsini . In addition, observations are taken at predicted quadrature phase with WiFeS using a mid-resolution $R \sim 7,000$ grating to perform radial velocity measurements at a precision of $\sim 2 \kms$. We use a cross-correlation method against RV standards observed every night, using bracketed Ne-Ar exposures and a selection of telluric lines for calibration. This is, however, dependent on stellar type and signal-to-noise of each individual target. This allows for the detection of radial velocity variations above $\sim 5 \kms$, and the exclusion of any targets showing large variations indicating that the transiting companion is a star. Details of these observations can be found in Table \ref{tab:specobs} and are described here:

\begin{itemize}

\item For \hatcur{22} we found an effective temperature of $4600 \pm 300 K$, $\loggstar$ of $4.8 \pm 0.3$ dex and metallicity of $\feh = -0.5 \pm 0.5$ dex, leading to the conclusion that this is a K-dwarf host star. Two measurements showed no clear variation at quadrature.

\item For \hatcur{23} we found an effective temperature estimate of $5900 \pm 300 K$, $\loggstar$ of $4.5 \pm 0.3$ dex and metallicity of $\feh = 0.0 \pm 0.5$ dex. We conclude that the host star is of F or G-type. A single radial velocity measurement with WiFeS was taken but later complemented by observations with the FEROS spectrograph (see Section \ref{sec:highspec}). 

\item \hatcur{24} was found to have an effective temperature of $5800 \pm 300 K$, $\loggstar$ of $3.4 \pm 0.3$ dex and metallicity of $\feh = -0.5 \pm 0.5$ dex. Based on this we concluded that the target is a G or F star, but the surface gravity suggested that this is a sub-giant. Three radial velocity measurements taken showed no significant variation in the covered orbital phase.

\end{itemize}

Having excluded clear eclipsing binaries and giant host stars, these targets were then promoted to the next steps in the follow-up campaign, leading to further higher radial velocity precision spectroscopy and photometric follow-up. 

\ifthenelse{\boolean{emulateapj}}{
    \begin{deluxetable*}{llrrrrr}
}{
    \begin{deluxetable}{llrrrrrrrr}
}
\tablewidth{0pc}
\tabletypesize{\scriptsize}
\tablecaption{
    Summary of spectroscopy observations
    \label{tab:specobs}
}
\tablehead{
    \multicolumn{1}{c}{Instrument}          &
    \multicolumn{1}{c}{UT Date(s)}             &
    \multicolumn{1}{c}{\# Spec.}   &
    \multicolumn{1}{c}{Res.}          &
    \multicolumn{1}{c}{S/N Range\tablenotemark{a}}           &
    \multicolumn{1}{c}{$\gamma_{\rm RV}$\tablenotemark{b}} &
    \multicolumn{1}{c}{RV Precision\tablenotemark{c}} \\
    &
    &
    &
    \multicolumn{1}{c}{$\Delta \lambda$/$\lambda$/1000} &
    &
    \multicolumn{1}{c}{(\kms)}              &
    \multicolumn{1}{c}{(\ms)}
}
\startdata
%
%
\sidehead{\textbf{\hatcur{22}}}\\
ESO~3.6\,m/HARPS & 2015 Feb--Apr & 4 & 115 & 8--18 & -7.370 & 15 \\
Euler~1.2\,m/CORALIE & 2015 Feb--Jun & 7 & 60 & 10--14 & -7.414 & 35 \\
ANU~2.3\,m/WiFeS & 2015 Feb 28 & 1 & 3 & 44 & $\cdots$ & $\cdots$ \\
ANU~2.3\,m/WiFeS & 2015 Feb--Mar & 2 & 7 & 63--83 & -7.7 & 4000 \\
MPG~2.2\,m/FEROS & 2015 Apr--Jun & 4 & 48 & 43--58 & -7.438 & 25 \\
\sidehead{\textbf{\hatcur{23}}}\\
ANU~2.3\,m/WiFeS & 2015 Jun 1 & 1 & 3 & 43 & $\cdots$ & $\cdots$ \\
ANU~2.3\,m/WiFeS & 2015 Jun 1 & 1 & 7 & 39 & -13.8 & 4000 \\
MPG~2.2\,m/FEROS & 2015 Jun 8--18 & 8 & 48 & 18--39 & -13.372 & 16 \\
\sidehead{\textbf{\hatcur{24}}}\\
ANU~2.3\,m/WiFeS & 2015 Feb 1 & 1 & 3 & 29 & $\cdots$ & $\cdots$ \\
ANU~2.3\,m/WiFeS & 2015 Feb 1--5 & 3 & 7 & 31--55 & -7.1 & 4000 \\
ESO~3.6\,m/HARPS & 2015 Apr 6--7 & 2 & 115 & 9--15 & -3.370 & 120 \\
AAT~3.9\,m/CYCLOPS2+UCLES \tablenotemark{d} & 2015 May 6--13 & 11 & 70 & 10--27 & -3.284 & 160 \\
Euler~1.2\,m/CORALIE & 2015 Jun 6--8 & 3 & 60 & 14--17 & -3.236 & 140 \\
MPG~2.2\,m/FEROS & 2015 Jun 17--21 & 4 & 48 & 42--60 & -3.259 & 43 \\
\enddata 
\tablenotetext{a}{
    S/N per resolution element near 5180\,\AA.
}
\tablenotetext{b}{
    For high-precision radial velocity observations included in the orbit determination this is the zero-point radial velocity from the best-fit orbit. For other instruments it is the mean value. We do not provide this quantity for the lower resolution WiFeS observations which were only used to measure stellar atmospheric parameters.
}
\tablenotetext{c}{
    For high-precision radial velocity observations included in the orbit
    determination this is the scatter in the radial velocity residuals from the
    best-fit orbit (which may include astrophysical jitter), for other
    instruments this is either an estimate of the precision (not
    including jitter), or the measured standard deviation. We do not
    provide this quantity for low-resolution observations from the
    ANU~2.3\,m/WiFeS.
}
\tablenotetext{d}{
    We excluded from the analysis two of the AAT~3.9\,m/CYCLOPS2+UCLES
    observations of \hatcur{24} which were taken during transit.
}
\ifthenelse{\boolean{emulateapj}}{
    \end{deluxetable*}
}{
    \end{deluxetable}
}

%
\setcounter{planetcounter}{1}
%
\ifthenelse{\boolean{emulateapj}}{
    \begin{figure*} [ht]
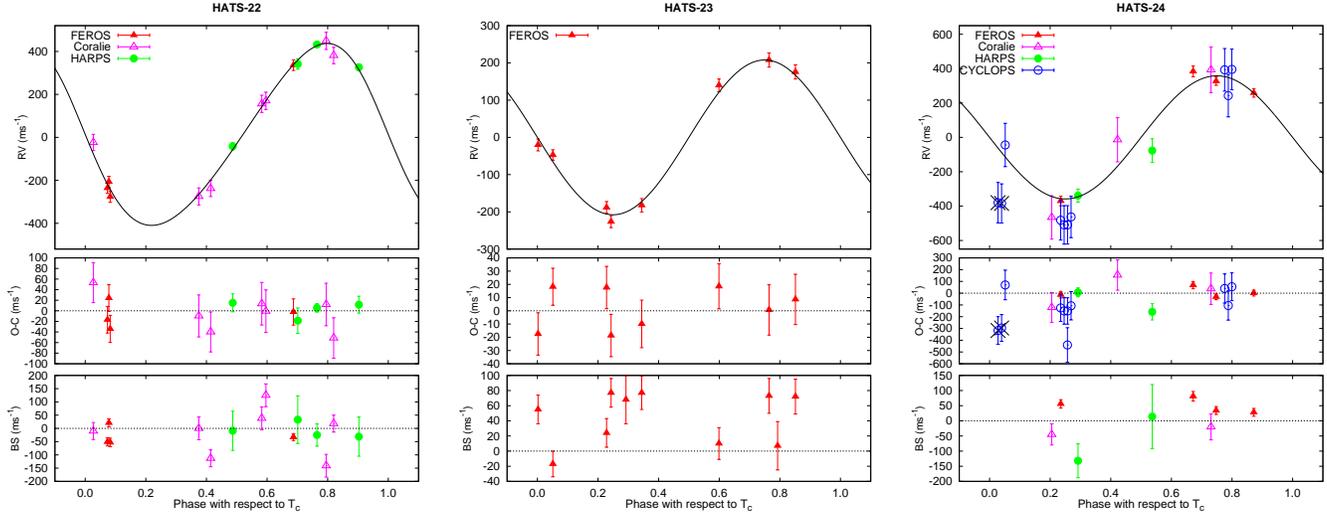

}{
    \begin{figure}[ht]
}
{
\centering
\setlength{\plotwidthtwo}{0.31\linewidth}
\includegraphics[width={\plotwidthtwo}]{\hatcurhtr{22}-rv-eccen.eps}
\hfil
\includegraphics[width={\plotwidthtwo}]{\hatcurhtr{23}-rv.eps}
\hfil
\includegraphics[width={\plotwidthtwo}]{\hatcurhtr{24}-rv.eps}
}
\caption{
    Phased high-precision radial velocity measurements for \hbox{\hatcur{22}{}} (left), \hbox{\hatcur{23}{}} (middle), and \hbox{\hatcur{24}{}} (right). The instruments used are labelled in the plots. For \hatcur{24} two observations marked with an X were obtained (partially) in transit and have been excluded from the analysis. In each case we display three panels. The top panel shows the phased measurements together with our best-fit model (see \reftabl{planetparam}) for each system where we show the RV jitter values for each case. Zero-phase corresponds to the time of mid-transit. The center-of-mass velocity has been subtracted. The second panel displays the velocity $O\!-\!C$ residuals from the best fit. The error bars include the jitter terms listed in \reftabl{planetparam} added in quadrature to the formal errors for each instrument. The third panel shows the bisector spans (BS). Note the different vertical scales of the panels. 
}
\label{fig:rvbis}
\ifthenelse{\boolean{emulateapj}}{
    \end{figure*}
}{
    \end{figure}
}
\subsubsection{High-precision spectroscopic observations}
\label{sec:highspec}

A full radial velocity characterisation covering a wide portion of the orbital phase of all of our targets is required in order to determine fundamental parameters such as mass and eccentricity of the orbits. Observations with the High Accuracy Radial Velocity Planet Searcher \cite[HARPS]{mayor:2003}, fed by the ESO 3.6m telescope at $R \sim 115,000$ were obtained for \hatcur{22} and \hatcur{24}, as well as $R \sim 60,000$ spectra using the CORALIE spectrograph \citep{queloz:2001} fed by the 1.2m Euler telescope, both located at La Silla Observatory (LSO), Chile. All three targets were also monitored for radial velocity measurements using the FEROS spectrograph \citep[][ $R \sim 48,000$]{kaufer:1998} fed by the MPG 2.2m telescope at LSO. The data reduction for all these spectra was performed using the method described in \citet{jordan:2014:hats4}, with modifications to accommodate the different formats of the FEROS and HARPS data. Additionally, eleven spectra of \hatcur{24} were also obtained at $R \sim 70,000$ with the CYCLOPS2 fibre-feed and the UCLES spectrograph on the 3.9m Anglo-Australian telescope (AAT) at SSO and the data were reduced using the methods described in \cite{addison:2013}. Further details about these observations can be found in Table \ref{tab:specobs}. The resulting data sets for all targets can be found in Table \ref{tab:rvs} at the end of the paper, and are shown in Figure \ref{fig:rvbis}, which includes radial velocity curves, best-fit models and bisector span (BS) estimates shown in the bottom panels for each target. All systems clearly show a radial velocity variation consistent with the detected orbital period from the photometric \lcs\, and no clear correlation between the radial velocity measurements and the bisector-spans, indicating the systems are likely bona fide transiting planets (see Section \ref{sec:blend}). 


\subsection{Photometric follow-up observations}
\label{sec:phot}

\begin{figure*}[!ht]
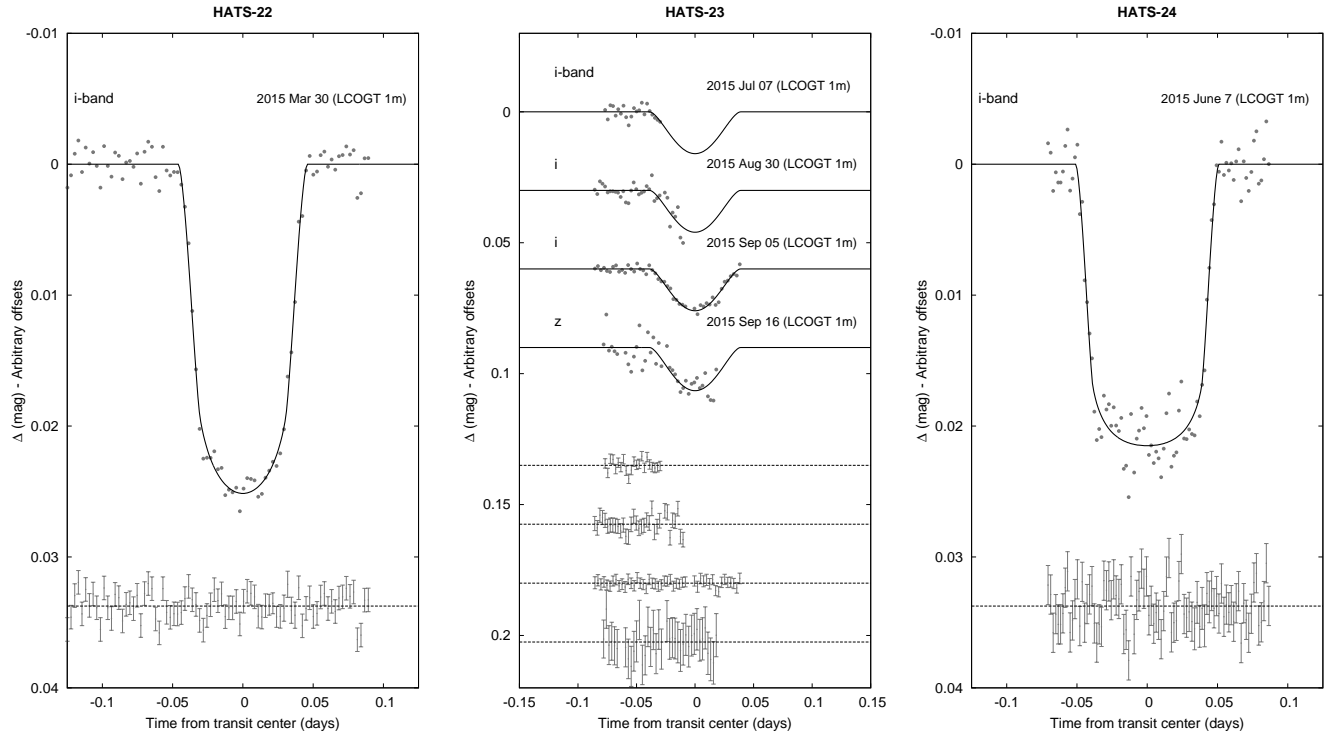

{
\centering
\setlength{\plotwidthtwo}{0.31\linewidth}
\includegraphics[width={\plotwidthtwo}]{\hatcurhtr{22}-lc-eccen.eps}
\hfil
\includegraphics[width={\plotwidthtwo}]{\hatcurhtr{23}-lc.eps}
\hfil
\includegraphics[width={\plotwidthtwo}]{\hatcurhtr{24}-lc.eps}
}
\caption{
    Unbinned transit \lcs{} for \hatcur{22} (left), \hatcur{23}
    (middle) and \hatcur{24} (right).  The light curves have been
    corrected for quadratic trends in time, and linear trends with up
    to three parameters characterizing the shape of the PSF, fitted
    simultaneously with the transit model.
    The dates of the events, filters and instruments used are
    indicated.  Light curves following the first are displaced
    vertically for clarity.  Our best fit from the global modeling
    described in \refsecl{globmod} is shown by the solid lines. The
    residuals from the best-fit model are shown below in the same
    order as the original light curves.  The error bars represent the
    photon and background shot noise, plus the readout noise.
}
\label{fig:lc}
\end{figure*}

The three candidates were all photometrically followed-up employing the Las Cumbres Observatory Global Network of telescopes \citep{brown:2013:lcogt}, specifically using the 1m sized telescopes of this network. These observations are undertaken to confirm the transit signal as well as refine the derived transit parameters from the HATSouth photometry. A single full transit of \hatcur{22} was observed in March 2015 using the \emph{i}-band filter in which 85 images at a 226 second cadence were obtained. A single transit of \hatcur{24} in June of the same year was also obtained, consisting of 90 images with 151 second cadence. Due to the grazing nature of \hatcur{23} a larger number of photometric follow-up observations were required. Two full and two partial transits of \hatcur{23} were observed between July and September 2015 (inclusive), the first three in the \emph{i}-band and the last using the \emph{z}-band filter. The \lcs\, for these high-precision photometric observations are shown in Figure \ref{fig:lc} along with the best-fit models. The photometric data were taken and reduced using the same strategy and methods described in \cite{penev:2013:hats1}, with details of setup in \citet{bayliss:2015}, using a customisable pipeline. This pipeline uses standard photometric reduction frames (master bias, darks, twilight flats) and the DAOPHOT aperture photometry package for flux extraction of target and comparison stars. A quadratic trend in time as well as variations correlated with PSF shape were fitted simultaneously with the transit shape to compensate for differential refraction effects due to airmass and poor seeing. The \emph{``V"-shaped} transit signal for \hatcur{23} is clearly indicative of the grazing nature of the planetary system and the consistent depth of the transits in both observed bands for this target also suggests that this is not an eclipsing binary system or a hierarchical triple system. The data from all photometric follow-up are available in electronic format in Table \ref{tab:phfu} and all photometric follow-up observations are also summarized in \reftabl{photobs}.

\setcounter{planetcounter}{1}
%



\subsection{Lucky imaging observations}
\label{sec:luckyimaging}

\begin{figure}
{
\centering

\includegraphics[width={\linewidth}]{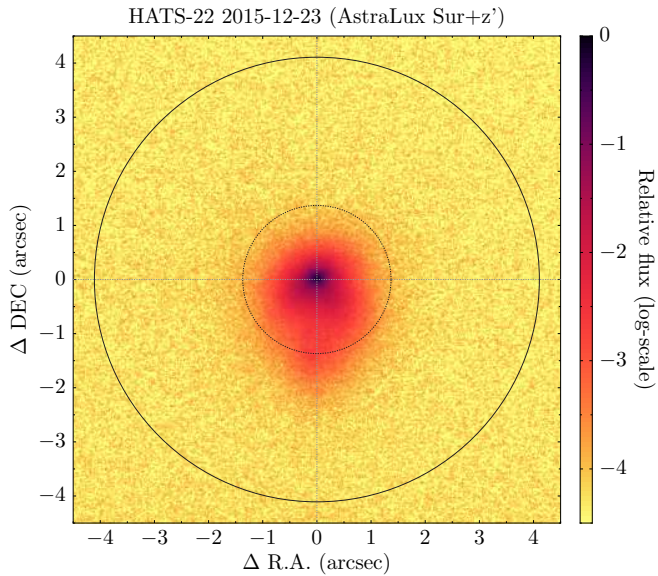}
}
\caption{
    \emph{Left} Lucky imaging observations of \hatcur{22} with the Astralux camera on the NTT telescope at La Silla in the $z'$ band. We show the 1" and 4" radii lines for reference as well as the fitted centre of the star from the PSF modelling process. We note that the asymmetrical shape of the PSF is due to an instrumental effect related to non-stable focus on the telescope through the night. Similar patterns can be seen in other observations of this kind at similar times during the night. 
}
\label{fig:lucky}
\end{figure}

High spatial resolution ``lucky'' imaging observations were made of \hatcur{22} using the Astralux camera \citep{hippler:2009} on the New Technology Telescope (NTT) at LSO on 2015/12/23. These observations are part of a campaign to detect potential companions for exoplanet host star candidates and place upper limits on magnitude contrasts. The data were taken using the SDSS z' filter, resulting in a set of $10^4$ images with an exposure time of 100 ms each. We used the Drizzle algorithm from \cite{fruchter:2002} to combine a set of the best 10\% of images acquired and the result of these can be found in Figure \ref{fig:lucky}, where we show the 1 and 4 arcsec radii lines for reference. A slightly asymmetric extended profile is visible on this image likely due to instrumental effects, confirmed by taking images of other targets on different nights that show a similar feature. As such, while we can confirm that there is no clear bright star in the vicinity of our target, we can not completely exclude the possibility that a faint close companion within 1" is not present. This is a generic problem with most confirmed transiting planets, and as such we further address this issue in Section \ref{sec:blend} where we perform a blend analysis that increases our confidence that this is indeed a planetary body companion.

%
%
\ifthenelse{\boolean{emulateapj}}{
    \begin{deluxetable*}{llrrrrl}
}{
    \begin{deluxetable}{llrrrrl}
}
\tablewidth{0pc}
\tablecaption{
    Light curve data for \hatcur{22}, \hatcur{23} and \hatcur{24}\label{tab:phfu}.
}
\tablehead{
    \colhead{Object\tablenotemark{a}} &
    \colhead{BJD\tablenotemark{b}} & 
    \colhead{Mag\tablenotemark{c}} & 
    \colhead{\ensuremath{\sigma_{\rm Mag}}} &
    \colhead{Mag(orig)\tablenotemark{d}} & 
    \colhead{Filter} &
    \colhead{Instrument} \\
    \colhead{} &
    \colhead{\hbox{~~~~(2,400,000$+$)~~~~}} & 
    \colhead{} & 
    \colhead{} &
    \colhead{} & 
    \colhead{} &
    \colhead{}
}
\startdata
   HATS-22 & $ 56443.36356 $ & $  -0.00627 $ & $   0.00463 $ & $ \cdots $ & $ r$ &         HS\\
   HATS-22 & $ 56363.07603 $ & $   0.00598 $ & $   0.00473 $ & $ \cdots $ & $ r$ &         HS\\
   HATS-22 & $ 56466.97802 $ & $  -0.00407 $ & $   0.00561 $ & $ \cdots $ & $ r$ &         HS\\
   HATS-22 & $ 56396.13657 $ & $   0.00730 $ & $   0.00502 $ & $ \cdots $ & $ r$ &         HS\\
   HATS-22 & $ 56438.64228 $ & $  -0.00929 $ & $   0.00545 $ & $ \cdots $ & $ r$ &         HS\\
   HATS-22 & $ 56424.47483 $ & $  -0.00012 $ & $   0.00467 $ & $ \cdots $ & $ r$ &         HS\\
   HATS-22 & $ 56315.85060 $ & $  -0.01903 $ & $   0.00429 $ & $ \cdots $ & $ r$ &         HS\\
   HATS-22 & $ 56443.36694 $ & $   0.01129 $ & $   0.00468 $ & $ \cdots $ & $ r$ &         HS\\
   HATS-22 & $ 56424.47573 $ & $   0.01945 $ & $   0.00459 $ & $ \cdots $ & $ r$ &         HS\\
   HATS-22 & $ 56386.69363 $ & $   0.01108 $ & $   0.00455 $ & $ \cdots $ & $ r$ &         HS\\

\enddata
\tablenotetext{a}{
    Either \hatcur{22}, \hatcur{23} or \hatcur{24}.
}
\tablenotetext{b}{
    Barycentric Julian Date is computed directly from the UTC time
    without correction for leap seconds.
}
\tablenotetext{c}{
    The out-of-transit level has been subtracted. For observations
    made with the HATSouth instruments (identified by ``HS'' in the
    ``Instrument'' column) these magnitudes have been corrected for
    trends using the EPD and TFA procedures applied {\em prior} to
    fitting the transit model. This procedure may lead to an
    artificial dilution in the transit depths. The blend factors for
    the HATSouth light curves are listed in
    Table~\ref{tab:planetparam}. For
    observations made with follow-up instruments (anything other than
    ``HS'' in the ``Instrument'' column), the magnitudes have been
    corrected for a quadratic trend in time, and for variations
    correlated with up to three PSF shape parameters, fit simultaneously
    with the transit.
}
\tablenotetext{d}{
    Raw magnitude values without correction for the quadratic trend in
    time, or for trends correlated with the seeing. These are only
    reported for the follow-up observations.
}
\tablecomments{
    This table is available in a machine-readable form in the online
    journal.  A portion is shown here for guidance regarding its form
    and content.
}
\ifthenelse{\boolean{emulateapj}}{
    \end{deluxetable*}
}{
    \end{deluxetable}
}

\section{Analysis}
\label{sec:analysis}

\subsection{Properties of the parent star}
\label{sec:stelparam}

We used the Zonal Atmospheric Stellar Parameter Estimator \citep[ZASPE;][]{brahm:2016:ZASPE} to model the stellar parameters of the host stars for all targets. ZASPE is capable of precise stellar atmospheric parameter estimation from high-resolution echelle spectra from FGK-type stars. It compares the observed spectrum with a grid of synthetic spectra by a least squares minimisation of the normalised continuum in only the most sensitive regions of the stellar spectrum. The complete FGK-type star parameter space is searched using this method. We note that we do not treat micro and macroturbulence as free parameters, but instead assume that these values are a function of atmospheric parameters and apply modifications to the synthetic spectra accordingly. To take into account the microturbulence dependence of the line widths, we computed an empirical relation between the microturbulence and the stellar parameters. In particular, we used the stellar parameters provided by the SweetCat \citep{santos:2013} catalogue to define a polynomial that delivers the microturbulence as function of $\teff$ and $\loggstar$. Then, the macroturbulence value used in the synthetization of each spectrum was obtained using that empirical function.
More details on this method can be found in \citet{brahm:2016:ZASPE}. This software used the combined spectra from the FEROS spectrograph taken for radial velocity purposes. We calculate an initial estimate of the effective temperature (\teff), the surface gravity (\logg), metallicity (\feh) and projected stellar rotational velocity of the stars (\vsini) and then use the Yonsei-Yale \citep[Y2;][]{yi:2001} isochrones to obtain the remaining physical parameters. We do not, however, search for the best isochrone using the \loggstar\, but instead use the stellar density \rhostar, which is well constrained by the photometric transit data and fitting routine. We then run the full set of parameters once again through a second iteration of ZASPE using the revised \loggstar\, to improve the results. We present the adopted results and an extensive set of host star parameters from several sources in Table \ref{tab:stellar}. We find \hatcur{22} to be a $V = \hatcurCCtassmveccen{22}$ magnitude solar metallicity K-type star with $\teff = \hatcurSMEteff{22}$ K and sub-solar mass and radius ($\mstar = \hatcurISOmlongeccen{22}$ \msun\, and $\rstar = \hatcurISOrlongeccen{22}$ \rsun\,). \hatcur{23} ($V = \hatcurCCtassmveccen{23}$) and \hatcur{24} ($V = \hatcurCCtassmveccen{24}$) are determined to have super-solar masses and radii ($\mstar = \hatcurISOmlongeccen{23}$ \msun ; $\rstar = \hatcurISOrlongeccen{23}$ \rsun\, and $\mstar = \hatcurISOmlongeccen{24}$ \msun ; $\rstar = \hatcurISOrlongeccen{24}$ \rsun\, respectively). \hatcur{23} is a G-type with a \teff\, of  \hatcurSMEteff{23} K and \feh\, of \hatcurSMEzfeh{23} whilst \hatcur{24} is a solar metallicity F-type star with determined $\teff = \hatcurSMEteff{24}$ K.

Distances to these stars were determined by comparing the measured broad-band photometry listed in Table \ref{tab:stellar} to the predicted magnitudes in each filter from the isochrones. We assumed a $R_{V} = 3.1$ extinction law from \citet{cardelli:1989} to determine the extinction and find these to be consistent within their uncertainties to reddening maps available on the NASA/IPAC infrared science archive \footnote{Publicly available at \\http://irsa.ipac.caltech.edu/applications/DUST/}. The locations of each star on an $\teffstar$--$\rhostar$ diagram
(similar to a Hertzsprung-Russell diagram) are shown in \reffigl{iso}.

\ifthenelse{\boolean{emulateapj}}{
    \begin{figure*}[!ht]
}{
    \begin{figure}[!ht]
}
{
\centering
\setlength{\plotwidthtwo}{0.31\linewidth}
\includegraphics[width={\plotwidthtwo}]{\hatcurhtr{22}-iso-rho-eccen.eps}
\hfil
\includegraphics[width={\plotwidthtwo}]{\hatcurhtr{23}-iso-rho.eps}
\hfil
\includegraphics[width={\plotwidthtwo}]{\hatcurhtr{24}-iso-rho.eps}
}
\caption{
    Model isochrones from \cite{\hatcurisocite{22}} for the measured
    metallicities of \hatcur{22} (left), \hatcur{23} (middle), and \hatcur{24} (right). We show models for ages of 0.2\,Gyr and 1.0 to 14.0\,Gyr in 1.0\,Gyr increments (ages increasing from left to right). The
    adopted values of $\teffstar$ and \rhostar\ are shown together with
    their 1$\sigma$ and 2$\sigma$ confidence ellipsoids.  The initial
    values of \teffstar\ and \rhostar\ from the first ZASPE and \lc\
    analyses are represented with a triangle.
}
\label{fig:iso}
\ifthenelse{\boolean{emulateapj}}{
    \end{figure*}
}{
    \end{figure}
}

%
%
\ifthenelse{\boolean{emulateapj}}{
    \begin{deluxetable*}{lcccl}
}{
    \begin{deluxetable}{lcccl}
}
\tablewidth{0pc}
\tabletypesize{\footnotesize}
\tablecaption{
    Stellar parameters for \hatcur{22}, \hatcur{23} and \hatcur{24}
    \label{tab:stellar}
}
\tablehead{
    \multicolumn{1}{c}{} &
    \multicolumn{1}{c}{\bf HATS-22} &
    \multicolumn{1}{c}{\bf HATS-23} &
    \multicolumn{1}{c}{\bf HATS-24} &
    \multicolumn{1}{c}{} \\
    \multicolumn{1}{c}{~~~~~~~~Parameter~~~~~~~~} &
    \multicolumn{1}{c}{Value}                     &
    \multicolumn{1}{c}{Value}                     &
    \multicolumn{1}{c}{Value}                     &
    \multicolumn{1}{c}{Source}
}
\startdata
\noalign{\vskip -3pt}
\sidehead{Astrometric properties and cross-identifications}
~~~~2MASS-ID\dotfill               & \hatcurCCtwomasseccen{22}  & \hatcurCCtwomass{23} & \hatcurCCtwomass{24} & \\
~~~~GSC-ID\dotfill                 & \hatcurCCgsceccen{22}      & \hatcurCCgsc{23}     & \hatcurCCgsc{24}     & \\
~~~~R.A. (J2000)\dotfill            & \hatcurCCraeccen{22}       & \hatcurCCra{23}    & \hatcurCCra{24}    & 2MASS\\
~~~~Dec. (J2000)\dotfill            & \hatcurCCdececcen{22}      & \hatcurCCdec{23}   & \hatcurCCdec{24}   & 2MASS\\
~~~~$\mu_{\rm R.A.}$ (\masy)              & \hatcurCCpmraeccen{22}     & \hatcurCCpmra{23} & \hatcurCCpmra{24} & UCAC4\\
~~~~$\mu_{\rm Dec.}$ (\masy)              & \hatcurCCpmdececcen{22}    & \hatcurCCpmdec{23} & \hatcurCCpmdec{24} & UCAC4\\
\sidehead{Spectroscopic properties}
~~~~$\teffstar$ (K)\dotfill         &  \hatcurSMEteff{22}   & \hatcurSMEteff{23} & \hatcurSMEteff{24} & ZASPE\tablenotemark{a}\\
~~~~$\feh$ (dex) \dotfill                  &  \hatcurSMEzfeh{22}   & \hatcurSMEzfeh{23} & \hatcurSMEzfeh{24} & ZASPE               \\
~~~~$\vsini$ (\kms)\dotfill         &  \hatcurSMEvsin{22}   & \hatcurSMEvsin{23} & \hatcurSMEvsin{24} & ZASPE                \\
~~~~$\vmac$ (\kms)\dotfill          &  $2.49$   & $4.00$ & $4.87$ & Assumed              \\
~~~~$\vmic$ (\kms)\dotfill          &  $0.58$   & $1.08$ & $1.56$ & Assumed              \\
~~~~$\gamma_{\rm RV}$ (\ms)\dotfill&  \hatcurRVgammaabs{22}  & \hatcurRVgammaabs{23} & \hatcurRVgammaabs{24} & CORALIE or FEROS\tablenotemark{b}  \\
\sidehead{Photometric properties}
~~~~$B$ (mag)\dotfill               &  \hatcurCCtassmBeccen{22}  & \hatcurCCtassmB{23} & \hatcurCCtassmB{24} & APASS\tablenotemark{c} \\
~~~~$V$ (mag)\dotfill               &  \hatcurCCtassmveccen{22}  & \hatcurCCtassmv{23} & \hatcurCCtassmv{24} & APASS\tablenotemark{c} \\
~~~~$g$ (mag)\dotfill               &  \hatcurCCtassmgeccen{22}  & \hatcurCCtassmg{23} & \hatcurCCtassmg{24} & APASS\tablenotemark{c} \\
~~~~$r$ (mag)\dotfill               &  \hatcurCCtassmreccen{22}  & \hatcurCCtassmr{23} & \hatcurCCtassmr{24} & APASS\tablenotemark{c} \\
~~~~$i$ (mag)\dotfill               &  \hatcurCCtassmieccen{22}  & \hatcurCCtassmi{23} & \hatcurCCtassmi{24} & APASS\tablenotemark{c} \\
~~~~$J$ (mag)\dotfill               &  \hatcurCCtwomassJmageccen{22} & \hatcurCCtwomassJmag{23} & \hatcurCCtwomassJmag{24} & 2MASS           \\
~~~~$H$ (mag)\dotfill               &  \hatcurCCtwomassHmageccen{22} & \hatcurCCtwomassHmag{23} & \hatcurCCtwomassHmag{24} & 2MASS           \\
~~~~$K_s$ (mag)\dotfill             &  \hatcurCCtwomassKmageccen{22} & \hatcurCCtwomassKmag{23} & \hatcurCCtwomassKmag{24} & 2MASS           \\
\sidehead{Derived properties}
~~~~$\mstar$ ($\msun$)\dotfill      &  \hatcurISOmlongeccen{22}   & \hatcurISOmlong{23} & \hatcurISOmlong{24} & YY+$\rhostar$+ZASPE \tablenotemark{d}\\
~~~~$\rstar$ ($\rsun$)\dotfill      &  \hatcurISOrlongeccen{22}   & \hatcurISOrlong{23} & \hatcurISOrlong{24} & YY+$\rhostar$+ZASPE         \\
~~~~$\loggstar$ (cgs)\dotfill       &  \hatcurISOloggeccen{22}    & \hatcurISOlogg{23} & \hatcurISOlogg{24} & YY+$\rhostar$+ZASPE         \\
~~~~$\rhostar$ (\gcmc) \tablenotemark{e}\dotfill       &  \hatcurLCrhoeccen{22}    & \hatcurLCrho{23} & \hatcurLCrho{24} & Light curves         \\
~~~~$\rhostar$ (\gcmc) \tablenotemark{e}\dotfill       &  \hatcurISOrhoeccen{22}    & \hatcurISOrho{23} & \hatcurISOrho{24} & YY+Light curves+ZASPE         \\
~~~~$\lstar$ ($\lsun$)\dotfill      &  \hatcurISOlumeccen{22}     & \hatcurISOlum{23} & \hatcurISOlum{24} & YY+$\rhostar$+ZASPE         \\
~~~~$M_V$ (mag)\dotfill             &  \hatcurISOmveccen{22}      & \hatcurISOmv{23} & \hatcurISOmv{24} & YY+$\rhostar$+ZASPE         \\
~~~~$M_K$ (mag,\hatcurjhkfilset{22})\dotfill &  \hatcurISOMKeccen{22} & \hatcurISOMK{23} & \hatcurISOMK{24} & YY+$\rhostar$+ZASPE         \\
~~~~Age (Gyr)\dotfill               &  $\cdots$ \tablenotemark{f}     & \hatcurISOage{23} & \hatcurISOage{24} & YY+$\rhostar$+ZASPE         \\
~~~~$A_{V}$ (mag)\dotfill               &  \hatcurXAveccen{22}     & \hatcurXAv{23} & \hatcurXAv{24} & YY+$\rhostar$+ZASPE         \\
~~~~Distance (pc)\dotfill           &  \hatcurXdistredeccen{22}\phn  & \hatcurXdistred{23} & \hatcurXdistred{24} & YY+$\rhostar$+ZASPE\\ [-1.5ex]
\enddata
\tablecomments{
For each system we adopt the class of model which has the highest Bayesian evidence from among those tested. For \hatcur{23} and \hatcur{24} the adopted parameters come from a fit in which the orbit is assumed to be circular. For \hatcur{22} the eccentricity is allowed to vary. 
}
\tablenotetext{a}{
    ZASPE = Zonal Atmospherical Stellar Parameter Estimator routine
    for the analysis of high-resolution spectra (Brahm et al.~2015, in
    preparation), applied to the FEROS spectra of \hatcur{22}, \hatcur{23} and \hatcur{24}. These
    parameters rely primarily on ZASPE, but have a small dependence
    also on the iterative analysis incorporating the isochrone search
    and global modeling of the data.
}
\tablenotetext{b}{
    This is based on CORALIE for \hatcur{22} and FEROS for \hatcur{23}
    and \hatcur{24}.  The error on $\gamma_{\rm RV}$ is determined
    from the orbital fit to the radial velocity measurements, and does not include
    the systematic uncertainty in transforming the velocities to the
    IAU standard system. The velocities have not been corrected for gravitational redshifts.
} \tablenotetext{c}{
    From APASS DR6  \citep{henden:2009} for as
    listed in the UCAC 4 catalog \citep{zacharias:2012:ucac4}.  
}
\tablenotetext{d}{
    \hatcurisoshort{22}+\rhostar+ZASPE = Based on the \hatcurisoshort{22}
    isochrones \citep{\hatcurisocite{22}}, \rhostar\ as a luminosity
    indicator, and the ZASPE results.
}
\tablenotetext{e}{
    In the case of $\rhostar$ we list two values. The first value is
    determined from the global fit to the light curves and radial velocity data,
    without imposing a constraint that the parameters match the
    stellar evolution models. The second value results from
    restricting the posterior distribution to combinations of
    $\rhostar$+$\teffstar$+$\feh$ that match to a \hatcurisoshort{22}
    stellar model.
}
\tablenotetext{f}{Omitted due to large uncertainty. Isochrone models (c.f. Figure \ref{fig:iso}) are unable to constrain age for this system.}
\ifthenelse{\boolean{emulateapj}}{
    \end{deluxetable*}
}{
    \end{deluxetable}
}

\subsection{Excluding blend scenarios}
\label{sec:blend}

In order to exclude blend scenarios we carried out an analysis
following \citet{hartman:2012:hat39hat41}. We model the
available photometric data (including light curves and catalog
broad-band photometric measurements) for each object as a blend
between an eclipsing binary star system and a third star along the
line of sight. The physical properties of the stars are constrained
using the Padova isochrones \citep{girardi:2000}, while we also
require that the brightest of the three stars in the blend have
atmospheric parameters consistent with those measured with ZASPE. We
also simulate composite cross-correlation functions (CCFs) and use
them to predict radial velocities and BSs for each blend scenario considered. For
\hatcur{22} all blend scenarios tested can be rejected with greater
than $3\sigma$ confidence, based on the photometry alone. Those models
which cannot be rejected with at least $5\sigma$ confidence would have
obviously double-lined spectra, and would also have BS variations in
excess of 1\,\kms. For \hatcur{23} all blend scenarios tested can be
rejected with greater than $3.3\sigma$ confidence based on the
photometry. Although some of the models which cannot be rejected with
at least $5\sigma$ confidence do predict low amplitude BS and radial velocity
variations, the simulated radial velocities do not reproduce the sinusoidal
variation with the orbital period that is clearly detected (Figure \ref{fig:rvbis}). For \hatcur{24} all blend
scenarios tested can be rejected with greater than $4\sigma$
confidence based on the photometry. Those that cannot be rejected with
at least $5\sigma$ confidence yield large amplitude radial velocity and BS
variations in excess of 1\,\kms. We conclude that all three objects
are transiting planet systems, however we cannot exclude the
possibility that one or more of these objects is an unresolved binary
stellar system with one component hosting a short period transiting
planet (see Section \ref{sec:luckyimaging}). For the remainder of the paper we assume that these are all
single stars with transiting planets, but we note that the radii, and
potentially the masses, of the planets would be larger than what we
infer here if subsequent observations reveal binary star companions.

\subsection{Global modeling of the data}
\label{sec:globmod}

We modeled the HATSouth photometry, the follow-up photometry, and the
high-precision radial velocity measurements following
\citet{pal:2008:hat7,bakos:2010:hat11,hartman:2012:hat39hat41}. We fit
\citet{mandel:2002} transit models to the light curves, allowing for a
dilution of the HATSouth transit depth as a result of blending from
neighboring stars and over-correction by the trend-filtering
method. To correct for systematic errors in the follow-up light curves we include in our model, for each event, a quadratic trend in time. Linear trends with up to three parameters describing the position and shape of the PSF are also included to compensate for any systematic effects due to poor guiding or PSF shape changes throughout the transit observation. We fit Keplerian orbits to the radial velocity curves allowing the zero-point
for each instrument to vary independently in the fit, and allowing for
radial velocity jitter which we also vary as a free parameter for each
instrument. We used a Differential Evolution Markov Chain Monte Carlo
procedure to explore the fitness landscape and to determine the
posterior distribution of the parameters.

We tried to both fit fixed circular orbit models and models with the eccentricity as a free parameter and then used the method of \citet{weinberg:2013} to estimate the Bayesian evidence for each scenario. We find a higher evidence for a non-circular orbital solution (by a factor of 80) for \hatcurb{22} and find the most likely eccentricity to be $e = \hatcurRVecceneccen{22}$. For \hatcurb{23} and \hatcurb{24} the fixed circular orbit models have the higher evidence; for \hatcur{23} the
circular model has an evidence that is $20$ times greater than the
free-eccentricity model, while for \hatcurb{24} the circular model has
an evidence that is $70$ times greater. We therefore adopt the
parameters from the circular orbit models for these two systems, placing 95\% confidence upper limits on their eccentricity as $e \hatcurRVeccentwosiglimeccen{23}$ and $e \hatcurRVeccentwosiglimeccen{24}$, respectively. The results of the fitting routines for each planet can be found in \reftabl{planetparam}.

The grazing nature of \hatcurb{23} naturally leads to a higher uncertainty on the determination of specific parameters typically constrained by the depth and shape of the transit. As a consequence of this, in Table \ref{tab:planetparam} we have indicated the best fit results (with the corresponding $1\sigma$ confidence) as well as lower limits for selected parameters. In particular, the poor constraint on the impact parameter from the photometric follow-up \lcs\, should be noted. This in turn affects the estimates of the orbital inclination, the planet radius and orbital separation. We also note that the larger uncertainties in the bottom \lc, for this planet are due to the fact that the data were taken in the z band. Whilst inclusion of these data in the transit fitting process has a minimal impact on the parameter estimation, the blend scenario analysis process (c.f. Section \ref{sec:blend}) benefited from this data set. The consistency between this partial transit with those observing in the i band limits the range of blend models which can fit the observations. Nevertheless, significantly higher precision multi-band follow-up would be required to improve the characterization of this system.



\section{Discussion}
\label{sec:discussion}

\begin{figure}
{
\centering
\includegraphics[width={\linewidth}]{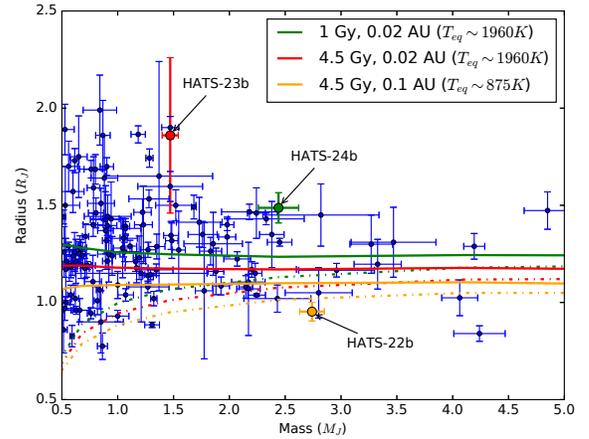}
}
\caption{
     Mass-radius relation for \hjs, defined as those planets with masses higher than $0.5 \mjup$ and periods shorter than 10 days. We show theoretical models for planet structures from \cite{fortney:2007} for each of the three planets announced in this paper for both no core (solid lines) and $100 \mearth$ core (dashed lines) scenarios. The new HATSouth planets are indicated. 4.5 Gy 0.1 AU models are shown for comparison with \hatcurb{22} (yellow lines). The 4.5Gy 0.02 AU models (red lines) are used as equivalent examples for \hatcurb{23} and the models for \hatcurb{24} (1 Gy 0.02 AU) is shown in green lines. 
}
\label{fig:massradius}
\end{figure}

\begin{figure}
{
\centering
\includegraphics[width={\linewidth}]{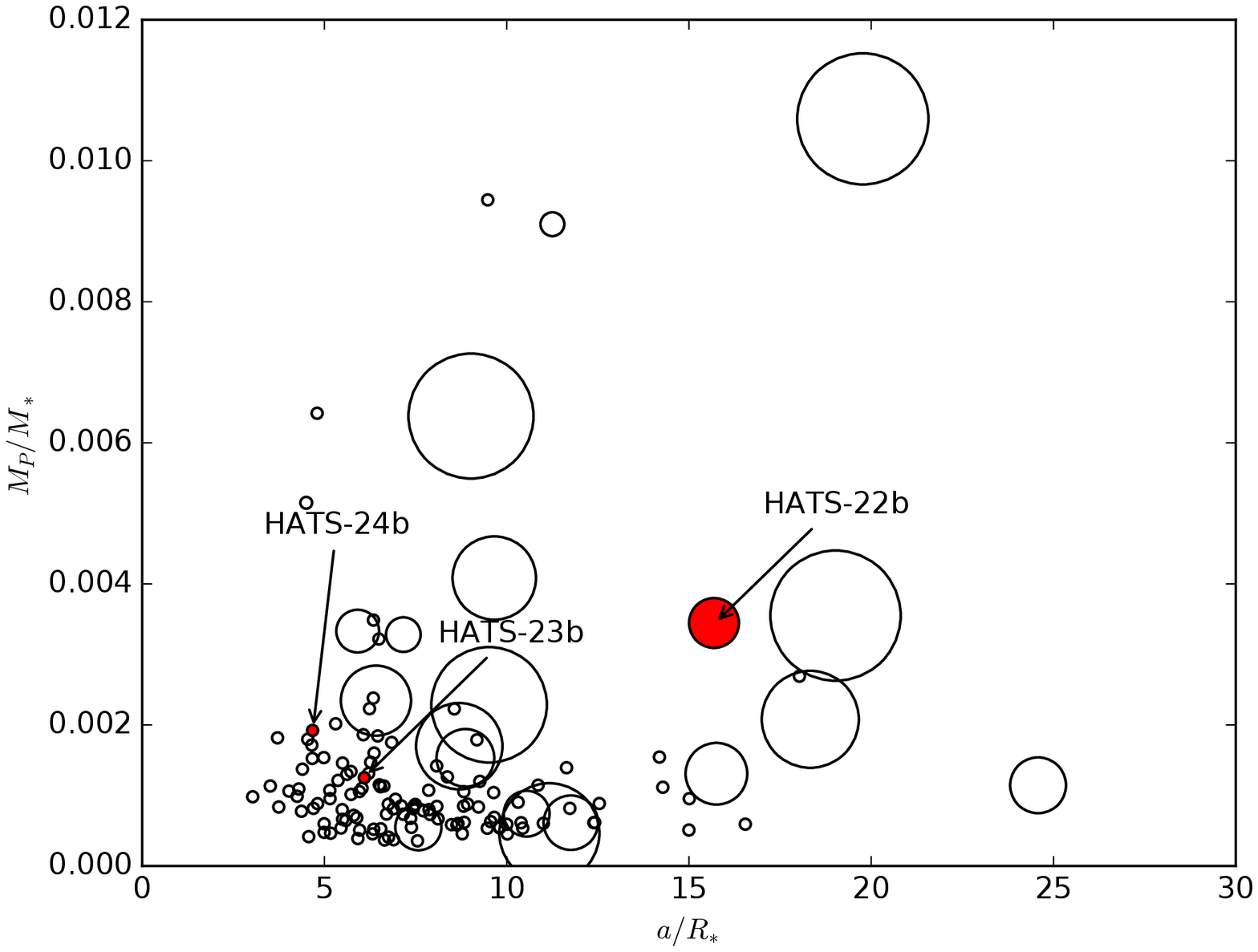}
}
\caption{
     Planet to star mass ratio as a function of the ratio between the orbital separation and the stellar radius for known \hjs. We have labelled our new discoveries in the plot, which are marked in filled red symbols. The dot sizes correspond to the planets' measured eccentricity ranging from zero (smallest size) up to a value of 0.562. 
}
\label{fig:arsmratio}
\end{figure}

In this work we report the discovery of three new moderately high-mass \hjs by the HATSouth survey: \hatcurb{22}, \hatcurb{23} and \hatcurb{24}. These planets add to the growing numbers of known \hjs and provide vital additional insights into the formation and distribution of short-period massive planets. In Figures \ref{fig:massradius} and \ref{fig:arsmratio} we show these discoveries in the context of known planets with masses higher than $0.5 \mjup$ and less than 10 day orbital periods \footnote{Previously known planets taken from the NASA Exoplanet Archive at http://exoplanetarchive.ipac.caltech.edu/ on 02/02/2017}.

\subsection{Mass-radius relation}
\label{sec:massradius}

In Figure \ref{fig:massradius} we plot a mass-radius relation for known \hjs, highlighting our three new discovered planets. Additionally, this plot contains a selection of predicted mass-radius relations from \cite{fortney:2007} that are closest to the conditions of each of these planets. For \hatcurb{23} and \hatcurb{24} (red and green curves respectively) we have selected those models that most closely matched the determined age and orbital separation of the planets (4.5 and 1 Gy respectively at 0.02 au separation) and with orbital periods of less than 10 days. These models assume a solar analogue host star and solar luminosity. For \hatcurb{22}, however, due to the lower luminosity of the host star, we have selected the curve that most closely matches the expected equilibrium temperature $T_{eq}$ for this planet, a 4.5 Gy model at 0.1 au, which is equivalent to a $T_{eq}$ of approximately 875K, thereby matching the model to the planet's conditions. In all cases, we plot the two extreme limits of planetary core mass: no core (solid line) and a $100 \mearth$ core model (dashed line). Despite the fact that none of the target's radii fall within their equivalent predicted mass-radius range from the model curves, \hatcurb{22} is consistent with a high-mass ($100 \mearth$) core at a 95\% confidence level. This places this planet among the twenty highest bulk density \hjs known to date, defined as those with masses higher than $0.5 \mjup$ and less than 10 day orbital periods. While we are not able to confidently make any conclusions regarding the structure of grazing transiting planet \hatcurb{23}, we note that even assuming the lower limit on the radius of this planet at $1.31 \mjup$ it is still more likely that this is an example of a low core mass \hj. 

Perhaps the most interesting case in this context is that of \hatcurb{24}. The radius of this planet is estimated more than $3 \sigma$ above the model line for a pure helium-hydrogen planet at approximately this age and orbital separation. The short period and young age of the host star leads to a high equilibrium temperature ($T_{eq} = 2067 \pm 39 K$), which puts it in a similar regime to other inflated \hjs (e.g. WASP-71b \citep{smith:2013} and HATS-35b \citep{valborro:2016}). As noted by \cite{marley:2007} and \cite{fortney:2007}, the physical properties of giant planets at young ages are uncertain, and the model shown does not include considerations on the formation mechanism. The somewhat inflated radius of this planet may be due to a combination of mechanisms discussed in Section \ref{sec:introduction}. Additionally, the apparent higher radius may be due to an ongoing evaporation of the upper layers of the planet's atmosphere as current models are unable to explain the radius of \hatcurb{24}. This factor, coupled with the fact that the host star is moderately bright ($V = 12.830 \pm 0.010$) and the large transit depth signal, makes this a good target for further wavelength dependent transmission studies with large-class telescopes. Assuming a hydrogen dominated atmosphere, the scale height $H$ for this planet is estimated at $314.8 \pm 32.8$ km, a value comparable to other inflated hot-Jupiters. Therefore, a transmission spectroscopic signal equivalent to 5 scale heights would be detectable at a level of $\sim$250ppm, well within the capability of current and future facilities. 

\subsection{Eccentricity and tidal circularisation}

A comparative study between the three announced planets and previously known \hjs is useful as the orbital eccentricity is an important parameter thought to be highly related to planetary migration both in disk interaction and planet scattering scenarios. As discussed in Section \ref{sec:introduction}, eccentric orbits are found preferentially for higher mass planets \citep[e.g.][]{southworth:2009} and a higher the orbital separation seems to correspond with non-zero eccentricity \citep{pont:2011}. This supports the fact that a high mass planet at a large separation should require more time for tidal circularisation. 

Our analysis (discussed in Section \ref{sec:globmod}) finds \hatcurb{22} to be the only planet of the three with a likely non-zero eccentricity of $e = \hatcurRVecceneccen{22}$. The distinction between \hatcurb{22} and the two other planets discovered in this work is made more evident in Figure \ref{fig:arsmratio}, in which we have plotted the planet to stellar mass ratio as a function of the ratio between the orbital separation and the stellar radius ($a/R_*$). These parameters are the dominant factors in determining the tidal circularisation timescale of planetary orbits \citep{duffell:2015,ogilvie:2014} in which the mass ratio is proportional to the circularisation timescale. This is due to the fact that more massive planets are more likely to carve a large gap in the protoplanetary disk, leading to less efficient circularisation due to disk interaction. Thus, these planets should remain in moderate non-zero eccentric orbits for a longer time. In Figure \ref{fig:arsmratio} the dot sizes correspond to the eccentricity values, ranging from zero to 0.562, where the concentration of low or zero eccentricity planets, can be found for low values of both plotted parameters. \hatcurb{22} can be seen occupying a region of parameter space clearly distinguished from that of \hatcurb{23} and \hatcurb{24}, suggesting that the eccentric orbit of \hatcurb{22} may indeed be a result of insufficient time for full tidal circularisation. However, several examples can still be found in this figure which are inconsistent with this picture. The case of CoRoT-27b \citep{parviainen:2014} is of a non-eccentric orbit planet that can be seen as the point with a mass ratio just under 0.01. Despite the moderate age of this system ($4.21 \pm 2.72$ Gyr), a planet with such a high mass ratio is still predicted to be found in an eccetric orbit. On the other hand, several cases can also be found with low values of both parameters plotted in Figure \ref{fig:arsmratio} that still presently show detectable orbital eccentricity, suggesting that there are other mechanisms, besides  tidal circularisation, determining the eccentricity distribution of exoplanets.

%
\ifthenelse{\boolean{emulateapj}}{
    \begin{deluxetable*}{lccc}
}{
    \begin{deluxetable}{lccc}
}
\tabletypesize{\scriptsize}
\tablecaption{Orbital and planetary parameters for \hatcurb{22}, \hatcurb{23} and \hatcurb{24}\label{tab:planetparam}}
\tablehead{
    \multicolumn{1}{c}{} &
    \multicolumn{1}{c}{\bf HATS-22b} &
    \multicolumn{1}{c}{\bf HATS-23b} &
    \multicolumn{1}{c}{\bf HATS-24b} \\ 
    \multicolumn{1}{c}{~~~~~~~~~~~~~~~Parameter~~~~~~~~~~~~~~~} &
    \multicolumn{1}{c}{Value} &
    \multicolumn{1}{c}{Value} &
    \multicolumn{1}{c}{Value}
}
\startdata
\noalign{\vskip -3pt}
\sidehead{\Lc{} parameters}
~~~$P$ (days)             \dotfill    & $\hatcurLCPeccen{22}$ & $\hatcurLCP{23}$ & $\hatcurLCP{24}$ \\
~~~$T_c$ (${\rm BJD}$)    
      \tablenotemark{a}   \dotfill    & $\hatcurLCTeccen{22}$ & $\hatcurLCT{23}$ & $\hatcurLCT{24}$ \\
~~~$T_{14}$ (days)
      \tablenotemark{a}   \dotfill    & $\hatcurLCdureccen{22}$ & $\hatcurLCdur{23}$ & $\hatcurLCdur{24}$ \\
~~~$T_{12} = T_{34}$ (days)
      \tablenotemark{a}   \dotfill    & $\hatcurLCingdureccen{22}$ & $\hatcurLCingdur{23}$ & $\hatcurLCingdur{24}$ \\
~~~$\arstar$              \dotfill    & $\hatcurPPareccen{22}$ & $\hatcurPPar{23}$ & $\hatcurPPar{24}$ \\
~~~$\zrstar$ \tablenotemark{b}             \dotfill    & $\hatcurLCzetaeccen{22}$\phn & $\hatcurLCzeta{23}$\phn & $\hatcurLCzeta{24}$\phn \\
~~~$\rpl/\rstar$          \dotfill    & $\hatcurLCrprstareccen{22}$ & $\hatcurLCrprstar{23}$ & $\hatcurLCrprstar{24}$ \\
~~~$b^2$                  \dotfill    & $\hatcurLCbsqeccen{22}$ & $\hatcurLCbsq{23}$ & $\hatcurLCbsq{24}$ \\
~~~$b^2$ lower limit \tablenotemark{c}     \dotfill    & $\cdots$ & $\hatcurLCbsqsiglowerlim{23}$ & $\cdots$ \\
~~~$b \equiv a \cos i/\rstar$
                          \dotfill    & $\hatcurLCimpeccen{22}$ & $\hatcurLCimp{23}$ & $\hatcurLCimp{24}$ \\
~~~$b$ lower limit \tablenotemark{c}
                          \dotfill    & $\cdots$ & $\hatcurLCimpsiglowerlim{23}$ & $\cdots$ \\
~~~$i$ (deg)              \dotfill    & $\hatcurPPieccen{22}$\phn & $\hatcurPPi{23}$\phn & $\hatcurPPi{24}$\phn \\
~~~$i$ upper limit (deg) \tablenotemark{c}    \dotfill    & $\cdots$\phn & $\hatcurPPitwosigupperlim{23}$\phn & $\cdots$\phn \\

\sidehead{HATSouth blend factors \tablenotemark{d}}
~~~Blend factor \dotfill & $\hatcurLCiblend{22}$ & $\hatcurLCiblend{23}$ & $\hatcurLCiblend{24}$ \\

\sidehead{Limb-darkening coefficients \tablenotemark{e}}
~~~$c_1,r$                  \dotfill    & $\hatcurLBireccen{22}$ & $\hatcurLBir{23}$ & $\hatcurLBir{24}$ \\
~~~$c_2,r$                  \dotfill    & $\hatcurLBiireccen{22}$ & $\hatcurLBiir{23}$ & $\hatcurLBiir{24}$ \\
~~~$c_1,i$                  \dotfill    & $\hatcurLBiieccen{22}$ & $\hatcurLBii{23}$ & $\hatcurLBii{24}$ \\
~~~$c_2,i$                  \dotfill    & $\hatcurLBiiieccen{22}$ & $\hatcurLBiii{23}$ & $\hatcurLBiii{24}$ \\
~~~$c_1,z$                  \dotfill    & $\cdots$ & $\hatcurLBiz{23}$ & $\cdots$ \\
~~~$c_2,z$                  \dotfill    & $\cdots$ & $\hatcurLBiiz{23}$ & $\cdots$ \\

\sidehead{radial velocity parameters}
~~~$K$ (\ms)              \dotfill    & $\hatcurRVKeccen{22}$\phn\phn & $\hatcurRVK{23}$\phn\phn & $\hatcurRVK{24}$\phn\phn \\
%
~~~$e$ \tablenotemark{f}               \dotfill    & $\hatcurRVecceneccen{22}$ & $\hatcurRVeccentwosiglimeccen{23}$ & $\hatcurRVeccentwosiglimeccen{24}$ \\
~~~$\omega$ (deg) \dotfill    & $\hatcurRVomegaeccen{22}$ & $\cdots$ & $\cdots$ \\
~~~$\sqrt{e}\cos\omega$               \dotfill    & $\hatcurRVrkeccen{22}$ & $\cdots$ & $\cdots$ \\
~~~$\sqrt{e}\sin\omega$               \dotfill    & $\hatcurRVrheccen{22}$ & $\cdots$ & $\cdots$ \\
~~~$e\cos\omega$               \dotfill    & $\hatcurRVkeccen{22}$ & $\cdots$ & $\cdots$ \\
~~~$e\sin\omega$               \dotfill    & $\hatcurRVheccen{22}$ & $\cdots$ & $\cdots$ \\
~~~radial velocity jitter FEROS (\ms) \tablenotemark{g}       \dotfill    & \hatcurRVjitterAeccen{22} & \hatcurRVjitter{23} & \hatcurRVjitterA{24} \\
~~~radial velocity jitter HARPS (\ms)        \dotfill    & \hatcurRVjitterCeccen{22} & $\cdots$ & \hatcurRVjitterC{24} \\
~~~radial velocity jitter CORALIE (\ms)        \dotfill    & \hatcurRVjitterCeccen{22} & $\cdots$ & \hatcurRVjitterB{24} \\
~~~radial velocity jitter CYCLOPS2+UCLES (\ms)        \dotfill    & $\cdots$ & $\cdots$ & \hatcurRVjitterD{24} \\

\sidehead{Planetary parameters}
~~~$\mpl$ ($\mjup$)       \dotfill    & $\hatcurPPmlongeccen{22}$ & $\hatcurPPmlong{23}$ & $\hatcurPPmlong{24}$ \\
~~~$\rpl$ ($\rjup$)       \dotfill    & $\hatcurPPrlongeccen{22}$ & $\hatcurPPrlong{23}$ & $\hatcurPPrlong{24}$ \\
~~~$\rpl$ lower limit ($\rjup$) \tablenotemark{c}      \dotfill    & $\cdots$ & $\hatcurPPrtwosiglowerlim{23}$ & $\cdots$ \\
~~~$C(\mpl,\rpl)$
    \tablenotemark{h}     \dotfill    & $\hatcurPPmrcorreccen{22}$ & $\hatcurPPmrcorr{23}$ & $\hatcurPPmrcorr{24}$ \\
~~~$\rhopl$ (\gcmc)       \dotfill    & $\hatcurPPrhoeccen{22}$ & $\hatcurPPrho{23}$ & $\hatcurPPrho{24}$ \\
~~~$\log g_p$ (cgs)       \dotfill    & $\hatcurPPloggeccen{22}$ & $\hatcurPPlogg{23}$ & $\hatcurPPlogg{24}$ \\
~~~$a$ (AU)               \dotfill    & $\hatcurPPareleccen{22}$ & $\hatcurPParel{23}$ & $\hatcurPParel{24}$ \\
~~~$T_{\rm eq}$ (K)        \dotfill   & $\hatcurPPteffeccen{22}$ & $\hatcurPPteff{23}$ & $\hatcurPPteff{24}$ \\
~~~$\Theta$ \tablenotemark{i} \dotfill & $\hatcurPPthetaeccen{22}$ & $\hatcurPPtheta{23}$ & $\hatcurPPtheta{24}$ \\
%
~~~$\log_{10}\langle F \rangle$ (cgs) \tablenotemark{j}
                          \dotfill    & $\hatcurPPfluxavglogeccen{22}$ & $\hatcurPPfluxavglog{23}$ & $\hatcurPPfluxavglog{24}$ \\ [-1.5ex]
\enddata
\tablenotetext{a}{
    Times are in Barycentric Julian Date calculated directly from UTC {\em without} correction for leap seconds.
    \ensuremath{T_c}: Reference epoch of
    mid transit that minimizes the correlation with the orbital
    period.
    \ensuremath{T_{14}}: total transit duration, time
    between first to last contact;
    \ensuremath{T_{12}=T_{34}}: ingress/egress time, time between first
    and second, or third and fourth contact.
}
\tablecomments{
For each system we adopt the class of model which has the highest Bayesian evidence from among those tested. For \hatcurb{23} and \hatcurb{24} the adopted parameters come from a fit in which the orbit is assumed to be circular. For \hatcurb{22} the eccentricity is allowed to vary. 
}
\tablenotetext{b}{
   Reciprocal of the half duration of the transit used as a jump parameter in our Markov chain Monte Carlo (MCMC) analysis in place of $\arstar$. It is related to $\arstar$ by the expression $\zrstar = \arstar(2\pi(1+e\sin\omega))/(P\sqrt{1-b^2}\sqrt{1-e^2})$ \citep{bakos:2010:hat11}.
}
\tablenotetext{c}{
    The grazing transits of \hatcurb{23} mean that we cannot place a strong upper limit on the impact parameter. For this system we also provide 95\% confidence lower limits on $b^{2}$, $b$ and $\rpl$, and the 95\% confidence upper limit on $i$.
}
\tablenotetext{d}{
    Scaling factor applied to the model transit that is fit to the HATSouth light curves. This factor accounts for dilution of the transit due to blending from neighboring stars and over-filtering of the light curve.  These factors are varied in the fit.
}
\tablenotetext{e}{
    Values for a quadratic law, adopted from the tabulations by
    \cite{claret:2004} according to the spectroscopic (ZASPE) parameters
    listed in \reftabl{stellar}.
}
\tablenotetext{f}{
    For fixed circular orbit models we list
    the 95\% confidence upper limit on the eccentricity determined
    when $\sqrt{e}\cos\omega$ and $\sqrt{e}\sin\omega$ are allowed to
    vary in the fit.
}
\tablenotetext{g}{
    Term added in quadrature to the formal radial velocity uncertainties for each
    instrument. This is treated as a free parameter in the fitting
    routine. In cases where the jitter is consistent with zero, we
    list its 95\% confidence upper limit.
}
\tablenotetext{h}{
    Correlation coefficient between the planetary mass \mpl\ and radius
    \rpl\ estimated from the posterior parameter distribution.
}
\tablenotetext{i}{
    The Safronov number is given by $\Theta = \frac{1}{2}(V_{\rm
    esc}/V_{\rm orb})^2 = (a/\rpl)(\mpl / \mstar )$
    \citep[see][]{hansen:2007}.
}
\tablenotetext{j}{
    Incoming flux per unit surface area, averaged over the orbit assuming a circular geometry.
}
\ifthenelse{\boolean{emulateapj}}{
    \end{deluxetable*}
}{
    \end{deluxetable}
}


\acknowledgements 

Development of the HATSouth project was funded by NSF MRI grant
NSF/AST-0723074, operations have been supported by NASA grants NNX09AB29G and NNX12AH91H, and
follow-up observations receive partial support from grant
NSF/AST-1108686.
A.J.\ acknowledges support from FONDECYT project 1130857, BASAL CATA PFB-06, and project IC120009 ``Millennium Institute of Astrophysics (MAS)'' of the Millenium Science Initiative, Chilean Ministry of Economy. R.B.\ and N.E.\ are supported by CONICYT-PCHA/Doctorado Nacional. R.B.\ and N.E.\ acknowledge additional support from project IC120009 ``Millenium Institute of Astrophysics  (MAS)'' of the Millennium Science Initiative, Chilean Ministry of Economy.  V.S.\ acknowledges support form BASAL CATA PFB-06.  M.R.\ acknowledges support from FONDECYT postdoctoral fellowship 3120097.
This work is based on observations made with ESO Telescopes at the La
Silla Observatory.
This paper also uses observations obtained with facilities of the Las
Cumbres Observatory Global Telescope.
Work at the Australian National University is supported by ARC Laureate
Fellowship Grant FL0992131.
We acknowledge the use of the AAVSO Photometric All-Sky Survey (APASS),
funded by the Robert Martin Ayers Sciences Fund, and the SIMBAD
database, operated at CDS, Strasbourg, France.
Operations at the MPG~2.2\,m Telescope are jointly performed by the
Max Planck Gesellschaft and the European Southern Observatory.  The
imaging system GROND has been built by the high-energy group of MPE in
collaboration with the LSW Tautenburg and ESO\@.  We thank R\'egis
Lachaume for his technical assistance during the observations at the
MPG~2.2\,m Telescope. We thank Helmut Steinle and Jochen Greiner for
supporting the GROND observations presented in this manuscript.
We are grateful to P.Sackett for her help in the early phase of the
HATSouth project.
This research has made use of the NASA/ IPAC Infrared Science Archive, which is operated by the Jet Propulsion Laboratory, California Institute of Technology, under contract with the National Aeronautics and Space Administration.
Observing time were obtained through proposals CN2013A-171,
CN2013B-55, CN2014A-104, CN2014B-57, CN2015A-51 and ESO 096.C-0544.

\bibliographystyle{apj}
\bibliography{hatsbib}

\clearpage
\LongTables

%
%
\tabletypesize{\scriptsize}
\ifthenelse{\boolean{emulateapj}}{
    \begin{deluxetable*}{lrrrrrl}
}{
    \begin{deluxetable}{lrrrrrl}
}
\tablewidth{0pc}
\tablecaption{
    Relative radial velocities and bisector spans for \hatcur{22} and
    \hatcur{23}.
    \label{tab:rvs}
}
\tablehead{
    \colhead{BJD} &
    \colhead{RV\tablenotemark{a}} &
    \colhead{\ensuremath{\sigma_{\rm RV}}\tablenotemark{b}} &
    \colhead{BS} &
    \colhead{\ensuremath{\sigma_{\rm BS}}} &
    \colhead{Phase} &
    \colhead{Instrument}\\
    \colhead{\hbox{(2,450,000$+$)}} &
    \colhead{(\ms)} &
    \colhead{(\ms)} &
    \colhead{(\ms)} &
    \colhead{(\ms)} &
    \colhead{} &
    \colhead{}
}
\startdata
\multicolumn{7}{c}{\bf HATS-22} \\
\hline\\
 $ 7072.74688 $ & $   432.36 $ & $     8.00 $ & $  -25.0 $ & $   42.0 $ & $   0.765 $ & HARPS \\
 $ 7075.81290 $ & $  -237.86 $ & $    17.00 $ & $ -113.0 $ & $   32.0 $ & $   0.414 $ & CORALIE \\
 $ 7077.72945 $ & $   381.14 $ & $    18.00 $ & $   18.0 $ & $   32.0 $ & $   0.820 $ & CORALIE \\
 $ 7078.70639 $ & $   -23.86 $ & $    17.00 $ & $  -10.0 $ & $   32.0 $ & $   0.027 $ & CORALIE \\
 $ 7109.73325 $ & $   170.14 $ & $    22.00 $ & $  125.0 $ & $   43.0 $ & $   0.596 $ & CORALIE \\
 $ 7118.66166 $ & $   -41.64 $ & $    17.00 $ & $   -9.0 $ & $   74.0 $ & $   0.487 $ & HARPS \\
 $ 7119.60792 $ & $   336.19 $ & $    10.00 $ & $  -33.0 $ & $   13.0 $ & $   0.687 $ & FEROS \\
 $ 7119.67550 $ & $   341.36 $ & $    24.00 $ & $   33.0 $ & $   90.0 $ & $   0.701 $ & HARPS \\
 $ 7120.62753 $ & $   326.36 $ & $    16.00 $ & $  -31.0 $ & $   74.0 $ & $   0.903 $ & HARPS \\
 $ 7179.52938 $ & $  -274.86 $ & $    21.00 $ & $    0.0 $ & $   43.0 $ & $   0.375 $ & CORALIE \\
 $ 7180.51051 $ & $   156.14 $ & $    22.00 $ & $   38.0 $ & $   43.0 $ & $   0.583 $ & CORALIE \\
 $ 7181.51430 $ & $   449.14 $ & $    22.00 $ & $ -141.0 $ & $   43.0 $ & $   0.795 $ & CORALIE \\
 $ 7187.55141 $ & $  -235.81 $ & $    11.00 $ & $  -51.0 $ & $   15.0 $ & $   0.073 $ & FEROS \\
 $ 7187.57284 $ & $  -206.81 $ & $    11.00 $ & $   21.0 $ & $   15.0 $ & $   0.078 $ & FEROS \\
 $ 7187.59426 $ & $  -276.81 $ & $    12.00 $ & $  -53.0 $ & $   16.0 $ & $   0.082 $ & FEROS \\

\cutinhead{\bf HATS-23}
 $ 7181.62180 $ & $  -182.31 $ & $    18.00 $ & $   77.0 $ & $   22.0 $ & $   0.344 $ & FEROS \\
 $ 7182.71745 $ & $   175.69 $ & $    19.00 $ & $   72.0 $ & $   23.0 $ & $   0.851 $ & FEROS \\
 $ 7183.66969 $ & \nodata      & \nodata      & $   68.0 $ & $   32.0 $ & $   0.292 $ & FEROS \\
 $ 7184.75104 $ & \nodata      & \nodata      & $    7.0 $ & $   32.0 $ & $   0.793 $ & FEROS \\
 $ 7185.72406 $ & $  -226.31 $ & $    16.00 $ & $   77.0 $ & $   19.0 $ & $   0.243 $ & FEROS \\
 $ 7186.85044 $ & $   207.69 $ & $    19.00 $ & $   73.0 $ & $   23.0 $ & $   0.764 $ & FEROS \\
 $ 7187.85244 $ & $  -188.31 $ & $    16.00 $ & $   24.0 $ & $   19.0 $ & $   0.228 $ & FEROS \\
 $ 7189.63018 $ & $   -47.31 $ & $    14.00 $ & $  -17.0 $ & $   17.0 $ & $   0.051 $ & FEROS \\
 $ 7190.81426 $ & $   139.69 $ & $    17.00 $ & $   10.0 $ & $   21.0 $ & $   0.599 $ & FEROS \\
 $ 7191.68515 $ & $   -20.31 $ & $    16.00 $ & $   55.0 $ & $   19.0 $ & $   0.002 $ & FEROS \\
    
\cutinhead{\bf HATS-24}
 $ 7118.75825 $ & $   -76.82 $ & $    69.00 $ & $   14.0 $ & $  106.0 $ & $   0.537 $ & HARPS \\
 $ 7119.77652 $ & $  -338.82 $ & $    38.00 $ & $ -132.0 $ & $   56.0 $ & $   0.292 $ & HARPS \\
 $ 7149.08745 $\tablenotemark{c} & $  -379.95 $ & $    40.30 $ & \nodata      & \nodata      & $   0.028 $ & CYCLOPS \\
 $ 7149.10350 $\tablenotemark{c} & $  -384.45 $ & $    27.60 $ & \nodata      & \nodata      & $   0.040 $ & CYCLOPS \\
 $ 7149.11954 $ & $   -44.25 $ & $    61.00 $ & \nodata      & \nodata      & $   0.052 $ & CYCLOPS \\
 $ 7150.09608 $ & $   393.15 $ & $    56.20 $ & \nodata      & \nodata      & $   0.776 $ & CYCLOPS \\
 $ 7150.11204 $ & $   242.95 $ & $    55.80 $ & \nodata      & \nodata      & $   0.788 $ & CYCLOPS \\
 $ 7150.12799 $ & $   395.45 $ & $    41.60 $ & \nodata      & \nodata      & $   0.799 $ & CYCLOPS \\
 $ 7152.06340 $ & $  -481.85 $ & $    33.80 $ & \nodata      & \nodata      & $   0.235 $ & CYCLOPS \\
 $ 7152.07872 $ & $  -509.25 $ & $    21.00 $ & \nodata      & \nodata      & $   0.246 $ & CYCLOPS \\
 $ 7152.09403 $ & $  -508.95 $ & $    17.90 $ & \nodata      & \nodata      & $   0.257 $ & CYCLOPS \\
 $ 7156.13898 $ & $  -798.25 $ & $    96.80 $ & \nodata      & \nodata      & $   0.257 $ & CYCLOPS \\
 $ 7156.15505 $ & $  -462.95 $ & $    48.40 $ & \nodata      & \nodata      & $   0.269 $ & CYCLOPS \\
 $ 7179.70211 $ & $   393.16 $ & $    71.00 $ & $  -20.0 $ & $   43.0 $ & $   0.731 $ & CORALIE \\
 $ 7180.63482 $ & $   -13.84 $ & $    62.00 $ & \nodata      & \nodata      & $   0.422 $ & CORALIE \\
 $ 7181.69032 $ & $  -465.84 $ & $    57.00 $ & $  -45.0 $ & $   35.0 $ & $   0.205 $ & CORALIE \\
 $ 7190.68048 $ & $   258.75 $ & $    25.00 $ & $   28.0 $ & $   13.0 $ & $   0.872 $ & FEROS \\
 $ 7191.75911 $ & $   383.75 $ & $    31.00 $ & $   81.0 $ & $   16.0 $ & $   0.672 $ & FEROS \\
 $ 7193.86712 $ & $  -370.25 $ & $    28.00 $ & $   56.0 $ & $   14.0 $ & $   0.235 $ & FEROS \\
 $ 7194.55840 $ & $   326.75 $ & $    24.00 $ & $   34.0 $ & $   13.0 $ & $   0.748 $ & FEROS \\

\enddata
\tablenotetext{a}{
    The zero-point of these velocities is arbitrary. An overall offset
    $\gamma_{\rm rel}$ fitted independently to the velocities from
    each instrument has been subtracted.
}
\tablenotetext{b}{
    Internal errors excluding the component of astrophysical jitter
    considered in \refsecl{globmod}.
}
\tablenotetext{c}{
    These observations were excluded from the analysis because the observations were (partially) obtained with the planet in transit, and thus may be affected by the Rossiter-McLaughlin effect. 
}
\ifthenelse{\boolean{rvtablelong}}{
    \tablecomments{
    }
}{
    \tablecomments{
    }
} 
\ifthenelse{\boolean{emulateapj}}{
    \end{deluxetable*}
}{
    \end{deluxetable}
}

\end{document}